\documentclass[aps,prb,reprint,onecolumn,10pt,superscriptaddress,nofootinbib,longbibliography]{revtex4-2}
\usepackage{tikz}
\usepackage{wrapfig}
\usepackage[T1]{fontenc}
\usepackage{lmodern,microtype,amsmath,amssymb,booktabs,array,xcolor}
\usepackage{hyperref}
\hypersetup{
	colorlinks=true,
	linkcolor=blue!45!black,
	citecolor=blue!45!black,
	urlcolor=blue!45!black
}

\usetikzlibrary{arrows.meta,calc}
\definecolor{ucolor}{RGB}{34,95,143}
\definecolor{tcolor}{RGB}{173,100,44}

\newcommand{\tr}{\operatorname{tr}}
\newcommand{\cZ}{\mathcal Z}

\makeatletter
\g@addto@macro\normalsize{\abovedisplayskip 6pt plus 2pt minus 2pt
  \belowdisplayskip 6pt plus 2pt minus 2pt
  \abovedisplayshortskip 3pt plus 1pt minus 1pt
  \belowdisplayshortskip 4pt plus 1pt minus 1pt
}
\newcommand{\compacttableofcontents}{\begingroup
  \let\l@subsection\@gobbletwo
  \let\l@subsubsection\@gobbletwo
  \def\l@f@section{\addpenalty{\@secpenalty}\addvspace{0.3em plus 1pt}}\tableofcontents
  \endgroup
}
\makeatother
\begin{document}

\title{Extensions of the Haldane gap proof: higher dimensions, near degeneracy, fermions and tensor-network states}

\author{Tarun Grover}

\affiliation{
	Department of Physics,
	University of California at San Diego,
	La Jolla, California 92093, USA
}

\begin{abstract}
	A recent remarkable paper by OpenAI showed that the spin-one Heisenberg chain with periodic boundary conditions has a spectral gap in the thermodynamic limit~\cite{haldane}. The proof elegantly deduces this gap from the Hamiltonian's spectrum on a \textit{finite system}. This basic idea is an inequality involving the purity of the original Gibbs state and a dual Gibbs state that is defined via space-time rotation. Here we consider a few generalizations of this purity bootstrap argument. We first extend it to arbitrary spatial dimensions. Our main result uses the space-time duality between symmetry generators and symmetry twists to obtain upper bounds on the splitting of
	nearly degenerate low-energy states, together with a nonzero gap above them, from finite-system information. As applications, we show that finite-system calculations for the 2+1-D toric code in a field and the 2+1-D transverse-field Ising model establish parameter regimes with four and two low-lying states whose splittings are bounded from above by $e^{-cL}$ and $e^{-cL^2}$	respectively for arbitrarily large system size $L$, while the gap above these states remains nonzero in the thermodynamic limit. Next, we extend it to fermionic systems and use a finite calculation to establish a symmetric mass generation phase in a $2+1$-D fermionic model.
	Finally, we apply the bootstrap to tensor-network states and establish exponential decay of connected correlations of all local operators in the symmetric square-lattice spin-one state introduced by Jian and Zaletel.
\end{abstract}

\maketitle

\compacttableofcontents
\section{Introduction}\label{sec:introduction}
Establishing a spectral gap in an interacting many-body system in the thermodynamic limit is a non-trivial task. Nonetheless, over time, several techniques have been developed, ranging from convergent expansions and perturbative stability arguments around solvable Hamiltonians~\cite{kirkwoodthomas1983,yarotsky2006,kennedytasaki1992,lapalevin2023}, through methods that relate quantum gaps to correlation decay in associated classical models~\cite{aizenmannachtergaele1994}, to finite-size criteria for frustration-free systems based on gaps of small subsystems~\cite{knabe1988} or overlaps between their ground-state spaces~\cite{nachtergaele1996,nachtergaelesimsyoung2018}. A recent remarkable result by OpenAI establishes a spectral gap for the spin-one Heisenberg chain with periodic boundary conditions in the thermodynamic limit~\cite{haldane}. The basic idea in this work is elegant and simple. It couples bounds on the purity of the physical Gibbs state and a `dual Gibbs state' defined via the spatial transfer matrix, i.e., by exchanging the roles of space and imaginary time. Notably, these inequalities allow spectral information from a \textit{finite system} to establish a non-zero gap in the thermodynamic limit. We refer to this method as the ``purity bootstrap''.

The goal of this paper is to demonstrate the power and generality of the purity bootstrap idea, especially when combined with other ingredients such as
symmetry twists. We extend it to higher dimensions, systems with nearly degenerate ground states, fermionic systems, and tensor-network
states. The most direct extension is to spatial dimensions $d>1$, which we discuss in Sec.~\ref{sec:higher-d}. A more interesting extension concerns systems in which, due to symmetry reasons, we expect several nearly degenerate low-energy states separated from the rest of the spectrum by a positive gap. For example, on a periodic square of linear size $L$, the ordered phase of the transverse-field Ising model (TFIM) breaks a discrete 0-form symmetry and is expected to have a splitting of order $e^{-cL^2}$. Similarly, for the $2+1$-dimensional toric code on a periodic square in a weak field, associated with a broken 1-form symmetry, the expected splitting is instead of order $e^{-cL}$. The presence of these nearly degenerate states prevents a direct application of the idea in Ref.~\cite{haldane}. In Sec.~\ref{sec:twists} we show that combining symmetry twists with the purity bootstrap allows us to establish both a positive, size-independent gap above the low-energy states and upper bounds on their splitting with the expected scaling in these examples. Both conclusions follow from finite calculations that control arbitrarily large systems.  The usage of symmetry twists is ultimately related to the fact that the insertion of a symmetry generator in the partition function corresponds to the insertion of a symmetry twist in its space-time rotated version. 

Next, in Sec.~\ref{sec:fermions} we extend the argument to fermionic systems where the most direct application of the method leads to a condition for a gap in the presence of anti-periodic boundaries. Employing symmetry twists associated with fermionic parity then also gives a finite-size certificate for periodic
boundaries. We consider an application of this argument to a $2+1$-dimensional Hamiltonian for symmetric mass generation (SMG)~\cite{wangyou2022}. Overall, the qualitative behavior of the aforementioned models is already well understood. Our goal is to illustrate the principles behind the generalizations shown here and we hope these examples will motivate
applications to less understood Hamiltonians.

Finally, in Sec.~\ref{sec:peps} we use the same concept to deduce properties of projected entangled-pair states (PEPS) from finite calculations. Their squared norms play the role of partition functions and the inverse correlation length associated with connected correlations plays the role of the gap. We illustrate the idea first with a state built from classical Ising weights, and then consider a more interesting application to the symmetric square-lattice spin-one state introduced by Jian and Zaletel in Ref.~\cite{jianzaletel2016}. We rigorously establish exponential decay of connected correlations of local operators in this state.

For pedagogical expositions of the analytic argument in OpenAI's paper (Ref.\cite{haldane}) and its extensions to other spin chains, see Ref.~\cite{tasaki2026notes}  (we also  review the original argument in Sec.~\ref{sec:one-d}). Reference~\cite{tasaki2026uniqueness} recently extended the results of Ref.~\cite{haldane} by establishing uniqueness of the translation-invariant ground state of the infinite spin-one
Heisenberg chain, including sufficiently small single-ion anisotropy. It also shows that periodic and several open boundary conditions yield the same bulk ground state.

Throughout, we use lowercase $d$ to denote the number of spatial dimensions.

\section{The purity bootstrap in one dimension}\label{sec:one-d}
In this section, we review the central idea of Ref.~\cite{haldane}. Let's consider a Hamiltonian on a periodic chain of size $L$ with eigenenergies $E_0(L)\le E_1(L)\le\cdots$, including multiplicities. Our goal is to find conditions on the Gibbs state at a finite size and inverse temperature that imply that $E_1(L)-E_0(L)>\Delta>0$ for every sufficiently large system.

As usual,  the partition function for a Hamiltonian $H$ at an inverse temperature $\beta$, $Z(L,\beta)= \tr e^{-\beta H} = \sum_b e^{-\beta E_b(L)}$, can be written as a path-integral with imaginary time as an additional dimension~\cite{feynman1948,kac1949,ginibre1968}.
However, there is nothing special about the imaginary-time direction of the path integral, and one may take other slices in the space-imaginary time to define analogs of Hamiltonians. For example, one may define a spatial transfer matrix where the original space plays the role of imaginary time along which the system evolves~\cite{suzuki1985,betsuyaku1985}. More recently, related exchanges of space and time have also been used even in real-time quantum circuits to study quantum chaos~\cite{bertinikosprosen2018} and dynamical phase transitions ~\cite{lugrover2021,ippolitirakovszkykhemani2022}.

The essential assumption in the proof is that there exists a \textit{Hermitian} spatial transfer matrix $X_\beta$, independent of $L$, which allows us to write
\begin{equation}
 Z(L,\beta)=\tr X_\beta^L=\sum_a|\lambda_a(\beta)|^L,
 \qquad L\ \text{even}.                         \label{eq:transfer}
\end{equation}
Correspondingly, one may define two probability distributions: one of them corresponding to the diagonal elements of the Gibbs state 
$w_b(L,\beta)=e^{-\beta E_b(L)}/Z(L,\beta)$ and the other one corresponding to the spatial transfer matrix, $v_a(L,\beta)=|\lambda_a(\beta)|^L/Z(L,\beta)$. The purities of these two distributions will play a crucial role in the argument:
\begin{equation}
 T(L,\beta)= \tr w^2 = \frac{Z(L,2\beta)}{Z(L,\beta)^2},\qquad
 S(L,\beta)= \tr v^2 = \frac{Z(2L,\beta)}{Z(L,\beta)^2}.       \label{eq:purities}
\end{equation}
Note that $S, T \leq 1$. The main trick below involves the following inequality relating the purity of a distribution $w$ with that of its normalized square $w^{(2)} = w^2/\tr w^2$: If a distribution $w$ has purity $1-u$, its normalized
square has purity $1-u'$ satisfying
\begin{equation}
 u'=\frac{2\sum_{a<b}w_a^2w_b^2}{(1-u)^2}
 \le\frac{u^2}{2(1-u)^2}=:f(u)    \label{eq:square}
\end{equation}

Since the procedure below will involve repeatedly doubling $L$ or $\beta$, a  useful observation is that $T(L, 2\beta)$ is the purity of the normalized square $w^{(2)}$, while $S(2L, \beta)$ is the purity of the normalized square $v^{(2)}$.

Let's assume that there exist some $L, \beta$ such that $S(L,\beta)\ge1-u$ and $T(2L,\beta)\ge1-u$ where $0 \leq u \leq 0.01$ (the cutoff 0.01 is a convenient, nonoptimal choice that makes the estimates below simple). Since we are interested in showing a spectral gap in the thermodynamic limit, we would like to say something about $S$ or $T$ at a larger $\beta$ or $L$. We will soon see that one can obtain bounds on $S(2L, 2 \beta)$ and $T(4L, 2\beta)$, i.e., both inverse temperature and lengths will get doubled. To that end, we start with the following observation:
\[
 S(L,2\beta)=\frac{S(L,\beta)^2T(2L,\beta)}{T(L,\beta)^2}
 \ge(1-u)^3\ge1-3u,
\]
where for the first inequality, we have used the aforementioned assumption that  $S(L,\beta)\ge1-u$ and $T(2L,\beta)\ge1-u$, and also the fact that $S, T \leq 1$ for any $L, \beta$. The second inequality is the Bernoulli inequality. We can't yet apply the inequality in Eq.~\eqref{eq:square}: $S(L,2 \beta)$ is simply the purity of the distribution $v_a(L,2 \beta)$, which is not, in general, the normalized square of $v_a(L,\beta)$.

Having doubled  $\beta$ at fixed $L$,  we next double $L$ at fixed inverse temperature $2 \beta$. Now we can use the fact that $S(2L, 2 \beta)$ is the purity of  the normalized square of the distribution whose purity is $S(L, 2 \beta)$. Since $S(L,2 \beta) \geq 1 - 3 u$,
Eq.~\eqref{eq:square} gives
\begin{equation} 
 1-S(2L,2\beta)\le\frac{(3u)^2}{2(1-3u)^2}
 \le5u^2. \label{eq:boundS(2L,2beta)}
\end{equation}
The last inequality uses only $1-3u\ge0.97$ and
$9/[2(0.97)^2]<5$ (recall that $0 \leq u \leq 0.01$). Thus we have doubled both arguments of $S$, while improving the bound on its deviation from one from $u$ to $5 u^2$. The remaining task is to double both arguments of $T$. To accomplish that, we follow steps similar to the ones above. We first write
\[
 T(4L,\beta)=\frac{T(2L,\beta)^2S(2L,2\beta)}{S(2L,\beta)^2}
 \ge(1-u)^2(1-5u^2)\ge(1-u)^3\ge1-3u.
\]
Here we used the bound in Eq.~\eqref{eq:boundS(2L,2beta)} and $5u^2\le u$.
Now, at this fixed length $4L$, we double $\beta$. This replaces the Gibbs distribution by its normalized square. Using Eq.~\eqref{eq:square} then gives $T(4L,2\beta)\ge1-5u^2$.
We have now proved the single rule that can be iterated:
\begin{equation}
 \begin{gathered}
 S(L,\beta),\ T(2L,\beta)\ge1-u\\
 \Longrightarrow\quad S(2L,2\beta),\ T(4L,2\beta)\ge1-5u^2,
 \qquad u\le0.01.
 \end{gathered}                                              \label{eq:one-round}
\end{equation}
The output has exactly the form of the input, with $(L,\beta)$ replaced
by $(2L,2\beta)$, which makes the iteration possible. We denote the starting length and inverse temperature by $L_0,\beta_0$. After $j$ rounds of iteration, $L_j=2^j L_0$ and $\beta_j=2^j\beta_0$.
Then $S(L_j,\beta_j)$ and $T(2L_j,\beta_j)$ deviate from unity by at most $u_j$, where $u_{j+1}=5u_j^2$ and $ 0 \leq u_0 \leq 0.01$. Solving this recursion  gives $u_j=\frac{1}{5}(5u_0)^{2^j}\le\frac{1}{5}\,20^{-2^j}$. Thus the deviation of purity from unity is exponentially small in $2^j$, while the inverse temperature grows only in proportion to $2^j$.

\paragraph{Gap in the thermodynamic limit.}
At length $2L_j$ and inverse temperature $\beta_j$, we have $T(2L_j,\beta_j)\ge1-u_j$. Since $T=\sum_b w_b^2\le w_0\sum_b w_b=w_0$, this gives $w_0\ge1-u_j$. Normalization then implies $w_1\le1-w_0\le
u_j$. Writing $\Delta_\ell=E_1(\ell)-E_0(\ell)$, we therefore obtain \begin{equation} \Delta_{2L_j}=\frac{1}{\beta_j}\log\frac{w_0}{w_1}\ge\frac{1}{\beta_j}\log\frac{1-u_j}{u_j}>\frac{\log20}{\beta_0}.
\label{eq:purity-to-gap} \end{equation} The last inequality uses $u_j/(1-u_j)<5u_j\le20^{-2^j}$ and $\beta_j=2^j\beta_0$. Thus one obtains a non-zero gap independent of the system size.

\paragraph{Lengths between doublings.}
The same gap bound holds for every even length $\ell\ge L_0$. To fill the lengths between doublings, we compare even powers of the transfer eigenvalues; see Appendix~\ref{app:one-d-lengths} for details.
\section{Higher dimensions}\label{sec:higher-d}
We now extend the above argument for certifying thermodynamic gaps from finite systems to higher dimensions. For simplicity, we will first work through two spatial dimensions.  Therefore, let's consider a Hamiltonian on a periodic rectangular lattice of size $L_x\times L_y$, with partition function $Z(L_x,L_y,\beta)=\tr e^{-\beta H_{L_x,L_y}}$. To reiterate, our goal is to find finite-size conditions that imply a gap on every sufficiently large square, with both spatial lengths growing together. All spatial lengths below are even and at least four.

We assume that there is a Hermitian spatial transfer matrix in each direction, so that
\begin{equation}
 Z(L_x,L_y,\beta) =\sum_a|\lambda_{x,a}(L_y,\beta)|^{L_x} =\sum_a|\lambda_{y,a}(L_x,\beta)|^{L_y}. \label{eq:2d-transfer}
\end{equation}
In the first expression, the eigenvalues are independent of $L_x$ when $L_y$ and $\beta$ are held fixed; in the second, they are independent of $L_y$ when $L_x$ and $\beta$ are held fixed. We assume that the partition function admits both of these representations for all allowed sizes and inverse temperatures, with convergent sums.

There are now three probability distributions: the normalized weights in each of these two sums, and the Gibbs weights. Their purities are
\begin{equation}
 \begin{gathered}
 S_x(L_x,L_y,\beta)=\frac{Z(2L_x,L_y,\beta)}{Z(L_x,L_y,\beta)^2},\,\,\,
 S_y(L_x,L_y,\beta)=\frac{Z(L_x,2L_y,\beta)}{Z(L_x,L_y,\beta)^2},\,\,\,
 T(L_x,L_y,\beta)=\frac{Z(L_x,L_y,2\beta)}{Z(L_x,L_y,\beta)^2}.
 \end{gathered}
 \label{eq:2d-purities}
\end{equation}
As before, all three purities lie between zero and one.
Doubling $L_x$, $L_y$, or $\beta$, with the other two arguments
fixed, replaces the corresponding probability distribution
by its normalized square. Thus Eq.~\eqref{eq:square} applies
in all three directions.

Let's assume that there exist $L,\beta$ such that
\begin{equation}
 S_x(L,L,\beta),\quad S_y(2L,L,\beta),\quad T(2L,2L,\beta)
 \ge1-u,\qquad 0\le u\le\frac1{98}.
 \label{eq:2d-input}
\end{equation}
The cutoff is again a convenient, generically nonoptimal choice.

As in one dimension, we want to double every argument of the three purities in Eq.~\eqref{eq:2d-input}, while improving the bound on their deviations from unity. Using the same partition-function identities and squaring inequality, we obtain the following update rule; see Appendix~\ref{app:2d-updates} for its derivation.

\begin{equation}
 \begin{gathered}
 S_x(L,L,\beta),\ S_y(2L,L,\beta),\ T(2L,2L,\beta)\ge1-u\\
 \Longrightarrow\quad S_x(2L,2L,2\beta),\ S_y(4L,2L,2\beta),\ T(4L,4L,2\beta)
 \ge1-49u^2,
 \qquad 0\le u\le1/98.
 \end{gathered}
 \label{eq:2d-round}
\end{equation}
The output has the same form as the input with $(L,\beta)$ replaced by $(2L,2\beta)$. We again denote the starting side length and inverse temperature by $L_0,\beta_0$. After $j$ rounds, let $L_j=2^jL_0$ and $\beta_j=2^j\beta_0$. The three purities $S_x(L_j,L_j,\beta_j)$, $S_y(2L_j,L_j,\beta_j)$, and $T(2L_j,2L_j,\beta_j)$ deviate from unity by at most $u_j$, where $u_0=u\le1/98$ and $u_{j+1}=49u_j^2$. These deviations remain below the initial cutoff, so the rule can be applied repeatedly. Solving the recursion gives
\begin{equation}
 u_j=\frac1{49}(49u_0)^{2^j}\le\frac1{49}\,2^{-2^j}.
 \label{eq:2d-error}
\end{equation}

\paragraph{Gap in the thermodynamic limit.}
The third purity gives $T(2L_j,2L_j,\beta_j)\ge1-u_j$. Using the same Gibbs-weight argument as in one dimension, we obtain
\[
 \Delta_{2L_j}\ge\frac1{\beta_j}\log\frac{1-u_j}{u_j}
 >\frac{\log2}{\beta_0}.
\]
Here $\Delta_\ell$ denotes the gap on the $\ell\times\ell$ square. The last inequality uses $u_j/(1-u_j)<49u_j\le2^{-2^j}$ and $\beta_j=2^j\beta_0$. Thus we have a positive gap bound independent of the size of these squares.

\paragraph{Squares between doublings.}
The same bound holds for every even side length $\ell\ge L_0$. As in one dimension, we compare eigenvalue powers, now in each spatial direction; Appendix~\ref{app:2d-lengths} gives the calculation. Both spatial lengths grow together, so the conclusion applies to a truly two-dimensional system.

\paragraph{A convenient finite-size test.}
For the applications, it is useful to obtain the three initial purity bounds from a single ratio,
\begin{equation}
 Q(L,\beta):=\frac{Z(2L,2L,2\beta)}{Z(L,L,\beta)^8} =S_x(L,L,\beta)^4S_y(2L,L,\beta)^2T(2L,2L,\beta). \label{eq:2d-cube}
\end{equation}
The factorization follows directly from the definitions in Eq.~\eqref{eq:2d-purities}. Since each purity is at most one, their product with these powers is no larger than any one of them. Hence $Q(L_0,\beta_0)\ge97/98$ implies all three starting bounds in Eq.~\eqref{eq:2d-input}, and is sufficient to establish $\Delta_\ell>\log2/\beta_0$ for every even $\ell\ge L_0$.

\paragraph{General dimensions.}
In $d$ spatial dimensions, there are $d$ spatial purities and one Gibbs purity. The key assumption is again a Hermitian spatial transfer representation along each of the $d$ spatial directions. We start with the $d+1$ purities obtained by doubling each spatial length in turn and then doubling $\beta$. Each successive doubling squares the preceding partition function, giving powers $2^d,2^{d-1},\ldots,1$ in their product. Their sum is $a=2^{d+1}-1$, which equals seven in two dimensions. The same calculation gives an intermediate purity at least $1-au$; squaring its weights reduces the bound on the deviation from unity to at most $a^2u^2/[2(1-au)^2]\le a^2u^2$ when $u\le1/(2a^2)$. This supplies the same iteration in all $d+1$ directions.

Writing $\cZ(L,\beta)=Z(L,\ldots,L,\beta)$ for the partition function on a hypercube, define
\begin{equation}
 Q(L,\beta)=\frac{\cZ(2L,2\beta)}{\cZ(L,\beta)^{2^{d+1}}}.
 \label{eq:cube}
\end{equation}
As in two dimensions, this ratio is the product of the successive purities with the powers just given, and is no larger than any of them. Iterating this bound and comparing intervening lengths in each spatial direction therefore gives
\begin{equation}
 Q(L_0,\beta_0)\ge1-\frac1{2a^2}
 \quad\Longrightarrow\quad
 \Delta_\ell>\frac{\log2}{\beta_0}
 \quad\text{for every even }\ell\ge L_0,
 \qquad a=2^{d+1}-1.
 \label{eq:dd}
\end{equation}
Here $\Delta_\ell$ is the gap on the $\ell^d$-site hypercube.

\subsection{Application to the Heisenberg bilayer}\label{sec:bilayer}
We consider the standard spin-$\tfrac12$ bilayer Heisenberg model on the square lattice~\cite{sandvikscalapino1994}:
\begin{equation}
H_{\rm bi}=J\sum_{\langle ij\rangle,a=1,2}\mathbf S_{i,a}\cdot\mathbf S_{j,a}
+J_\perp\sum_i\mathbf S_{i,1}\cdot\mathbf S_{i,2},
\qquad J,J_\perp>0,\qquad g=J_\perp/J.
\label{eq:bilayer-overview}
\end{equation}
Each pair of vertically aligned spins forms a rung. At $J=0$, its ground state is a singlet, and its three triplet states lie an energy $J_\perp$ above it. The finite test establishes a unique ground state and a size-independent gap throughout $0\le J/J_\perp\le1/10$ on sufficiently large even periodic squares.

To apply the test, we regard each rung as one site. A basis adapted to time reversal makes the rung Hamiltonian real and writes each bond interaction using the same real matrices at its two ends. This gives Hermitian spatial transfers in both directions. Bounds on the $4\times4$ partition function and a trial energy on $8\times8$ then supply the finite information needed in Eq.~\eqref{eq:2d-cube}. Appendix~\ref{app:bilayer} provides the details.

\section{Gapped systems with nearly degenerate ground states}\label{sec:twists}

 So far, we have considered a gap above a single ground state. In a system that spontaneously breaks a discrete symmetry, we instead expect several ground states in
the thermodynamic limit which, in the absence of any other gapless degrees of freedom, will be generically separated from the rest of the spectrum by a finite gap. On a finite system, these states will be nearly degenerate with splitting that is exponentially small in the system's volume, while the gap to the remaining states
stays nonzero. We now want to extend the purity bootstrap approach to establish these thermodynamic properties using spectral information from finite systems.  Similar nearly degenerate ground states also occur when higher-form symmetries are spontaneously broken, as in the toric-code application at the end of this section (Sec.~\ref{sec:toric}).

\subsection{Symmetry twists and the proof for Ising symmetry}\label{sec:ising-argument}
Let's consider a local Hamiltonian $H_L$ on a periodic chain, commuting with $U=\prod_iX_i$. Its two symmetry sectors have eigenvalues $k=\pm1$. Our goal is to show that there is one low-energy state in each sector, that the difference between their energies tends to zero as $L$ grows, and that a nonzero gap separates these two states from all other states.

To use the argument of Sec.~\ref{sec:one-d}, we will combine the spectra for two boundary conditions. We denote the original Hamiltonian by $H_L^{(t=+1)}$ and introduce $H_L^{(t=-1)}$ by applying a symmetry twist at the closing seam. The twist acts with $X$ on one side of each interaction crossing the seam; for example, $-JZ_LZ_1$ becomes $+JZ_LZ_1$. Both Hamiltonians commute with $U$. We project onto a fixed sector and define
\begin{equation}
 P_k=\frac{I+kU}{2},\qquad
 \mathcal Z_k(L,\beta)=
 \sum_{t=\pm1}\tr\!\left(P_k e^{-\beta H_L^{(t)}}\right).
 \label{eq:physical-twist-sum}
\end{equation}
Here $k$ labels a symmetry sector and $t$ labels a boundary condition. Thus $\mathcal Z_k$ sums the Gibbs weights of all energies in sector $k$ for both boundary conditions, including multiplicities.

\begin{wrapfigure}{r}{0.40\textwidth}
\centering
\begin{tikzpicture}[
 font=\footnotesize, >=Latex,
 outline/.style={draw=black!55,line width=0.5pt,fill=black!2},
 defect/.style={draw=ucolor,line width=1.5pt}
]
\draw[outline] (0,0) rectangle (1.65,1.45);
\draw[outline] (3.10,0) rectangle (4.75,1.45);
\draw[defect] (0,0.725)--(1.65,0.725);
\draw[defect] (3.10,0.725)--(4.75,0.725);
\node at (0.825,1.72) {$U$ insertion};
\node at (3.925,1.72) {Spatial twist};
\draw[->,draw=black!50] (1.95,0.725)--(2.50,0.725);
\node at (2.225,1.15) {$x\leftrightarrow\tau$};
\draw[->] (0,-0.20)--(0.65,-0.20) node[right] {$x$};
\draw[->] (-0.20,0)--(-0.20,0.65) node[above] {$\tau$};
\draw[->] (3.10,-0.20)--(3.75,-0.20) node[right] {$\tau$};
\draw[->] (2.90,0)--(2.90,0.65) node[above] {$x$};
\end{tikzpicture}
\caption{Exchanging space and imaginary time turns an insertion of the symmetry generator $U$ into a spatial symmetry twist~\cite{cardy1986boundary,zuber1986discrete,shao2023tasi}. Opposite edges of the rectangle are identified.}
\label{fig:symmetry-twist}
\end{wrapfigure}
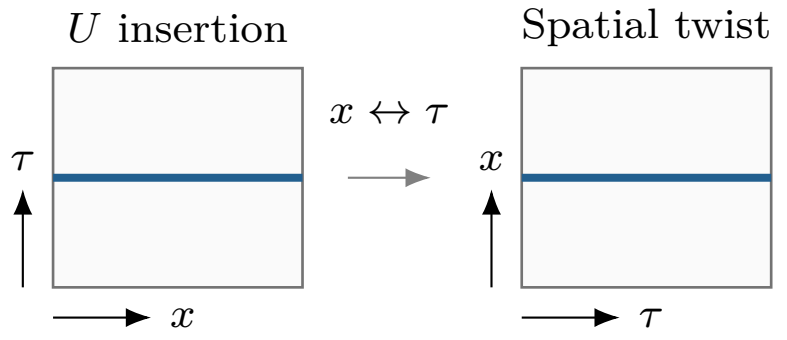
The motivation for summing over boundary twists can be understood
from space-time rotation (Fig.~\ref{fig:symmetry-twist}). First, since
spontaneous symmetry breaking produces two nearly degenerate low-lying
states, the test in Sec.~\ref{sec:one-d} for a gap above a single state
cannot succeed. Therefore, one may first try studying the partition
function in a given symmetry sector. In the Euclidean path integral,
inserting the global symmetry $U$ in the thermal trace is equivalent
to twisting the boundary condition around the imaginary-time cycle.
Exchanging space and imaginary time therefore maps a symmetry
projection to a \textit{weighted sum} over spatial twists in the dual
description. Although projecting onto a fixed symmetry sector isolates
a single state, it can spoil the interpretation of
$Z(2L,\beta)/Z(L,\beta)^2$ as the purity of the dual Gibbs state: replacing $Z$ by the projected trace can make this
expression exceed one. For example, in the classical Ising chain
$H_L=-J\sum_i Z_iZ_{i+1}$ with $J>0$, the projected untwisted partition
function is $\frac12[(2\cosh\beta J)^L+(2\sinh\beta J)^L]$, for which
this ratio exceeds one.

Summing over the original spatial twists, on the other hand,
implements a symmetry projection in the rotated description. Under
the conditions stated below, this gives the Hermitian spatial transfer
needed for the argument in the $k=+1$ sector. If this sector has a gap
and the added twisted states stay a finite energy above its lowest
state, the combined spectrum retains a gap. Therefore, summing over
twists while projecting onto a given symmetry sector may allow one to
apply the purity bootstrap.

For this idea to work, we require two properties. First, we assume that there are Hermitian transfer operators $A_0,A_1$, depending on $\beta$ but independent of $L$, such that
\begin{equation}
 \mathcal Z_k=\tr A_0^L+k\tr A_1^L.
 \label{eq:ising-explicit-R}
\end{equation}
The two terms come from the identity and $U$ in the projector: $\tr A_0^L=\frac12\sum_{t=\pm1}\tr(e^{-\beta H_L^{(t)}})$ and $\tr A_1^L=\frac12\sum_{t=\pm1}\tr(Ue^{-\beta H_L^{(t)}})$. We require these representations at every even $L\ge4$ and positive $\beta$, with convergent traces. For notational convenience, we introduce two orthogonal copies of the Hilbert space associated with the spatial transfer and define $X_\beta=A_0\oplus A_1$, which acts as $A_0$ on the first copy and $A_1$ on the second.
Defining $R_k=I\oplus kI$ on the same enlarged space allows us to write $\mathcal Z_k=\tr(X_\beta^L R_k)$. Second, we assume that, within each sector $k$, the lowest energy for $t=+1$ is no higher than that for $t=-1$. We give sufficient local conditions below and their proofs in Appendix~\ref{app:symmetry}.

Under these assumptions, we now outline the proof; see Appendix~\ref{app:ising-bounds} for details.

\emph{First, we show that the combined $k=+1$ spectrum has a gap.} Since $\mathcal Z_{+1}=\tr X_\beta^L$ with $X_\beta$ Hermitian, we can use the argument of Sec.~\ref{sec:one-d}. We define $S_{+1},T_{+1}$ by Eq.~\eqref{eq:purities}, replacing $Z$ by $\mathcal Z_{+1}$.  For even $L$, $S_{+1}$ is the purity of the density matrix $(A_0^L\oplus A_1^L)/(\tr A_0^L+\tr A_1^L)$ on the enlarged transfer space, whereas $S_{-1}$ generally has no
such interpretation because of the relative minus sign. In contrast, the temporal ratio $T_k=\mathcal Z_k(L,2\beta)/\mathcal Z_k(L,\beta)^2$ remains a valid purity of the normalized Gibbs density matrix formed from the
combined spectra of both boundary conditions $t=\pm1$, within either symmetry sector $k = \pm 1$. If $S_{+1}(L_0,\beta_0),T_{+1}(2L_0,\beta_0)\ge0.99$ at some even $L_0\ge4$, using the argument in Sec.~\ref{sec:one-d}, we obtain a gap greater than $(\log20)/\beta_0$ above the lowest state in the combined spectrum of
the two boundary conditions $t=\pm1$, both restricted to the symmetry sector $k=+1$, for every even $L\ge L_0$.

\emph{Next, we show that $\mathcal Z_{-1}/\mathcal Z_{+1}$ approaches one as $S_{+1}$ approaches one.} We use the definition of $S_{+1}$ and the identities $\mathcal
Z_{\pm1}=\tr A_0^L\pm\tr A_1^L$ to write
\begin{equation}
\begin{aligned}
S_{+1} =\frac{\tr A_0^{2L}+\tr A_1^{2L}}
{(\tr A_0^L+\tr A_1^L)^2} \le\frac{(\tr A_0^L)^2+(\tr A_1^L)^2}
{(\tr A_0^L+\tr A_1^L)^2}
=\frac12\left[1+\left(\frac{\mathcal Z_{-1}}{\mathcal Z_{+1}}\right)^2\right].
\end{aligned}
\label{eq:ising-purity-comparison}
\end{equation}
The inequality follows because $L$ is even: each $\tr A_a^L$ is a sum of nonnegative eigenvalue powers, so its square contains all the terms in $\tr A_a^{2L}$ plus
nonnegative cross terms. Also, $\mathcal Z_{-1}>0$ because it is a sum of partition functions in the $k=-1$ sector, while $\mathcal Z_{-1}\le\mathcal Z_{+1}$ because $
\tr A_1^L\ge0$. Therefore, $\sqrt{2S_{+1}-1}\le
\frac{\mathcal Z_{-1}}{\mathcal Z_{+1}}\le1$, and as $S_{+1}$ approaches one, the ratio must approach one as well.

\emph{We now show that the combined $k=-1$ spectrum also has a gap.} Using the definition of the temporal purity, we obtain
\[
\begin{aligned}
T_{-1}(L,\beta)
&=T_{+1}(L,\beta)
\frac{\mathcal Z_{-1}(L,2\beta)}{\mathcal Z_{+1}(L,2\beta)}
\left[\frac{\mathcal Z_{+1}(L,\beta)}{\mathcal Z_{-1}(L,\beta)}\right]^2\\
&\ge T_{+1}(L,\beta)\sqrt{2S_{+1}(L,2\beta)-1}.
\end{aligned}
\]
Here the squared bracket is at least one, and we bounded the other ratio using the preceding result at inverse temperature $2\beta$. The bounds from Sec.~\ref{sec:one-d} on the two
purities on the right therefore give an exponentially small deviation of $T_{-1}$ from one as the scales increase. Applying the Gibbs-weight argument leading to Eq.~\eqref{eq:purity-to-gap} to these bounds on $T_{-1}$ gives a size-independent gap in the combined $k=-1$ spectrum. 

\emph{Next, we show that the lowest energies for $k=+1$ and $k=-1$ are close, including both boundary conditions $t=\pm1$ in each sector.} Let $E_{0,k}$ be the lowest energy in the combined spectrum for sector $k$, with the dependence on $L$ left implicit. At the same $L,\beta$, its
Gibbs weight is $e^{-\beta E_{0,k}}/\mathcal Z_k$. As shown in Sec.~\ref{sec:one-d}, this weight is at least the purity $T_k$, so
$T_k\mathcal Z_k\le e^{-\beta E_{0,k}}\le\mathcal Z_k.$
Dividing the bounds for the two sectors gives
\[
T_{-1}\frac{\mathcal Z_{-1}}{\mathcal Z_{+1}}
\le e^{-\beta(E_{0,-1}-E_{0,+1})}
\le\frac{1}{T_{+1}}\frac{\mathcal Z_{-1}}{\mathcal Z_{+1}}.
\]
Both bounding expressions approach one: the partition-function ratio does so by the preceding argument, and both temporal purities do so by their bounds.  Taking logarithms and dividing by $\beta$ therefore shows that the difference between the lowest energies tends to zero. Appendix~\ref{app:ising-bounds} quantifies this and gives an upper bound that decays exponentially with $L$ [Eq.~\eqref{eq:multiplet}].

\emph{Finally, we show that these two low lying states belong to the original, i.e. untwisted ($t= +1$), Hamiltonian.}  We have proved a size-independent gap above the lowest state in each combined sector spectrum, and that the difference between these two lowest energies vanishes as $L$
grows. By our assumption,  the lowest untwisted energy is no higher than the lowest twisted energy in each sector. Therefore, both of these nearly degenerate states belong to the untwisted Hamiltonian. Thus, for
sufficiently large $L$, the original Hamiltonian has a nearly degenerate pair of lowest states separated from the rest of its spectrum by a size-independent gap.

More explicitly, writing $W_\ell=E_1(\ell)-E_0(\ell)$ for their splitting and $\Gamma_\ell=E_2(\ell)-E_1(\ell)$ for the gap above them, the bounds are
\begin{equation}
W_\ell\le\frac{12}{5\beta_0}e^{-(\log20)\ell/(2L_0)},
\qquad
\Gamma_\ell>\frac{\log8}{\beta_0}
\quad(\ell\ge L_0,\ \ell\ \mathrm{even}).
\label{eq:multiplet}
\end{equation}
Here $L_0,\beta_0$ are the starting length and inverse temperature in the finite purity test.

\subsection{Sufficient conditions on the Hamiltonian}\label{sec:ising-local}
We now give sufficient conditions for the representation
$\mathcal Z_k=\tr A_0^L+k\tr A_1^L$, where $A_0,A_1$ are Hermitian and independent of $L$.
Consider a translationally invariant Hamiltonian
$H_L=\sum_{i=1}^L(b_i+v_{i,i+1})$ with periodic boundary conditions,
commuting with the Ising symmetry $U=\prod_i X_i$.
We assume that the local operators satisfy
\[
b=b^{\mathsf T}\ \text{is real},\qquad [b,X]=0,\qquad
-v=\sum_\alpha F_\alpha\otimes F_\alpha,\qquad
XF_\alpha X=s_\alpha F_\alpha,\quad s_\alpha=\pm1,
\]
where the $F_\alpha$ are real and $v$ is Hermitian.
For example, the Hamiltonian
\begin{equation}
H_L=-J\sum_i Z_iZ_{i+1}-g\sum_i X_iX_{i+1}
-h\sum_i X_i,\qquad J,g>0,\quad h\ne0,
\label{eq:ising-interacting-chain}
\end{equation}
satisfies these conditions with
$b=-hX$, $F_1=\sqrt J\,Z$, and $F_2=\sqrt g\,X$.
Here $s_1=-1$ and $s_2=+1$. The boundary twist reverses the
closing $ZZ$ interaction but leaves the $XX$ interaction unchanged.

We want to prove that $E_{0,k,t=+1}\le E_{0,k,t=-1}$ for every $L$ and $k=\pm1$. We choose a single-site basis that diagonalizes both $b$ and $X$, which is possible because $[b,X]=0$. If every off-diagonal entry of $v$ in the corresponding two-site basis is real and nonpositive, then the required ordering holds. This requires checking only the fixed two-site matrix $v$.

Appendix~\ref{sec:ising-checks} constructs the transfer operators, and Appendix~\ref{app:ising-bounds} proves the energy ordering and the bounds in Eq.~\eqref{eq:multiplet}.

\subsection{A splitting exponentially small in the system volume}\label{sec:volume-splitting}
In two dimensions, we can sometimes strengthen the splitting bound from
$e^{-cL}$ to $e^{-cL^2}$, which is the expected scaling of splitting due to spontaneous breaking of discrete zero-form symmetries. We sketch the idea here and apply it to the square-lattice TFIM in Sec.~\ref{sec:tfim}. As above, we first define $\mathcal Z_k(L_x,L_y,\beta)$ by summing over the four
boundary twists at fixed symmetry sector $k=\pm1$. When the finite
purity test succeeds, combining the
two-dimensional bounds of Sec.~\ref{sec:higher-d} with the sector
comparison above first gives a gap within each sector and an
$e^{-cL}$ bound on the splitting. To strengthen this bound to $e^{-cL^2}$,
we define the relative difference between the two partition functions:

\[
\delta(L_x,L_y,\beta)
:=1-\frac{\mathcal Z_{-1}}{\mathcal Z_{+1}}
=\frac{2\tr A_1^{L_x}}{\mathcal Z_{+1}}.
\]
For even $L_x$ at fixed $L_y,\beta$, the inequality
$\tr A_1^{2L_x}\le(\tr A_1^{L_x})^2$ gives
\[
\delta(2L_x,L_y,\beta)\le
\frac{\delta(L_x,L_y,\beta)^2}{2S_x(L_x,L_y,\beta)},
\]
where $S_x=\mathcal Z_{+1}(2L_x,L_y,\beta)/
\mathcal Z_{+1}(L_x,L_y,\beta)^2$.
The same argument along the $y$ direction leads to another squaring of $\delta$. Thus, if
$S_x(L,L,\beta),S_y(2L,L,\beta)>1/2$,
\[
\delta(2L,2L,\beta)\le\delta(L,L,\beta)^4.
\]
Repeated fourth powers give an exponent growing as
$4^j=L_j^2/L_0^2$, where $L_j=2^j L_0$.

To turn this into a bound on the energy splitting, we must also
suppress the excited-state contributions to $\mathcal Z_{\pm1}$. Choosing $\beta$ proportional to $L^2$, the gap and Gibbs-weight bounds already established  then show that the excited-state contributions are exponentially small
in $L^2$ relative to the ground state terms. We must also check that the spatial purities remain above
$1/2$ as we lower the temperature. For the TFIM they indeed do, so
the two spatial squarings and this cooling step together give
$W_L\le Ce^{-cL^2}$ on an $L\times L$ square lattice, while the gap above
the pair stays nonzero.
Appendix~\ref{app:tfim} gives the details.
In passing we note that in contrast to the TFIM, for the $2+1$-dimensional perturbed toric code, the two spatial directions compare different loop sectors, so the aforementioned 
idea does not apply and our splitting bound is instead
$e^{-cL}$, as expected based on physical grounds (Sec.~\ref{sec:toric}).

 For the TFIM in $d$ spatial dimensions, whenever the corresponding finite purity test succeeds, the same argument gives
$W_L\le Ce^{-cL^d}$, with $C,c>0$ independent of $L$. Doubling all $d$ lengths raises the bound on $\delta$ to the power $2^d$, while the gap within each sector allows us to choose $\beta\propto L^d$.

The importance of a gap above the low-lying pair for constructing symmetry-broken infinite-volume ground states
was emphasized by Koma and Tasaki~\cite{komatasaki1994}.
For sufficiently weak transverse fields, a splitting exponentially small in volume and a gap above the pair were established relatively recently by Lapa and
Levin~\cite{lapalevin2023} using a polymer expansion. The purity bootstrap provides a substantially different approach to these conclusions.

\subsection{General finite Abelian symmetries.}
The proof above extends to any finite Abelian group $G$. We again project onto each symmetry sector and sum over the boundary twists associated with $G$.  The Ising signs are
replaced by phases. Under the analogous Hermitian transfer representation, and provided the lowest untwisted energy is no higher than the lowest twisted energy in each symmetry
sector, the finite purity test gives $|G|$ low states with exponentially small spread and a gap above them.  See Appendix~\ref{sec:abelian} for details in one dimension. We expect the extension to higher dimensions to follow the same steps as in the Ising case.

\subsection{Application to the \texorpdfstring{$2+1$}{2+1}-dimensional transverse-field Ising model}\label{sec:tfim}
We consider the square-lattice Hamiltonian with periodic boundaries,
\begin{equation}
H=-J\sum_{\langle ij\rangle}Z_iZ_j-h\sum_iX_i,
\qquad J>0,\quad h\ge0,
\label{eq:tfim-overview}
\end{equation}
where $X_i,Z_i$ are Pauli matrices. The two polarized states are exactly degenerate at $h=0$. The finite calculation establishes that, throughout $0\le h/J\le1.8$, the ordered energies on sufficiently large even periodic squares obey
\[
W_L=E_1(L)-E_0(L)\le Ce^{-cL^2},\qquad
\Gamma_L=E_2(L)-E_1(L)\ge\Delta>0,
\]
with $C,c,\Delta$ independent of $L$.

We apply the symmetry argument to $U=\prod_iX_i$, summing over the four boundary twists at fixed $U=\pm1$. These combined partition functions satisfy the required transfer conditions in both directions, and the lowest state in each symmetry sector belongs to the untwisted Hamiltonian. Complete spectra on $4\times4$ and a trial state on $8\times8$ initialize the purity bounds. The two spatial squarings and cooling to $\beta\propto L^2$ in Sec.~\ref{sec:volume-splitting} then give the upper bound $W_L\le Ce^{-cL^2}$. Appendix~\ref{app:tfim} provides the details.

\subsection{Application to the \texorpdfstring{$2+1$}{2+1}-dimensional toric code in a field}\label{sec:toric}
We place a spin-$\tfrac12$ on every edge of an $L\times L$ periodic square lattice and consider
\begin{equation}
\begin{gathered}
H_{\rm TC}=-J_e\sum_v A_v-J_m\sum_p B_p-h\sum_eX_e,\\
A_v=\prod_{e\ni v}X_e,\qquad B_p=\prod_{e\in\partial p}Z_e,
\qquad J_e,J_m>0,\quad h\ge0.
\end{gathered}
\label{eq:toric-overview}
\end{equation}
Here $v,p,e$ label vertices, plaquettes, and edges. The four ground states are exactly degenerate at $h=0$. For $J_e=J_m$ and $0\le h/J_m\le1/7$, the finite test gives
\[
E_3(L)-E_0(L)\le Ce^{-cL},\qquad
E_4(L)-E_3(L)\ge\Delta>0
\]
on sufficiently large even periodic squares, with $C,c,\Delta$ independent of $L$. The argument also applies to any fixed $J_e>0$, with the gap bound depending on $J_e$.

The key step is to map the four conserved loop sectors, with all $A_v=+1$, to the four boundary conditions of an Ising model. Physical states correspond to its even spin-flip sector, which contains the ground state for each boundary condition. The purity bounds compare these four spectra and give one low state in each physical sector. States with $A_v=-1$ cost at least $4J_e$ above the lowest energy and hence remain outside this group. Appendix~\ref{app:toric} gives the duality and the finite test using $4\times4$ and $8\times8$ systems. The $e^{-cL}$ dependence of our upper bound agrees with the expected amplitude for an anyon to tunnel around a noncontractible cycle of length $L$~\cite{kitaev2003}, in contrast to the $e^{-cL^2}$ dependence for the TFIM.

\section{Fermions}\label{sec:fermions}
Our goal again is to use partition functions of finite systems to prove a nonzero gap in the thermodynamic limit. For fermions, we start with antiperiodic boundary conditions in space. This is because the ordinary fermionic thermal trace imposes antiperiodic boundary conditions in imaginary time. Choosing antiperiodic boundaries also in space therefore lets us use an ordinary
trace after exchanging the two directions. After establishing a finite condition for a gap with antiperiodic boundaries, we employ the symmetry-twist introduced in Sec.~\ref{sec:twists}, with fermion parity as the symmetry, to establish a gap with periodic boundaries from the same finite information. Of course, we still need conditions on the Hamiltonian to make the corresponding spatial transfer Hermitian.

For a chain, we denote the physical partition function by $Z_{\rm A}(L,\beta)$, where A indicates the antiperiodic boundary condition. As in Sec.~\ref{sec:one-d}, we want two expressions for this same quantity:
\begin{equation}
 \begin{split}
 Z_{\rm A}(L,\beta)
 &=\tr e^{-\beta H_{\rm A}(L)}=\sum_b e^{-\beta E_b(L)}\\
 &=\tr X_\beta^L
 =\sum_a|\lambda_a(\beta)|^L,\qquad L\ \text{even}.
 \end{split}
 \label{eq:fermion-two-representations}
\end{equation}
The first expression follows by diagonalizing the physical Hamiltonian. Our task is to obtain the second, with $X_\beta$ Hermitian and independent of $L$. Once we have it, we define
\[
 T_{\rm A}(L,\beta)=\frac{Z_{\rm A}(L,2\beta)}{Z_{\rm A}(L,\beta)^2},
 \qquad
 S_{\rm A}(L,\beta)=\frac{Z_{\rm A}(2L,\beta)}{Z_{\rm A}(L,\beta)^2}.
\]
These are the same two purities as before. Once the spatial representation is proved, the numerical test requires only the finite partition functions in these ratios; we do not need to compute $X_\beta$ or its eigenvalues. Thus $S_{\rm A}(L_0,\beta_0),T_{\rm A}(2L_0,\beta_0)\ge0.99$ at an even starting length $L_0\ge4$ and inverse temperature $\beta_0$ gives $\Delta_\ell>(\log20)/\beta_0$ for every even $\ell\ge L_0$, by Sec.~\ref{sec:one-d}. We now give sufficient conditions for the spatial representation and show why they work.

\subsection{Conditions for a Hermitian spatial transfer}
We first consider a chain of identical cells, each with an even number $M$ of complex fermion modes. A cell can be a site; later we will use entire columns or rows for a $2+1$-dimensional system. We write the fermions as Majoranas, $\gamma^{2a-1}=c_a+c_a^\dagger$ and $\gamma^{2a}=-i(c_a-c_a^\dagger)$, and consider
\begin{equation}
 H_{\rm A}(L)=\sum_{j=1}^L b_j-\frac{it}{2}
 \sum_{j=1}^L\sum_{a=1}^{2M}\gamma_j^a\gamma_{j+1}^a,
 \qquad \gamma_{L+1}^a=-\gamma_1^a,\qquad t\ge0.
 \label{eq:fermionclass}
\end{equation}
Here $b_j$ is the same Hermitian, parity-preserving interaction on each cell. The trace defining $Z_{\rm A}(L,\beta)$ includes both values of total fermion parity.

To prove Hermiticity of the spatial transfer, we will use the reality of the thermal Majorana correlations within one cell. A sufficient condition is an antiunitary $\mathcal C$ satisfying
\begin{equation}
 \mathcal C\gamma^a\mathcal C^{-1}=\gamma^a,
 \qquad \mathcal Cb\mathcal C^{-1}=b.
 \label{eq:fermion-local}
\end{equation}
Applying $\mathcal C$ to a time-ordered thermal trace complex-conjugates its value and leaves every factor unchanged, so that trace is real. Real quartic Majorana interactions satisfy this condition.

The construction, as detailed in Appendix~\ref{app:fermion-construction}, starts from the fermionic path integral and expands the hopping between cells, while keeping the interaction within each cell exact. The resulting single-cell traces define the spatial transfer. The antiunitary condition makes the block corresponding to an even (odd) number of hopping insertions real symmetric (imaginary antisymmetric).
Both blocks are therefore Hermitian, proving
Eq.~\eqref{eq:fermion-two-representations}.

\subsection{Extension to two spatial dimensions}
For a $2+1$-dimensional system, we want two spatial representations of the same $Z_{\rm A}(L_x,L_y,\beta)$, with antiperiodic boundaries in both directions. We label sites by $(x,y)$ and write $\gamma_{x,y}^r$ for the Majorana modes at a site, with $r$ running over those modes. We first group each column $x$ into one cell. Its Hamiltonian $b_x$ contains the interactions and hopping within that column, including the antiperiodic closing bond in $y$. As in the chain, we assume identical columns and write $b$ for their common cell Hamiltonian.

We consider Hamiltonians whose hopping between neighboring columns has the form
\[
 v_{x,x+1}=-\frac{it}{2}\sum_{y=1}^{L_y}\sum_r
 \eta_{y,r}\gamma_{x,y}^r\gamma_{x+1,y}^r,
 \qquad t\ge0,\quad \eta_{y,r}=\pm1.
\]
The sum over $y$ collects all hopping terms connecting columns $x$ and $x+1$. The signs $\eta_{y,r}$ are fixed coefficients of the Hamiltonian, independent of $x$. Antiperiodic boundaries in this direction mean $\gamma_{L_x+1,y}^r=-\gamma_{1,y}^r$. In the earlier chain notation, the cell label is now $j=x$, and its Majorana label $a$ is the pair $(y,r)$.

To obtain a Hermitian transfer in $x$, we assume that one column has an antiunitary satisfying
\begin{equation}
 \mathcal Cb\mathcal C^{-1}=b,\qquad
 \mathcal C\gamma_y^r\mathcal C^{-1}=\eta_{y,r}\gamma_y^r.
 \label{eq:fermion-signed-local}
\end{equation}
Here we suppress $x$ because these operators act within a single column. The calculation in Appendix~\ref{app:fermion-signed} then gives a Hermitian $X_x(L_y,\beta)$, independent of the number of columns $L_x$. We then group rows into cells and require the corresponding hopping form and antiunitary condition, with $x$ and $y$ interchanged. This gives a Hermitian $X_y(L_x,\beta)$, independent of $L_y$. Thus
\[
 Z_{\rm A}(L_x,L_y,\beta)
 =\tr X_x(L_y,\beta)^{L_x}
 =\tr X_y(L_x,\beta)^{L_y}.
\]
For even lengths, these are precisely the spatial representations used in Sec.~\ref{sec:higher-d}. Its three purities therefore give a finite test for a gap as both spatial lengths grow.

\subsection{Obtaining a gap with periodic spatial boundaries}\label{sec:fermion-periodic}
We now want to show that the same finite information also gives a unique ground state and a gap with periodic spatial boundaries. We first consider a chain. Let $H_{\rm P}(L)$ be the Hamiltonian in Eq.~\eqref{eq:fermionclass} with $\gamma_{L+1}^a=\gamma_1^a$, and let $Z_{\rm P}(L,\beta)=\tr e^{-\beta H_{\rm P}(L)}$, again including both fermion parities.  Fermion parity acts as $\gamma_j^a\mapsto-\gamma_j^a$. Applying this transformation to one end of each hopping interaction crossing the closing seam reverses its sign, interchanging periodic
and antiperiodic boundaries. Thus the boundary change is a symmetry twist.

The construction in Appendix~\ref{app:fermion-construction} separates even and odd numbers of hopping insertions into two Hermitian operators $K_{0,\beta},K_{1,\beta}$. Changing the boundary condition reverses the odd contribution, giving
\begin{equation}
 Z_{\rm A}(L,\beta)=\tr K_{0,\beta}^L+\tr K_{1,\beta}^L,
 \qquad Z_{\rm P}(L,\beta)=\tr K_{0,\beta}^L-\tr K_{1,\beta}^L.
 \label{eq:fermion-periodic-traces}
\end{equation}
For even $L$, both traces on the right are nonnegative. Since $Z_{\rm P}$ is a physical partition function, it is positive, giving $0<Z_{\rm P}/Z_{\rm A}\le1$.

We next want to bound this ratio using the antiperiodic spatial purity. As in Sec.~\ref{sec:twists}, we calculate directly
\[
 \begin{split}
 S_{\rm A}(L,\beta)
 &=\frac{\tr K_{0,\beta}^{2L}+\tr K_{1,\beta}^{2L}}
 {(\tr K_{0,\beta}^{L}+\tr K_{1,\beta}^{L})^2}\\
 &\le\frac{(\tr K_{0,\beta}^{L})^2+(\tr K_{1,\beta}^{L})^2}
 {(\tr K_{0,\beta}^{L}+\tr K_{1,\beta}^{L})^2}
 =\frac12\left[1+\left(\frac{Z_{\rm P}(L,\beta)}{Z_{\rm A}(L,\beta)}\right)^2\right].
 \end{split}
\]
The inequality uses $\sum_i a_i^2\le(\sum_i a_i)^2$ for the nonnegative eigenvalues of $K_{s,\beta}^L$. The squared ratio is at most the ratio itself, so rearranging gives
\begin{equation}
 2S_{\rm A}(L,\beta)-1
 \le\frac{Z_{\rm P}(L,\beta)}{Z_{\rm A}(L,\beta)}\le1.
 \label{eq:fermion-periodic-ratio}
\end{equation}
Thus the partition functions approach one another as $S_{\rm A}$ approaches one.

We can now bound the periodic Gibbs purity $T_{\rm P}(L,\beta)=Z_{\rm P}(L,2\beta)/Z_{\rm P}(L,\beta)^2$. Substituting the partition-function ratios gives
\begin{equation}
 \begin{split}
 T_{\rm P}(L,\beta)
 &=T_{\rm A}(L,\beta)
 \frac{Z_{\rm P}(L,2\beta)}{Z_{\rm A}(L,2\beta)}
 \left[\frac{Z_{\rm A}(L,\beta)}{Z_{\rm P}(L,\beta)}\right]^2\\
 &\ge T_{\rm A}(L,\beta)\,[2S_{\rm A}(L,2\beta)-1].
 \end{split}
 \label{eq:fermion-periodic-purity}
\end{equation}
The squared bracket is at least one, and the other ratio is bounded by Eq.~\eqref{eq:fermion-periodic-ratio}. Both purities on the right are  controlled by the bounds already proved for $S_{\rm A}$ and $T_{\rm A}$.

Suppose $S_{\rm A}(L_0,\beta_0),T_{\rm A}(2L_0,\beta_0)\ge0.99$ at some even starting length $L_0\ge4$. The same Gibbs-weight argument as in Sec.~\ref{sec:twists} then gives
\begin{equation}
 \Delta_{\rm P}(\ell)>\frac{\log10}{\beta_0},
 \qquad\ell\ge L_0,\quad\ell\ \mathrm{even}.
 \label{eq:fermion-periodic-gap}
\end{equation}
Here $\Delta_{\rm P}$ is the difference between the two lowest periodic energies, including multiplicities. Thus the periodic ground state is unique and the gap remains positive as the system grows. Appendix~\ref{app:fermion-gap} gives the estimates.

With more than one spatial direction, we change one boundary at a time. A sufficient condition is that Eq.~\eqref{eq:fermion-signed-local} hold for the cell Hamiltonian in each transfer direction with every choice of transverse periodic or antiperiodic boundaries. In particular, if the antiunitary preserves each internal interaction separately, reversing the sign of a closing hopping term preserves this condition.

\subsection{Application to symmetric mass generation}\label{sec:smg}
This application is motivated by the phenomenon of symmetric mass generation (SMG), which is a trivially gapped state of fermions whose existence necessarily requires interactions. See Ref.~\cite{wangyou2022} for a review. Each site of our model has four complex fermion modes $c_{i,a,\sigma}$, with layer $a=1,2$ and spin $\sigma=\uparrow,\downarrow$. We use a square-lattice $\pi$-flux version of the bilayer honeycomb Hamiltonian studied in Refs.~\cite{houyou2023,heyouxu2026}:
\begin{equation}
\begin{gathered}
H_{\rm f}=-it\sum_{i\to j,a,\sigma}\nu_{ij}
\bigl(c_{i,a,\sigma}^\dagger c_{j,a,\sigma}-c_{j,a,\sigma}^\dagger c_{i,a,\sigma}\bigr)
+J\sum_i\mathbf S_{i,1}\cdot\mathbf S_{i,2},\\
\mathbf S_{i,a}=\frac12\sum_{\sigma,\sigma'}
c_{i,a,\sigma}^\dagger\boldsymbol\sigma_{\sigma\sigma'}c_{i,a,\sigma'},
\qquad t\ge0,\quad J>0,
\end{gathered}
\label{eq:smg-overview}
\end{equation}
where $\boldsymbol\sigma$ denotes the three Pauli matrices. Bonds are oriented in the positive coordinate directions; for $i=(x,y)$, we take $\nu_{ij}=(-1)^y$ on horizontal bonds and $\nu_{ij}=1$ on vertical bonds, giving flux $\pi$ through each plaquette.

At $t=0$, the ground state is a product of on-site singlets, with one fermion in each layer. The finite test proves that a unique symmetric ground state and a size-independent gap persist throughout $0\le t/J\le1/35$ on sufficiently large even squares, with either periodic or antiperiodic spatial boundaries. The trace includes all occupation states at zero chemical potential. A particle--hole transformation verifies the Hermitian transfer conditions in both directions, allowing bounds on $4\times4$ and $8\times8$ systems to establish the antiperiodic gap. Comparing boundary conditions one direction at a time, as in Sec.~\ref{sec:fermion-periodic}, then gives the periodic gap from the same finite input. Uniqueness and the symmetries make the ground state a spin singlet with average density one fermion per layer per site. Appendix~\ref{app:smg} provides the details.

\section{Tensor-network states}\label{sec:peps}
We now use the same idea to study a wave function rather than a Gibbs state. Our goal is to establish exponential decay of connected correlations in a two-dimensional projected entangled-pair state (PEPS), using its norm on finite lattices. The norm will play  the role of the partition function from the previous sections, while the two spatial directions will replace space and imaginary time. We will obtain a bound on correlation functions that remains valid when both lengths grow together. We first describe the method and then apply it to the square-lattice spin-one PEPS constructed by Jian and Zaletel~\cite{jianzaletel2016}, who provided substantial evidence for its featureless character. In Sec.~\ref{sec:spin1-peps}, we will provide a rigorous proof using purity bootstrap that their state has exponentially decaying connected correlations of all local operators in the thermodynamic limit.

\subsection{The norm and its two transfer operators}
A PEPS specifies the amplitude of a spin configuration by multiplying local tensors and summing their shared indices. On the square lattice, each tensor $A^s_{lrdu}$ has one physical index $s$, labeling a state of the spin, and four virtual indices $l,r,d,u=1,\ldots,D$, one for each neighboring site. Explicitly, the wave function has the form
\[
|\Psi\rangle
=\sum_{\{s_i\}}
\left[
\sum_{\text{shared virtual indices}}
\prod_i A^{s_i}_{l_i r_i d_i u_i}
\right]
|\{s_i\}\rangle.
\]
Here $i$ labels lattice sites and each bond carries one summed virtual index, shared by the tensors at its two ends. For simplicity, we restrict to translationally-invariant PEPS defined by the same local tensor at every site and for every system size.  The squared norm of the PEPS on a periodic $L_x\times L_y$ rectangle is given by:
\begin{equation}
 \mathcal N(L_x,L_y)=\langle\Psi_{L_x,L_y}|\Psi_{L_x,L_y}\rangle
 \label{eq:peps-norm}
\end{equation}
 One way to calculate this norm is by doing it one column at a time which can then be recast as the trace of a matrix. In our notation, a column has fixed $x$ and contains $L_y$ sites, and there are $L_x$ columns. Let's fix a physical configuration $\boldsymbol s$ of one column and sum over the virtual indices on its vertical
bonds, while leaving the virtual indices on the horizontal bonds connecting this column to its left and right neighbors unsummed. These indices then define a $D^{L_y}\times D^{L_y}$ matrix $K_{\boldsymbol s}$. Closing the $L_x$ columns gives an
amplitude $\tr(K_{\boldsymbol s_1}\cdots
K_{\boldsymbol s_{L_x}})$. Summing its absolute square over physical configurations therefore gives
\begin{equation}
 \mathcal N(L_x,L_y)=\tr\mathbb T_x(L_y)^{L_x},
 \qquad
 \mathbb T_x(L_y)=\sum_{\boldsymbol s}
 K_{\boldsymbol s}\otimes\overline{K_{\boldsymbol s}}.
 \label{eq:peps-transfer}
\end{equation}
The overbar denotes complex conjugation and the two factors come from the ket and the bra. Equivalently, we can regard $\mathbb T_x$ as a linear map on $D^{L_y}\times D^{L_y}$ matrices $Y$ acting as $Y\mapsto\sum_{\boldsymbol s}K_{\boldsymbol s}YK_{\boldsymbol s}^\dagger$. This action makes it clear that $\mathbb T_x$ maps positive semidefinite matrices to positive semidefinite matrices.

One may repeat the above way of writing $\mathcal N(L_x, L_y)$ by switching the $x$ and $y$ directions, similar to how we wrote the partition function in two different ways in the previous sections. Analogously, the condition for the bootstrap to work is that both the column transfer and the row transfer are Hermitian:
\begin{equation}
 \mathcal N(L_x,L_y)=\tr\mathbb T_x(L_y)^{L_x}
 =\tr\mathbb T_y(L_x)^{L_y},\qquad
 \mathbb T_x(L_y)^\dagger=\mathbb T_x(L_y),\quad
 \mathbb T_y(L_x)^\dagger=\mathbb T_y(L_x).
 \label{eq:peps-two-transfers}
\end{equation}
Hermiticity of the two transfer operators is not automatic and must be verified for each PEPS. We also assume that the norms are nonzero on the even rectangles considered below. As an aside, a sufficient condition for the Hermiticity of both transfers is a real tensor satisfying
$A^s_{lrdu}=A^s_{rldu}=A^s_{lrud}$.

\subsection{From a finite norm ratio to decay of correlations}
We now want a finite test that bounds the correlation length in the two-dimensional thermodynamic limit. Suppose for the moment that $\mathbb T_x(L_y)$ has a unique leading eigenvalue $\lambda_0>0$, with $|\lambda_a|<\lambda_0$ for every other eigenvalue. On an infinite cylinder of circumference $L_y$, connected correlations between observables with $r$ columns
between their supports contain only factors $(\lambda_a/\lambda_0)^r$ with $a\ne0$. They therefore decay as $e^{-r\Delta_x(L_y)}$ or faster, where $\Delta_x(L_y):=-\log\max_{a\ne0}|\lambda_a/\lambda_0|$. This ``transfer gap'' plays the role of the energy gap in the earlier sections and its inverse bounds the correlation length.  We will find a sufficient condition on finite PEPS norms that guarantees a unique leading eigenvalue and a positive lower bound on $
\Delta_x(L_y)$ independent of $L_y$. Under this condition, we will also bound the correlation prefactor independently of $L_y$ and show
that the bound survives when both lengths grow.

\paragraph{A gap for the transfer operators.}
Since the transfer operators are Hermitian and $L_x,L_y$ are even, we can define the density matrices $\rho_x=\mathbb T_x(L_y)^{L_x}/\mathcal N(L_x,L_y)$
and $\rho_y=\mathbb T_y(L_x)^{L_y}/\mathcal N(L_x,L_y)$ on the corresponding doubled virtual spaces. Their purities are
\begin{equation}
 S_x(L_x,L_y)=\tr\rho_x^2=\frac{\mathcal N(2L_x,L_y)}{\mathcal N(L_x,L_y)^2},\qquad
 S_y(L_x,L_y)=\tr\rho_y^2=\frac{\mathcal N(L_x,2L_y)}{\mathcal N(L_x,L_y)^2}.
 \label{eq:peps-purities}
\end{equation}
 We now show that the following finite norm bound is enough to keep $\Delta_x(L_y)$ above a positive constant as $L_y$ grows, and likewise
in the other direction. We denote the starting side length by $L_0$:
\begin{equation}
 Q_{\rm PEPS}(L_0):=\frac{\mathcal N(2L_0,2L_0)}{\mathcal N(L_0,L_0)^4}
 =S_x(L_0,L_0)^2S_y(2L_0,L_0)\ge0.99,
 \qquad L_0\ge4\ \text{even}.
 \label{eq:peps-test}
\end{equation}
Since both purities are at most one, each is then at least $0.99$. The argument below only requires these two purity bounds.

Doubling the corresponding length squares the normalized weights, exactly as in Sec.~\ref{sec:one-d}. All the identities used there apply with $Z(L,\beta)$ replaced by $\mathcal N(L_x,L_y)$, $S$ by $S_x$, and $T$ by $S_y$. Equation~\eqref{eq:one-round} therefore doubles both lengths while replacing a deviation $u$ by $5u^2$. After $j$ rounds, with $L_j=2^j L_0$, we have
\[
 S_x(L_j,L_j),\ S_y(2L_j,L_j)\ge1-u_j,
 \qquad u_j=\frac15\,20^{-2^j}.
\]
To cover even transverse sizes between doublings, we apply Eq.~\eqref{eq:interpolation} with the axes interchanged:
\begin{equation}
S_x(L_j,L_y)\ge
\bigl[S_x(L_j,L_j)^2S_y(2L_j,L_j)\bigr]^{L_y/(2L_j)}
\ge1-3u_j,
\qquad L_j\le L_y\le2L_j. \label{eq:peps-interpolation}
\end{equation}

We first show that $\mathbb T_x(L_y)$ has a unique eigenvalue of largest absolute value. Its normalized weights $w_a=|\lambda_a|^{L_j}/
\mathcal N(L_j,L_y)$ sum to one, so Eq.~\eqref{eq:peps-interpolation} gives
\[
1-3u_j\le S_x(L_j,L_y)=\sum_a w_a^2\le\max_a w_a.
\]
The largest weight therefore exceeds $1/2$ (since $u_j \le 0.01$ for any $j$) and cannot be tied with another weight. We label its eigenvalue $\lambda_0$. Next, we show that $\lambda_0$ is positive. Equation~\eqref{eq:peps-transfer} gives
\[
\mathcal N(L_x,L_y)=\sum_a\lambda_a^{L_x}\ge0
\]
for odd lengths $L_x$ as well. If $\lambda_0$ were negative, its contribution would dominate this sum and make it negative for
sufficiently large odd $L_x$, contradicting positivity of the squared norm.

Finally, $w_0\ge1-3u_j$ and normalization give $w_a\le3u_j$ for $a\ne0$. Using $L_j=2^j L_0$, we obtain
\begin{equation}
\begin{gathered}
\left|\frac{\lambda_a}{\lambda_0}\right|^{L_j}
=\frac{w_a}{w_0}
\le\frac{3u_j}{1-3u_j}<20^{-2^j}\\
\Longrightarrow\quad
\left|\frac{\lambda_a}{\lambda_0}\right|<20^{-1/L_0}\quad(a\ne0),
\qquad \Delta_x(L_y)>\frac{\log20}{L_0}.
\end{gathered}
\label{eq:peps-eigenvalue-ratio}
\end{equation}
The bound holds for every even $L_y\ge L_0$. Exchanging the axes gives the same bound for the row transfer.

\paragraph{Decay of correlations on a cylinder.}
We have established a transfer gap that is independent of the
circumference $L_y$. This controls the exponential decay with
distance, but does not yet give a uniform bound on physical
correlations. The reason is that correlation functions also
involve transfer operators with physical observables inserted,
whose contributions could potentially grow with $L_y$.
We now show that these contributions are bounded by the norms
of the physical observables, independently of $L_y$.

We fix $L_y$ and take $L_x\to\infty$. Let
$\widehat{\mathbb T}=\mathbb T_x(L_y)/\lambda_0$ and
$P=|R\rangle\langle R|$ be its leading eigenprojector.
We can choose $|R\rangle$ to represent a positive semidefinite matrix $R$.
To see this, start with a positive matrix having nonzero overlap with the leading eigenvector.
Such a matrix exists because positive matrices span the Hermitian matrices.
Repeated applications of $\widehat{\mathbb T}$ preserve positivity and converge to a nonzero matrix in the leading eigenspace.
We normalize $R$ by $\tr R^2=1$.
For a Hermitian observable $O$ supported on $a$ columns,
we denote by $\mathcal E_O$ the transfer through these
columns with $O$ inserted, normalized by $\lambda_0^a$.
Explicitly,
\[
\mathcal E_O=\lambda_0^{-a}
\sum_{s,t}\langle s|O|t\rangle
K_t\otimes\overline{K_s},
\qquad
\mathcal E_I=\widehat{\mathbb T}^{\,a},
\]
where $K_s$ denotes the matrix associated with a physical
configuration $s$ of these $a$ columns.

For two observables $O_1,O_2$ separated by $r$ columns between their respective supports, the infinite-cylinder correlation function is
\[
\langle O_1O_2\rangle
=\langle R|\mathcal E_{O_1}
\widehat{\mathbb T}^{\,r}
\mathcal E_{O_2}|R\rangle.
\]
Since $\langle O_i\rangle
=\langle R|\mathcal E_{O_i}|R\rangle$, subtracting the
product of expectations removes the leading eigenprojector:
\begin{equation}
\langle O_1O_2\rangle-\langle O_1\rangle\langle O_2\rangle
=\langle R|\mathcal E_{O_1}
(\widehat{\mathbb T}^{\,r}-P)
\mathcal E_{O_2}|R\rangle.
\label{eq:peps-correlation-identity}
\end{equation}
The transfer gap bounds $\|\widehat{\mathbb T}^{r}-P\|$ by $20^{-r/L_0}$. Positivity also bounds the lengths of the vectors $\mathcal E_{O_1}^\dagger|R\rangle$ and $\mathcal E_{O_2}|R\rangle$ by $\|O_1\|$ and $\|O_2\|$, independently of $L_y$; Appendix~\ref{app:peps-cylinder} proves these bounds. Here $\|\cdot\|$ is the operator norm, the largest absolute eigenvalue for a Hermitian matrix. Consequently,
\begin{equation}
|\langle O_1O_2\rangle-\langle O_1\rangle\langle O_2\rangle|
\le\|O_1\|\,\|O_2\|\,20^{-r/L_0}.
\label{eq:peps-cylinder-correlation}
\end{equation}
This holds on every infinite cylinder with even circumference $L_y\ge L_0$, with a decay length at most $L_0/\log20$.

\paragraph{Taking both spatial lengths to infinity.}
So far, we have taken $L_x\to\infty$ at fixed $L_y$, leaving only the leading transfer eigenvector. We now want the same correlation bound
on growing periodic squares, $L_x=L_y=L\to\infty$. On a finite square, the subleading eigenvalues also contribute. The transfer gap
suppresses each contribution, but their number grows with $L$. We will now use the purity bound to control their total weight and show that the expectation values of $O_1$, $O_2$, and $O_1O_2$ on the
square approach those on the infinite cylinder of circumference $L$. This will let us transfer the bound on their connected correlation
from the cylinder to the square.
The purity bound makes the total subleading weight exponentially small in $L$. Appendix~\ref{app:peps-square} uses it to compare the square and cylinder expectations. For every even $L\ge L_0$, this gives
\begin{equation}
 |\langle O_1O_2\rangle_{L\times L} -\langle O_1\rangle_{L\times L}\langle O_2\rangle_{L\times L}| \qquad\le\|O_1\|\,\|O_2\| \left(20^{-r/L_0}+5\,20^{-L/(4L_0)}\right). \label{eq:peps-square-correlation}
\end{equation}
Here the two disjoint supports must together fit within $L/2$ consecutive columns. Every fixed pair of local observables satisfies this condition for sufficiently large $L$. Exchanging the axes gives the corresponding bound for separation in rows.

\paragraph{Uniqueness of the thermodynamic limit.}
We also want local expectation values to approach one definite answer as the lattice grows. The same square--cylinder comparison, applied in both directions, shows that expectations on different growing rectangles approach one another. Appendix~\ref{app:peps-limit} gives the convergence proof. Thus all local expectations, and hence the reduced density matrix on any fixed region, converge to one common limit as both even side lengths tend to infinity. This state selected by periodic rectangles satisfies the exponential correlation bound in Eq.~\eqref{eq:peps-cylinder-correlation}.

\paragraph{Symmetry breaking.}
If the finite-square states preserve a symmetry, their thermodynamic limit preserves it as well. An order parameter whose expectation vanishes by symmetry has a two-point function equal to its connected correlation. The exponential bound therefore excludes spontaneous symmetry breaking of any such zero-form symmetry.

\subsection{A simple example: a state with classical Ising weights}
Consider the spin-$\tfrac12$ state~\cite{verstraete2006criticality}
\begin{equation}
 |\Psi_{L_x,L_y}(x)\rangle
 =2^{-L_xL_y/2}\sum_{\{s_i=\pm1\}}
 \prod_{\langle ij\rangle}\sqrt{1+xs_i s_j}\,|\{s_i\}\rangle,
 \qquad 0<x<1.
 \label{eq:peps-ising-state}
\end{equation}
Its squared amplitudes are classical ferromagnetic Ising weights: $1+xs_i s_j=e^{Ks_i s_j}/\cosh K$ with $x=\tanh K$. Its norm is consequently
\[
 \mathcal N(L_x,L_y)=2^{-L_xL_y}\sum_{\{s_i\}}
 \prod_{\langle ij\rangle}(1+xs_i s_j).
\]
This state has a particularly simple PEPS representation. Let $B$ be the real positive square root of the two-by-two matrix with entries $\sqrt{1+xss'}$. Then
$A^s_{lrdu}=2^{-1/2}B_{sl}B_{sr}B_{sd}B_{su}$
reproduces Eq.~\eqref{eq:peps-ising-state}, because summing a shared bond index gives $(BB^{\mathsf T})_{ss'}=\sqrt{1+xss'}$. This tensor is real and unchanged by exchanging any of its virtual indices, so both norm transfers are Hermitian.

We now check the finite test at $x=1/10$ and $L_0=4$. Summing the $2^{16}$ spin configurations on the $4\times4$ torus gives, in exact rational arithmetic,
$1.002537<\mathcal N(4,4)<1.002538.$
For a lower bound on $\mathcal N(8,8)$, expand its product of bond factors. The spin sum removes every term in which a site occurs an odd number of times. Every surviving term is nonnegative. Keeping just the empty selection of bonds and the four bonds around each of the 64 plaquettes gives $\mathcal N(8,8)\ge1+64x^4$. Hence
\[
 Q_{\rm PEPS}(4)\ge\frac{1+64(1/10)^4}{(1.002538)^4}
 >0.996>0.99.
\]
The finite test therefore proves Eq.~\eqref{eq:peps-square-correlation} with $L_0=4$, and exponential decay of all local connected correlations in the thermodynamic limit. Only the finite spin sum and the displayed lower bound entered this conclusion.

\subsection{Application to the Jian--Zaletel spin-one PEPS}\label{sec:spin1-peps}
We now apply the finite norm test to the square-lattice spin-one
state of Jian and Zaletel~\cite{jianzaletel2016}.
Lieb--Schultz--Mattis--Oshikawa--Hastings theorems~\cite{liebschultzmattis1961,oshikawa2000,hastings2004} do not forbid
a featureless paramagnet with one spin one per unit cell and
the symmetries of a square lattice. Constructing featureless states that preserve both internal and lattice symmetries is nevertheless nontrivial~\cite{kimchi2013,parameswaran2013,kim2016}.
For spin one on the square lattice, the usual AKLT construction, with
one virtual spin half per bond, gives physical spin $S=z/2$,
where $z$ is the coordination number, and hence $S=2$ on the
square lattice~\cite{aklt1987}. Jian and Zaletel constructed a spin-one PEPS
with spin-rotation and square-lattice symmetries and found strong
numerical evidence for short-range correlations. We are not
aware of a rigorous proof of exponential decay of correlations
in this state in the thermodynamic limit; here we establish
this using the purity bootstrap. The state can be constructed from independent spin-one AKLT chains on every horizontal row and every vertical column. At each site, we project the two spin ones onto their total-spin-one subspace, leaving one physical spin one.

The resulting tensor has Hermitian norm transfers in both directions. We obtain bounds on its $12\times12$ and $24\times24$ norms.
One round of the doubling inequalities then gives purities
above $0.99$ at the starting scale $L_0=24$, as required in
Sec.~\ref{sec:peps}. Consequently, for bounded local Hermitian observables separated by $r$ columns between their supports, the thermodynamic limit obeys
\[
|\langle O_1O_2\rangle-\langle O_1\rangle\langle O_2\rangle|
\le\|O_1\|\,\|O_2\|\,20^{-r/24}.
\]
Thus the correlation length is at most $24/\log20<8.012$ lattice spacings.
This limit preserves spin and lattice symmetries, and the bound excludes N\'eel, quadrupolar, and dimer long-range order. Appendix~\ref{app:jz} gives the local tensor, the finite calculations, and the bound on finite squares. Whether the local parent Hamiltonian constructed in Appendix~\ref{app:jz-parent} is gapped in the thermodynamic limit remains open.

\begin{acknowledgments}
I thank Hal Tasaki for helpful comments on a draft of this paper.
The author used ChatGPT Astra and Codex (OpenAI) extensively to understand the original OpenAI proof~\cite{haldane}, explore extensions, implement numerical calculations and develop technical arguments. ChatGPT was also used to draft technical explanations, especially in the appendices, which the author substantially rewrote to make them more accessible. The author developed the main ideas behind the extensions to higher dimensions, symmetry twists, and fermions, while AI was used to refine the arguments and work through their details. For tensor-network states, the author proposed the initial direction, while AI developed much of the bootstrap argument connecting
finite PEPS norms to exponential decay of correlations in the
thermodynamic limit. Codex was also used extensively to develop
various variational bounds for the applications.
\end{acknowledgments}

\appendix
\section{Details of the doubling proof}\label{app:doubling}
This appendix provides the details of the calculations from Secs.~\ref{sec:one-d} and~\ref{sec:higher-d}. We derive the two-dimensional update rule
and prove that the gap bounds in one and two dimensions also hold at even lengths between successive doublings.

\subsection{Lengths between doublings in one dimension}\label{app:one-d-lengths}
Section~\ref{sec:one-d} gives a uniform gap bound at lengths $2L_j = 2^{j+1} L_0$. Following Ref.~\cite{haldane}, we now show that the same bound holds for every even length $\ell\ge L_0$. Let's choose $j$ such
that $L_j\le\ell\le2L_j$. At fixed inverse temperature, $Z(\ell,\beta)=\sum_a|\lambda_a(\beta)|^\ell$, so changing the length changes only the powers of the same nonnegative numbers. We can therefore compare partition
functions at different lengths using  $\sum_a x_a^q\le(\sum_a x_a^p)^{q/p}$ for $x_a\ge0$ and $q\ge p>0$.
Applying this inequality at inverse temperature $\beta_j$ to lengths $L_j\le\ell$, and at inverse temperature $2\beta_j$ to lengths $\ell\le2L_j$, gives \[ Z(\ell,\beta_j)\le Z(L_j,\beta_j)^{\ell/
	L_j},\qquad Z(\ell,2\beta_j)\ge Z(2L_j,2\beta_j)^{\ell/(2L_j)}. \]

We therefore have an upper bound on the denominator of $T(\ell,\beta_j)$ and a lower bound on its numerator. Dividing these bounds gives \begin{equation} \begin{aligned} T(\ell,
\beta_j)&\ge\left[\frac{Z(2L_j,2\beta_j)}{Z(L_j,\beta_j)^4}\right]^{\ell/(2L_j)}\\ &=\left[T(2L_j,\beta_j)S(L_j,\beta_j)^2\right]^{\ell/(2L_j)}\ge(1-u_j)^3\ge1-3u_j. \end{aligned}
\label{eq:interpolation} \end{equation} Here we used the bounds on $S$ and $T$ obtained by iteration. We also used $\ell/(2L_j)\le1$ (raising a number between zero and one to such a power can only increase it).

The same Gibbs-weight argument as before now gives \begin{equation} \Delta_\ell\ge\frac{1}{\beta_j}\log\frac{1-3u_j}{3u_j}>\frac{\log20}{\beta_0}. \label{eq:1d} \end{equation} The last inequality
follows from $3u_j/(1-3u_j)<5u_j\le20^{-2^j}$ and $\beta_j=2^j\beta_0$. Since the intervals $[L_j,2L_j]$ cover every length $\ell\ge L_0$, this proves the gap bound for every sufficiently large even ring.

\subsection{The two-dimensional update rule}\label{app:2d-updates}
Here we derive Eq.~\eqref{eq:2d-round} of Sec.~\ref{sec:higher-d}, starting from the three purity bounds in Eq.~\eqref{eq:2d-input}. We first improve $S_x$, then $S_y$, and finally $T$.

To obtain a bound on $S_x(2L,2L,2\beta)$, we first bound $S_x(L,2L,2\beta)$. Its denominator is $Z(L,2L,2\beta)^2$, which can be bounded as follows. Using $\sum_i a_i^2\le(\sum_i a_i)^2$ for nonnegative $a_i$, we obtain $Z(2L_x,L_y,\beta)\le Z(L_x,L_y,\beta)^2$, and similarly when doubling $L_y$ or $\beta$. Therefore, $Z(L,2L,2\beta)\le Z(L,2L,\beta)^2\le Z(L,L,\beta)^4$. Using this upper bound on the denominator of $S_x(L,2L,2\beta)$, we obtain
\begin{equation}
 \begin{split}
 S_x(L,2L,2\beta)
 &=\frac{Z(2L,2L,2\beta)}{Z(L,2L,2\beta)^2}
 \ge\frac{Z(2L,2L,2\beta)}{Z(L,L,\beta)^8}\\
 &=S_x(L,L,\beta)^4S_y(2L,L,\beta)^2T(2L,2L,\beta)\\
 &\ge(1-u)^7\ge1-7u.
 \end{split}
 \label{eq:2d-Rx}
\end{equation}
The equality in the second line  can be checked by substituting the definitions of the three purities. Intuitively, the exponents $4,2,1$ arise because each successive doubling squares the preceding partition function. By our assumption, all three purities are at least $1-u$, giving the exponent $4+2+1=7$; the last step is Bernoulli's inequality.

We now double the first length in $S_x(L,2L,2\beta)$, holding the second length at $2L$ and the inverse temperature at $2\beta$. As in the 1d proof, at this stage we can use the fact that $S_x(2L, 2L, 2 \beta)$ is the purity of  the normalized square of the distribution whose purity is $S_x(L, 2L, 2 \beta)$. Therefore, Eq.~\eqref{eq:square} gives
\begin{equation}
 1-S_x(2L,2L,2\beta) \le\frac{[1-S_x(L,2L,2\beta)]^2}{2S_x(L,2L,2\beta)^2} \le\frac{49u^2}{2(1-7u)^2}\le49u^2. \label{eq:2d-x-update}
\end{equation}
The last inequality uses $u\le1/98$, which gives $1-7u\ge13/14$ and hence $2(1-7u)^2>1$. Thus all three arguments of $S_x$ have been doubled, while the bound on its deviation from unity has improved from $u$ to $49u^2$. Since $49u^2\le u$ in this range, this new purity is also at least $1-u$ and can be used in the next step.

The remaining task is to improve $S_y$ and $T$ using the same steps. For $S_y$, start at $(2L,L,\beta)$ and double the second length, then time, then the first length. The three purities along this sequence are $S_y(2L,L,\beta)$, $T(2L,2L,\beta)$, and $S_x(2L,2L,2\beta)$, all now known to be at least $1-u$. The same calculation gives
\begin{equation}
 \begin{split}
 S_y(4L,L,2\beta)
 &\ge\frac{Z(4L,2L,2\beta)}{Z(2L,L,\beta)^8}\\
 &=S_y(2L,L,\beta)^4T(2L,2L,\beta)^2S_x(2L,2L,2\beta)\\
 &\ge(1-u)^7\ge1-7u.
 \end{split}
 \label{eq:2d-Ry}
\end{equation}
For the first inequality we used $Z(4L,L,2\beta)\le Z(2L,L,\beta)^4$, obtained by doubling time and then the first length; the equality uses the three known purities. We next double the second length from $L$ to $2L$, keeping the first length at $4L$ and the inverse temperature at $2\beta$. Equation~\eqref{eq:square} then gives, with the same estimate as in Eq.~\eqref{eq:2d-x-update},
\begin{equation}
 1-S_y(4L,2L,2\beta)\le49u^2.
 \label{eq:2d-y-update}
\end{equation}

To improve $T$, start instead at $(2L,2L,\beta)$ and double time, the first length, and the second length, in that order. The three purities are now $T(2L,2L,\beta)$, $S_x(2L,2L,2\beta)$, and $S_y(4L,2L,2\beta)$. Both spatial purities have just been bounded, so the same calculation gives
\begin{equation}
 \begin{split}
 T(4L,4L,\beta)
 &\ge\frac{Z(4L,4L,2\beta)}{Z(2L,2L,\beta)^8}\\
 &=T(2L,2L,\beta)^4S_x(2L,2L,2\beta)^2S_y(4L,2L,2\beta)\\
 &\ge(1-u)^7\ge1-7u.
 \end{split}
 \label{eq:2d-Rt}
\end{equation}
Here the denominator bound is $Z(4L,4L,\beta)\le Z(2L,2L,\beta)^4$, obtained by doubling both spatial lengths. At this fixed spatial size $4L\times4L$, we double $\beta$. This squares the Gibbs weights and gives $1-T(4L,4L,2\beta)\le49u^2$. Together these three bounds prove Eq.~\eqref{eq:2d-round}.

\subsection{Squares between doublings in two dimensions}\label{app:2d-lengths}

 Section~\ref{sec:higher-d} gives a uniform gap bound on squares of side $2L_j=2^{j+1}L_0$. We now show that the same bound holds on every square with even side length $\ell\ge L_0$. For a given $\ell$, choose $j$ such that $L_j\le\ell\le2L_j$. Changing one spatial length while keeping the other length and inverse temperature fixed changes only the powers of the same nonnegative numbers in Eq.~\eqref{eq:2d-transfer}. Applying $\sum_a x_a^q\le(\sum_a x_a^p)^{q/p}$ separately in the first and second spatial directions gives
\begin{equation}
 \begin{split}
 Z(\ell,\ell,\beta_j)
 &\le Z(L_j,\ell,\beta_j)^{\ell/L_j}
 \le Z(L_j,L_j,\beta_j)^{(\ell/L_j)^2},\\
 Z(\ell,\ell,2\beta_j)
 &\ge Z(2L_j,\ell,2\beta_j)^{\ell/(2L_j)}
 \ge Z(2L_j,2L_j,2\beta_j)^{(\ell/(2L_j))^2}.
 \end{split}
 \label{eq:2d-interp-bounds}
\end{equation}
In the first line we compare $\ell$ with the smaller length $L_j$, obtaining an upper bound. In the second we compare it with the larger length $2L_j$, obtaining a lower bound. Dividing the bound on the numerator of the Gibbs purity by the square of the bound on its denominator gives
\begin{equation}
 T(\ell,\ell,\beta_j)
 \ge\left[\frac{Z(2L_j,2L_j,2\beta_j)}
 {Z(L_j,L_j,\beta_j)^8}\right]^{(\ell/(2L_j))^2}
 \ge(1-u_j)^7\ge1-7u_j.
 \label{eq:2d-interpolation}
\end{equation}
The ratio in brackets has the same factorization into three purities as in Eq.~\eqref{eq:2d-Rx}, with $L,\beta$ replaced by $L_j,\beta_j$. It therefore lies between $(1-u_j)^7$ and one. Raising it to the power $(\ell/(2L_j))^2\le1$ can only increase it.

The Gibbs-weight argument now gives
\begin{equation}
 \Delta_\ell\ge\frac1{\beta_j}\log\frac{1-7u_j}{7u_j}
 >\frac{\log2}{\beta_0}.
 \label{eq:2d-gap}
\end{equation}
Here we used $7u_j/(1-7u_j)<49u_j\le2^{-2^j}$, which holds since $u_j\le1/98$. These intervals cover every even $\ell\ge L_0$. Thus checking the three bounds in Eq.~\eqref{eq:2d-input} at a single starting scale guarantees a gap on every larger even square. Both spatial lengths grow together, so the conclusion applies to a truly two-dimensional system.
\section{Details for symmetry twists and near degeneracy}\label{app:symmetry}
Here we give detailed proofs of the results in Sec.~\ref{sec:twists}: the spatial transfer construction, the gap and splitting bounds, and the extension to finite Abelian groups.

\subsection{The spatial transfer operators}\label{sec:ising-checks}
We use the local Hamiltonian and the conditions in Sec.~\ref{sec:ising-local} to prove the representation $\mathcal Z_k=\tr A_0^L+k\tr A_1^L$ with Hermitian $A_0,A_1$ independent of $L$.
We expand the bond interactions in imaginary time, keeping the
on-site evolution under $b$ exact. For a short time step,
\[
e^{-\delta\tau H_L}
=e^{-\delta\tau\sum_i b_i}
\left[I+\delta\tau\sum_{i,\alpha}
(F_\alpha)_i(F_\alpha)_{i+1}\right]+O(\delta\tau^2).
\]
Multiplying these steps and taking $\delta\tau\to0$ gives an
exact time-ordered expansion in bond insertions~\cite{aizenmannachtergaele1994,haldane}. Each insertion has a time and a
label $\alpha$, and places the same matrix $F_\alpha$ at both ends
of a bond. We denote the ordered list of insertions on bond
$(i,i+1)$ by $\omega_i$. For fixed bond histories, the many-body
trace factors into single-site traces involving the histories
$\omega_{i-1},\omega_i$ and the evolution under $b$.
An insertion of $U^a$, $a=0,1$, adds $X^a$ to every single-site trace.

The symmetry restricts which histories contribute. Define the
parity of a bond history by
$q(\omega)=\prod_{(\tau,\alpha)\in\omega}s_\alpha=\pm1,$
with $q(\varnothing)=1$. We call a history even when $q=1$
and odd when $q=-1$. For the Hamiltonian in Eq.~\eqref{eq:ising-interacting-chain}, this is simply
the parity of the number of $Z$ insertions; the $X$ insertions
do not affect it. Conjugating the single-site trace by $X$
leaves $b$ and $X^a$ unchanged but multiplies each
$F_\alpha$ by $s_\alpha$. The trace therefore equals itself
times $q(\omega_{i-1})q(\omega_i)$ and vanishes unless $q(\omega_{i-1})=q(\omega_i)$. Applying this at every site shows that all bond histories
in a nonzero contribution have the same parity. We next average over the two boundary conditions. The twist
multiplies a history by $q(\omega_L)$, since it changes the sign
of every odd insertion on the closing bond. Thus the average
multiplies each term by $\frac{1+q(\omega_L)}2$. It eliminates odd histories on the closing bond and hence, by
the preceding selection rule, on every bond. The two averaged
traces therefore contain only even bond histories.

We can now construct $A_0,A_1$. Their indices $\omega,\omega'$
are even histories on the left and right bonds of one site.
Merge their insertions into a single time-ordered list
$(\tau_1,\alpha_1),\ldots,(\tau_r,\alpha_r)$, with
$\tau_1<\cdots<\tau_r$, and define
\begin{equation}
A_a(\omega,\omega')
=\tr\!\left[X^a e^{-(\beta-\tau_r)b}F_{\alpha_r}
e^{-(\tau_r-\tau_{r-1})b}\cdots
F_{\alpha_1}e^{-\tau_1b}\right],
\qquad a=0,1.
\label{eq:ising-direct-kernel}
\end{equation}
For empty histories this is $\tr(X^a e^{-\beta b})$.
Exchanging $\omega$ and $\omega'$ does not change the merged
time-ordered list, so
$A_a(\omega,\omega')=A_a(\omega',\omega)$.
Since all matrices in the trace are real, $A_0,A_1$ are real
symmetric and hence Hermitian.

Finally, multiplying these kernels around the ring and
integrating over the shared bond histories gives
\begin{equation}
\begin{split}
\frac12\sum_{t=\pm1}\tr(U^a e^{-\beta H_L^{(t)}})
&=\int_{\rm even}d\omega_1\cdots d\omega_L\,
A_a(\omega_L,\omega_1)\cdots
A_a(\omega_{L-1},\omega_L)\\
&=\tr A_a^L,\qquad a=0,1.
\end{split}
\label{eq:ising-history-contraction}
\end{equation}
Here the integration sums over insertion numbers and labels
and integrates their ordered times, including the empty history.
The factorial suppression of the time-ordered integrals
ensures convergence and square-integrability of the kernels. Thus the relevant
operator powers have well-defined traces. Crucially, the operators
$A_0,A_1$ depend on $\beta$ and the local Hamiltonian but not
on $L$. Using $P_k=(I+kU)/2$ now gives
$\mathcal Z_k=\tr A_0^L+k\tr A_1^L$, as required.

\subsection{The gap and splitting for Ising symmetry}\label{app:ising-bounds}
\paragraph{A gap in the combined $k=+1$ spectrum.}
We can now apply Sec.~\ref{sec:one-d} to $\mathcal Z_{+1}=\tr(A_0\oplus A_1)^L$. Suppose $S_{+1}(L_0,\beta_0),T_{+1}(2L_0,\beta_0)\ge0.99$, where the even integer $L_0\ge4$ is the starting length and $\beta_0$ is the starting inverse temperature. Let $E_{0,k}(\ell),E_{1,k}(\ell)$ be the two lowest energies in sector $k$ among both boundary conditions $t=\pm1$, including multiplicities. Equation~\eqref{eq:1d} gives
\[
 E_{1,+1}(\ell)-E_{0,+1}(\ell)>\frac{\log20}{\beta_0}
 \quad(\ell\ge L_0,\ \ell\ \mathrm{even}).
\]
We also need the purity bounds from that argument. After $j$ rounds, the length and inverse temperature are $L_j,\beta_j$, with
$L_j=2^j L_0,\qquad \beta_j=2^j\beta_0,\qquad u_j=\frac15\,20^{-2^j}.$
Section~\ref{sec:one-d} gives $S_{+1}(L_j,\beta_j)\ge1-u_j$ and $S_{+1}(L_j,2\beta_j)\ge1-3u_j$. For any even length $\ell\ge L_0$, choose $j$ so that $L_j\le\ell\le2L_j$. At fixed inverse temperature, increasing the even length raises the nonnegative eigenvalue weights to a higher power and renormalizes them, which can only increase their purity.\footnote{For normalized weights $p_i$, raising them to a power $r\ge1$ multiplies each weight by $p_i^{r-1}/\sum_m p_m^r$. This factor increases with $p_i$, so probability moves from smaller weights to larger ones. Each such move increases the sum of squares: for $x\ge y$ and $\delta\ge0$, $(x+\delta)^2+(y-\delta)^2-x^2-y^2=2\delta(x-y)+2\delta^2\ge0$.}
Combining this with Eq.~\eqref{eq:interpolation} gives the bounds we will use below:
\begin{equation}
 S_{+1}(\ell,\beta_j)\ge1-u_j,\qquad
 S_{+1}(\ell,2\beta_j)\ge1-3u_j,\qquad
 T_{+1}(\ell,\beta_j)\ge1-3u_j.
 \label{eq:twist-inputs}
\end{equation}

\paragraph{Proof that the ratio $\mathcal Z_{-1}/\mathcal Z_{+1}$ approaches one exponentially with system size.}
We now use the bounds on $S_{+1}$ to compare the partition functions. Section~\ref{sec:ising-argument} gave
\[
S_{+1}\le\frac12\left[1+\left(\frac{\mathcal Z_{-1}}{\mathcal Z_{+1}}\right)^2\right].
\]
Since $0<\mathcal Z_{-1}/\mathcal Z_{+1}\le1$, its square is at most the ratio itself. Rearranging gives
\begin{equation}
 0\le1-\frac{\mathcal Z_{-1}(L,\beta)}{\mathcal Z_{+1}(L,\beta)}
 \le2\bigl[1-S_{+1}(L,\beta)\bigr].
 \label{eq:ising-sector-comparison}
\end{equation}
Applying the bounds on $S_{+1}$ at $\beta_j$ and $2\beta_j$ in Eq.~\eqref{eq:twist-inputs}, we obtain
\begin{equation}
 1-2u_j\le
 \frac{\mathcal Z_{-1}(\ell,\beta_j)}{\mathcal Z_{+1}(\ell,\beta_j)}
 \le1,\qquad
 1-6u_j\le
 \frac{\mathcal Z_{-1}(\ell,2\beta_j)}{\mathcal Z_{+1}(\ell,2\beta_j)}
 \le1.
 \label{eq:ising-ratio-scales}
\end{equation}
Both deviations from one are bounded by multiples of $u_j$, which decays exponentially with $L_j$. We will use these bounds to obtain the $k=-1$ gap and the splitting between the two lowest energies.

\paragraph{A gap in the combined $k=-1$ spectrum.}
For every even length $\ell\in[L_j,2L_j]$, we now bound $T_{-1}(\ell,\beta_j)$ using $T_{+1}(\ell,\beta_j)$ and the partition-function ratios. Recalling $T_k(L,\beta)=\mathcal Z_k(L,2\beta)/\mathcal Z_k(L,\beta)^2$, we have
\begin{equation}
 \begin{split}
 T_{-1}(\ell,\beta_j)
 &=T_{+1}(\ell,\beta_j)
 \frac{\mathcal Z_{-1}(\ell,2\beta_j)}{\mathcal Z_{+1}(\ell,2\beta_j)}
 \left[\frac{\mathcal Z_{+1}(\ell,\beta_j)}
 {\mathcal Z_{-1}(\ell,\beta_j)}\right]^2\\
 &\ge(1-3u_j)(1-6u_j)\ge1-9u_j.
 \end{split}
 \label{eq:twisted-purity}
\end{equation}
The squared bracket is at least one; the other factors are bounded by Eqs.~\eqref{eq:twist-inputs} and \eqref{eq:ising-ratio-scales}.

Applying the Gibbs-weight argument of Eq.~\eqref{eq:purity-to-gap} gives
\begin{equation}
 E_{1,-1}(\ell)-E_{0,-1}(\ell) \ge\frac1{\beta_j}\log\frac{T_{-1}(\ell,\beta_j)}{1-T_{-1}(\ell,\beta_j)} \ge\frac1{\beta_j}\log\frac{1-9u_j}{9u_j} >\frac{\log10}{\beta_0}. \label{eq:twist-gap}
\end{equation}
For the last inequality, we used $u_j\le0.01$ and
\[
\frac{9u_j}{1-9u_j}<10u_j
=2 \cdot 20^{-2^j}\le10^{-2^j},
\qquad \beta_j=2^j\beta_0.
\]
Together with the $k=+1$ result, this bounds both sector gaps below by $(\log10)/\beta_0$, independently of $\ell$.

\paragraph{An exponentially small difference between the two lowest energies.}
We now bound $E_{0,-1}(\ell)-E_{0,+1}(\ell)$. Since $\mathcal Z_{-1}(L,\beta)\le\mathcal Z_{+1}(L,\beta)$ for every $\beta$, taking $-\beta^{-1}\log$ and then $\beta\to\infty$ at fixed $L$ gives $E_{0,-1}(L)\ge E_{0,+1}(L)$. For the upper bound, we recall from Sec.~\ref{sec:ising-argument} that
$T_k\mathcal Z_k\le e^{-\beta E_{0,k}}\le\mathcal Z_k.$
Dividing the lower bound for $k=-1$ by the upper bound for $k=+1$ gives
\[
 e^{-\beta_j[E_{0,-1}(\ell)-E_{0,+1}(\ell)]}
 \ge T_{-1}(\ell,\beta_j)
 \frac{\mathcal Z_{-1}(\ell,\beta_j)}{\mathcal Z_{+1}(\ell,\beta_j)}
 \ge(1-9u_j)(1-2u_j).
\]
Taking logarithms therefore yields
\begin{equation}
 \begin{split}
 0\le E_{0,-1}(\ell)-E_{0,+1}(\ell)
 &\le\frac{-\log(1-9u_j)-\log(1-2u_j)}{\beta_j}
 <\frac{12u_j}{\beta_j}.
 \end{split}
 \label{eq:twist-band}
\end{equation}
The last estimate follows from $-\log(1-x)\le x/(1-x)$:
\[
-\log(1-9u_j)-\log(1-2u_j)
\le\left(\frac9{0.91}+\frac2{0.98}\right)u_j<12u_j.
\]
To express this bound in terms of the length, we substitute $u_j=\frac15\,20^{-2^j}$ and $2^j=L_j/L_0$. Since $\ell\le2L_j$ and $\beta_j\ge\beta_0$,
\[
\frac{12u_j}{\beta_j}
=\frac{12}{5\beta_j}e^{-(\log20)L_j/L_0}
\le\frac{12}{5\beta_0}e^{-(\log20)\ell/(2L_0)}.
\]
Thus the two lowest energies differ by an amount exponentially small in $\ell$, with the starting scale $L_0,\beta_0$ held fixed.

\paragraph{Both low states belong to the untwisted Hamiltonian.}
We want to prove that $E_{0,k,t=+1}\le E_{0,k,t=-1}$ for every $L$ and $k=\pm1$. We choose a single-site basis that diagonalizes both $b$ and $X$, which is possible because $[b,X]=0$. If every off-diagonal entry of $v$ in the corresponding two-site basis is real and nonpositive, then the required ordering holds. This requires checking only the fixed two-site matrix $v$.

To prove the implication for every $L$, recall that the twist replaces the closing bond by $v^{(-1)}=(I\otimes X)v(I\otimes X)$. In this product basis, the twist changes
only signs of off-diagonal entries. Consider any normalized state $\psi$ in sector $k$, and replace its coefficients $\psi_\sigma$ in this basis by their absolute values
to obtain $|\psi|$. This preserves both its norm and its sector. For each bond, an off-diagonal entry $-c$, with $c\ge0$, contributes under either boundary condition
\[
\pm2c\,\operatorname{Re}(\psi_\sigma^*\psi_{\sigma'})
\ge-2c\,|\psi_\sigma|\,|\psi_{\sigma'}|.
\]
The right-hand side is the corresponding untwisted contribution for $|\psi|$. The on-site term $b$ is diagonal, and all diagonal contributions are unchanged, so summing over the Hamiltonian gives
\begin{equation}
\langle\psi,H_L^{(t)}\psi\rangle
\ge\langle|\psi|,H_L^{(t=+1)}|\psi|\rangle
\ge E_{0,k,t=+1}.
\label{eq:sectors}
\end{equation}
Minimizing over $\psi$ in sector $k$ proves the required ordering. For  example, $v=-JZ\otimes Z-gX\otimes X$ has diagonal $XX$ contributions and off-diagonal entries $-J$ in this basis, so the additional condition holds for $J>0$.

We therefore have $E_{0,k}=E_{0,k,t=+1}$: both low states belong to the original Hamiltonian. Every other untwisted state also occurs in its sector's combined spectrum and
hence lies more than $(\log10)/\beta_0$ above that sector's lowest energy. Since the two lowest energies differ by less than $12u_j/\beta_j\le0.12/\beta_0$, they form the
lowest pair in the original spectrum.

Writing $W_\ell=E_1(\ell)-E_0(\ell)$ for their splitting and $\Gamma_\ell=E_2(\ell)-E_1(\ell)$ for the gap above them, we have $\Gamma_\ell>(\log10)/\beta_0-W_\ell>(\log8)/\beta_0$. Combining this with the exponential bound just proved gives Eq.~\eqref{eq:multiplet}.
These bounds cover every even $\ell\ge L_0$, since each lies in an interval $[L_j,2L_j]$.

\subsection{More general finite Abelian symmetries}\label{sec:abelian}
We now want to show the presence of one low-lying state in each of the $m$ sectors of a $\mathbb Z_m$ symmetry. Consider a periodic nearest-neighbor chain $H_L=\sum_i(b_i+v_{i,i+1})$, with symmetry generator $U=\prod_i V_i$. Each $V_i$ is the same unitary $V$ acting on site $i$, with $V^m=I$, and each local term commutes with $U$. Sector $k$ consists of states with $U$ eigenvalue $e^{2\pi i k/m}$, where $k=0,\ldots,m-1$.

As for the Ising case, we combine boundary conditions at fixed $k$. Boundary condition $t$ replaces the closing bond by
\[
v_{L,1}^{(t)}=V_1^{-t}v_{L,1}V_1^t,\qquad t=0,\ldots,m-1.
\]
The conjugation acts on site $1$ only. We call the resulting Hamiltonian $H_L^{(t)}$, with $t=0$ untwisted. For $m=2$ and $V=X$, $k,t=0,1$ correspond to the Ising labels $+1,-1$.

We project onto sector $k$ and sum its partition functions over these boundary conditions:
\begin{equation}
 P_k=\frac1m\sum_{a=0}^{m-1}e^{-2\pi i ka/m}U^a,
 \qquad
 \mathcal Z_k(L,\beta)=\sum_{t=0}^{m-1}
 \tr\!\left(P_k e^{-\beta H_L^{(t)}}\right).
 \label{eq:abelian-projector}
\end{equation}
The sum over $a$ selects the symmetry eigenvalue while the sum over $t$ combines boundary conditions.

We need the counterparts of the two Ising assumptions. First, for each insertion $U^a$ in the projector, we require a Hermitian spatial transfer operator $A_a$ satisfying
\begin{equation}
 \tr A_a^L=\frac1m\sum_{t=0}^{m-1}
 \tr\!\left(U^a e^{-\beta H_L^{(t)}}\right),
 \qquad a=0,\ldots,m-1.
 \label{eq:abelian-transfer-blocks}
\end{equation}
These operators may depend on $\beta$ but not on $L$; the traces must converge for every even $L\ge4$ and $\beta>0$.

We now form the direct sum of the $m$ transfer spaces and define
\[
 X_\beta=\bigoplus_{a=0}^{m-1}A_a,
 \qquad R_k=\bigoplus_{a=0}^{m-1}e^{-2\pi i ka/m}I.
\]
Substituting Eq.~\eqref{eq:abelian-transfer-blocks} into the projector formula gives
\begin{equation}
 \mathcal Z_k=\sum_{a=0}^{m-1}e^{-2\pi i ka/m}\tr A_a^L
             =\tr(X_\beta^L R_k).
 \label{eq:twist}
\end{equation}
Thus $X_\beta$ is Hermitian, and $R_k$ is unitary and commutes with $X_\beta$. For $m=2$, this gives $R_0=I\oplus I$ and $R_1=I\oplus(-I)$, exactly as in the Ising construction.

Second, we require the lowest energy in each sector to occur with the original boundary condition: $E_{0,k,t=0}\le E_{0,k,t}$ for every $t$. As in Eq.~\eqref{eq:sectors}, nonpositive off-diagonal entries of the untwisted local terms in a product basis diagonalizing $V_i$ suffice for this ordering. We assume all $m$ sectors are nonempty at every allowed length.

We first apply the Ising argument to $\mathcal Z_0=\tr X_\beta^L$. Defining $S_0,T_0$ by Eq.~\eqref{eq:purities}, the finite test $S_0(L_0,\beta_0),T_0(2L_0,\beta_0)\ge0.99$ at an even starting length $L_0\ge4$ gives a gap in the combined $k=0$ spectrum and purity bounds at larger sizes.

We next compare $\mathcal Z_k$ with $\mathcal Z_0$. The new feature is that the coefficients in Eq.~\eqref{eq:twist} are complex phases. For even $L$, each $\tr A_a^L$ is nonnegative, and expanding the squared modulus gives
\[
 \begin{aligned}
 |\mathcal Z_k|^2
 &=\sum_a(\tr A_a^L)^2
 +2\sum_{a<b}\cos\!\left(\frac{2\pi k(a-b)}m\right)
                  \tr A_a^L\tr A_b^L\\
 &\ge2\sum_a(\tr A_a^L)^2-\mathcal Z_0^2
 \ge2\sum_a\tr A_a^{2L}-\mathcal Z_0^2
 =(2S_0-1)\mathcal Z_0^2.
 \end{aligned}
\]
We used $\cos\theta\ge-1$, followed by $\tr A_a^{2L}\le(\tr A_a^L)^2$. The last equality recalls $S_0=\sum_a\tr A_a^{2L}/\mathcal Z_0^2$ (in the Ising case $S_0$ was called $S_{+1}$). Also, $\mathcal Z_k>0$ is a sum of sector partition functions, and the triangle inequality gives $\mathcal Z_k\le\mathcal Z_0$. Dividing by $\mathcal Z_0^2$ and using $(\mathcal Z_k/\mathcal Z_0)^2\le\mathcal Z_k/\mathcal Z_0$ therefore yields
\begin{equation}
0\le1-\frac{\mathcal Z_k}{\mathcal Z_0}
\le2(1-S_0).
\label{eq:invisible}
\end{equation}

Applying this comparison at $\beta$ and $2\beta$, Eqs.~\eqref{eq:twisted-purity}--\eqref{eq:twist-band} give a gap within each combined spectrum and exponentially small differences between their lowest energies. The assumed energy ordering puts all $m$ low states in the original Hamiltonian.

Every finite Abelian group has the form $G=\mathbb Z_{m_1}\times\cdots\times\mathbb Z_{m_r}$. We use one sector and twist label per factor, multiplying the corresponding projectors and the phases in $R_k$. Under the same conditions, the argument gives $g=|G|$ low states, with Eq.~\eqref{eq:multiplet} applying to
$W_L=E_{g-1}(L)-E_0(L),\qquad \Gamma_L=E_g(L)-E_{g-1}(L).$
This conclusion assumes all $g$ sectors are nonempty and together exhaust the physical spectrum. If an application uses traces covering only part of that spectrum, the remaining states need a separate lower energy bound, as in the toric-code example in Sec.~\ref{sec:toric}.

\section{Details of the fermionic transfer construction}\label{app:fermions}
Here we prove the spatial-transfer representation and periodic-boundary gap stated in Sec.~\ref{sec:fermions}. We start from the fermionic path integral and keep track of the signs that arise when its fields are grouped by cell.

\subsection{Writing the partition function as a spatial trace}\label{app:fermion-construction}
We now want to obtain the spatial trace from the usual fermionic path integral~\cite[Sec.~6.10]{shankar2017}. Let $\chi_j^a(\tau)$ be the anticommuting Grassmann fields associated with the Majoranas. For the Hamiltonian in Eq.~\eqref{eq:fermionclass}, we write
\begin{equation}
 Z_{\rm A}(L,\beta)
 =\int_{\rm AP}\mathcal D\chi\,
 \prod_{j=1}^L e^{-S_b[\chi_j]}
 \prod_{j=1}^L
 \exp\!\left[\frac{it}{2}\int_0^\beta d\tau
                 \sum_{a=1}^{2M}\chi_j^a(\tau)\chi_{j+1}^a(\tau)\right].
 \label{eq:fermion-path-integral}
\end{equation}
Here $S_b[\chi_j]$ is the action of an isolated cell, including its imaginary-time kinetic term and interaction $b$. The subscript AP specifies antiperiodic boundaries in time and space:
$\chi_j^a(\beta)=-\chi_j^a(0),\qquad \chi_{L+1}^a(\tau)=-\chi_1^a(\tau).$
As usual, this compact notation means the limit of the time-sliced coherent-state integral.

We expand the hopping exponentials in the second product, retaining all orders, while leaving the factors $e^{-S_b[\chi_j]}$ intact. We can then collect the fields belonging to each cell and integrate them. Factoring hopping terms in this way is also used in Grassmann tensor constructions~\cite{akiyamakadoh2021}.

\paragraph{Separating the two ends of a bond.}
We first fix the bond connecting cells $j$ and $j+1$. An order-$N$ term in its hopping exponential describes $N$ hopping events. We denote their list by
\[
 \omega=\bigl((\tau_1,a_1),\ldots,(\tau_N,a_N)\bigr),
 \qquad 0<\tau_1<\cdots<\tau_N<\beta.
\]
Here $\tau_r$ is the imaginary time of event $r$, and $a_r\in\{1,\ldots,2M\}$ specifies its Majorana label. Each event inserts $\chi_j^{a_r}(\tau_r)\chi_{j+1}^{a_r}(\tau_r)$, so there are $2N$ fields in total, with $N$ at each endpoint.

We want to collect these fields by cell. The reordering gives
\[
 \begin{aligned}
 &[\chi_j^{a_N}(\tau_N)\chi_{j+1}^{a_N}(\tau_N)]\cdots
   [\chi_j^{a_1}(\tau_1)\chi_{j+1}^{a_1}(\tau_1)]\\
 &\quad=(-1)^{N(N-1)/2}
 \underbrace{\chi_j^{a_N}(\tau_N)\cdots\chi_j^{a_1}(\tau_1)}_{\Phi_\omega[\chi_j]}
 \underbrace{\chi_{j+1}^{a_N}(\tau_N)\cdots\chi_{j+1}^{a_1}(\tau_1)}_{\Phi_\omega[\chi_{j+1}]}.
 \end{aligned}
\]
This defines $\Phi_\omega[\chi_j]$ as the product of the $N$ fields at cell $j$; the other endpoint uses the same times and Majorana labels. We set $\Phi_\varnothing=1$ for an empty list. The sign counts the $0+1+\cdots+(N-1)=N(N-1)/2$ interchanges needed to group the fields. After taking out the positive weight $(t/2)^N$, the coefficient is therefore
\begin{equation}
 i^N(-1)^{N(N-1)/2}
 =\begin{cases}1,&N\text{ even},\\ i,&N\text{ odd}.
 \end{cases}
 \label{eq:fermion-bond-phase}
\end{equation}
We include $(t/2)^N$, the sum over labels, and the ordered time integrals in a positive measure $d\mu(\omega)$.

\paragraph{Obtaining the transfer from one cell.}
For each bond $(j,j+1)$, we denote by $\omega_j$ the ordered list of hopping times and Majorana labels appearing in a term of the expansion of its hopping exponential. Cell $j$ receives the fields $\Phi_{\omega_{j-1}}[\chi_j]$ from bond $(j-1,j)$ and $\Phi_{\omega_j}[\chi_j]$ from bond $(j,j+1)$. Their total number is $N(\omega_{j-1})+N(\omega_j)$, where $N(\omega)$ denotes the length of a list. Since $b$ preserves parity, the single-cell integral vanishes if this sum is odd. Consequently all bond lists around the ring must have the same parity $s=0$ or $1$. For lists of parity $s$, we define
\begin{equation}
 K_{s,\beta}(\omega_{j-1},\omega_j)
 =i^s\int_{\mathrm{AP\ in\ time}}\!\mathcal D\chi_j\,
 e^{-S_b[\chi_j]}\Phi_{\omega_{j-1}}[\chi_j]\Phi_{\omega_j}[\chi_j].
 \label{eq:fermion-kernels}
\end{equation}
The first argument labels the incoming bond list and the second the outgoing list. Since the cells are identical, the same kernel $K_{s,\beta}$ is used at every $j$. The integral is the unnormalized, fermion-time-ordered thermal correlation of the inserted Majoranas; with no insertions it equals $\tr e^{-\beta b}$. These identities also define the integral through finite-dimensional operator traces.

We next close the ring. Moving $\Phi_{\omega_L}[\chi_1]$ to the first cell gives $(-1)^s$, since this group and the remaining fields both have parity $s$. The antiperiodic spatial seam supplies another $(-1)^s$, so the two signs cancel. Multiplying the cell contributions and integrating the shared bond lists gives an ordinary trace:
\begin{equation}
 \begin{split}
 Z_{\rm A}(L,\beta)
 &=\sum_{s=0}^1\int\prod_{j=1}^L d\mu_s(\omega_j)
             \prod_{j=1}^L K_{s,\beta}(\omega_{j-1},\omega_j)\\
 &=\tr K_{0,\beta}^L+\tr K_{1,\beta}^L
 =\tr X_\beta^L,\qquad X_\beta=K_{0,\beta}\oplus K_{1,\beta}.
 \end{split}
 \label{eq:assembly}
\end{equation}
Here $\omega_0=\omega_L$ and $d\mu_s$ restricts the lists to parity $s$. With periodic spatial boundaries, the closing sign would remain, giving $\tr K_{0,\beta}^L-\tr K_{1,\beta}^L$ instead.

\paragraph{Showing that the transfer is Hermitian.}
The integral in Eq.~\eqref{eq:fermion-kernels} is real by Eq.~\eqref{eq:fermion-local}. Exchanging $\Phi_\omega[\chi_j]$ and $\Phi_{\omega'}[\chi_j]$ gives $(-1)^{N(\omega)N(\omega')}=(-1)^s$. Thus $K_{0,\beta}$ is real symmetric, while $K_{1,\beta}$ is imaginary antisymmetric. Both obey
\begin{equation}
 K_{s,\beta}(\omega',\omega)
 =\overline{K_{s,\beta}(\omega,\omega')}.
 \label{eq:fermion-kernel-hermitian}
\end{equation}
The factor of $i$ for odd fermionic products also appears in Majorana reflection positivity~\cite{jaffejanssens2016}.
We interpret $K_{s,\beta}$ as an ordinary integral operator on functions of $\omega$, with inner product $\int d\mu_s\,\overline f g$. Its kernel is bounded by $2^M e^{\beta\|b\|}$ and the total history measure is $e^{Mt\beta}$, so the squared kernel is integrable and the even power traces converge. Here and below, $\|\cdot\|$ denotes the operator norm for matrices; for a Hermitian matrix such as $b$, this is the largest absolute value of its eigenvalues. We have obtained a Hermitian $X_\beta$, independent of $L$, completing Eq.~\eqref{eq:fermion-two-representations}.

\subsection{Signed hopping between columns}\label{app:fermion-signed}
We use the hopping signs and antiunitary condition in Eq.~\eqref{eq:fermion-signed-local}.
We now check that the signed hopping gives a Hermitian transfer. Each event in a history $\omega$ specifies a time $\tau$, a row $y$, and a Majorana label $r$. We write
$\eta(\omega)=\prod_{(\tau,y,r)\in\omega}\eta_{y,r},$
with $\eta(\varnothing)=1$. Let $G_\beta(\omega,\omega')$ denote the integral in Eq.~\eqref{eq:fermion-kernels}, now over the fields of one column. Including the factors $(t/2)^{N(\omega)}$ in the history measure, the hopping signs and Eq.~\eqref{eq:fermion-signed-local} give
\[
 K_{s,\beta}(\omega,\omega')=i^s\eta(\omega')G_\beta(\omega,\omega'),
 \qquad
 \overline{G_\beta(\omega,\omega')}
 =\eta(\omega)\eta(\omega')G_\beta(\omega,\omega').
\]
Exchanging the two Grassmann products still gives $(-1)^s$. Consequently,
\[
 \overline{K_{s,\beta}(\omega,\omega')}
 =(-i)^s\eta(\omega)G_\beta(\omega,\omega')
 =i^s\eta(\omega)G_\beta(\omega',\omega)
 =K_{s,\beta}(\omega',\omega).
\]

This proves Hermiticity for the column transfer. The same calculation applies to rows.

\subsection{The gap with periodic boundaries}\label{app:fermion-gap}
To state the resulting gap, suppose $S_{\rm A}(L_0,\beta_0),T_{\rm A}(2L_0,\beta_0)\ge0.99$ at some even starting length $L_0\ge4$. As in Eq.~\eqref{eq:twist-inputs}, let $L_j=2^j L_0$, $\beta_j=2^j\beta_0$, and $u_j=20^{-2^j}/5$. For every even $\ell\in[L_j,2L_j]$, we have
\[
 T_{\rm A}(\ell,\beta_j)\ge1-3u_j,
 \qquad S_{\rm A}(\ell,2\beta_j)\ge1-3u_j.
\]
Equation~\eqref{eq:fermion-periodic-purity} therefore gives $T_{\rm P}(\ell,\beta_j)\ge(1-3u_j)(1-6u_j)\ge1-9u_j$. Applying the Gibbs-weight argument of Eq.~\eqref{eq:purity-to-gap}, exactly as in Eq.~\eqref{eq:twist-gap}, yields
\begin{equation}
 \Delta_{\rm P}(\ell)\ge\frac1{\beta_j}\log\frac{1-9u_j}{9u_j}
 >\frac{\log10}{\beta_0},
 \qquad\ell\ge L_0,\quad\ell\ \mathrm{even}.
\end{equation}
Here $\Delta_{\rm P}$ is the difference between the two lowest periodic energies, including multiplicities. Thus the periodic ground state is unique and the gap remains positive as the system grows.

\section{Details of the PEPS proof}\label{app:peps}
We complete the correlation proof of Sec.~\ref{sec:peps}. We first bound correlations on infinite cylinders, then on finite squares. Finally, we show that local expectation values converge to a unique limit as both lengths grow.

\subsection{Correlations on an infinite cylinder}\label{app:peps-cylinder}
The transfer gap already controls how correlations decrease with separation. We now bound the remaining factors independently of the cylinder circumference $L_y$. Combining Eq.~\eqref{eq:peps-correlation-identity} with $\|\widehat{\mathbb T}^{\,r}-P\|\le20^{-r/L_0}$ and applying Cauchy--Schwarz gives
\[
|\langle O_1O_2\rangle-\langle O_1\rangle\langle O_2\rangle|
\le
\|\mathcal E_{O_1}^\dagger(R)\|_F\,
\|\mathcal E_{O_2}(R)\|_F\,20^{-r/L_0},
\]
where $\|B\|_F^2=\sum_{i,j}|B_{ij}|^2=\tr(B^\dagger B)$ defines the Frobenius norm. Recall that $R$ is the positive semidefinite matrix representing the leading transfer eigenvector, normalized by $\tr R^2=1$. We want to bound the two norms on the right by $\|O_1\|$ and $\|O_2\|$, where $\|O\|$ is the operator norm, or the largest absolute eigenvalue of a Hermitian $O$.

For a Hermitian observable $O$ supported on $a$ columns, choose an orthonormal physical basis of eigenstates of $O$, with eigenvalues $o_s$. Let $K_s$ be the matrix of amplitudes for state $s$ of these columns, with rows and columns labeling the left and right virtual boundaries. The transfer with $O$ inserted acts on $R$ as
\[
\mathcal E_O(R)=\lambda_0^{-a}\sum_s o_s K_s R K_s^\dagger,
\qquad \mathcal E_I(R)=R.
\]
We can now bound the norm directly:
\[
\begin{aligned}
\|\mathcal E_O(R)\|_F^2
&=\lambda_0^{-2a}\sum_{s,t}o_so_t\,
\tr(K_s R K_s^\dagger K_t R K_t^\dagger)\\
&\le\|O\|^2\lambda_0^{-2a}\sum_{s,t}
\tr(K_s R K_s^\dagger K_t R K_t^\dagger)\\
&=\|O\|^2\tr[\mathcal E_I(R)^2]
=\|O\|^2.
\end{aligned}
\]
Each trace is nonnegative because it is the trace of a product of two positive semidefinite matrices, $K_s R K_s^\dagger$ and $K_t R K_t^\dagger$. The inequality therefore follows from $o_so_t\le\|O\|^2$.
Replacing each $K_s$ by $K_s^\dagger$ gives the same bound for $\mathcal E_O^\dagger(R)$, since Hermiticity of $\mathcal E_I=\widehat{\mathbb T}^{\,a}$ gives $\mathcal E_I^\dagger(R)=R$.
Substituting these bounds into the correlation inequality proves Eq.~\eqref{eq:peps-cylinder-correlation}, with decay length and prefactor independent of the even circumference $L_y\ge L_0$.

\subsection{From cylinders to finite squares}\label{app:peps-square}
We now bound the difference between the expectation of an observable on a square and on the cylinder of the same circumference. Let $L\ge L_0$ be even and divide the square into two halves of $m=L/2$ columns, with the Hermitian observable $O$ acting only in one half. Here $\mathcal E_O$ includes all $m$ columns in that half, with the identity on unused sites. The other half contributes $\widehat{\mathbb T}^{m}$, where $\widehat{\mathbb T}=\mathbb T_x(L)/\lambda_0$. Thus
\[
 \langle O\rangle_{L\times L}
 =\frac{\tr(\widehat{\mathbb T}^{m}\mathcal E_O)}{\tr\widehat{\mathbb T}^{L}},
 \qquad
 \langle O\rangle_{\mathrm{cyl},L}=\tr(P\mathcal E_O),
 \qquad P=|R\rangle\langle R|.
\]
We write $\epsilon_L:=1-1/\tr\widehat{\mathbb T}^{L}$ for the fraction of the square's norm contributed by subleading eigenvalues. Subtracting the two expectations and applying Cauchy--Schwarz gives
\[
 \begin{aligned}
 |\langle O\rangle_{L\times L}-\langle O\rangle_{\mathrm{cyl},L}|
 &=\left|\frac{\tr[(\widehat{\mathbb T}^{m}-P)\mathcal E_O]}{\tr\widehat{\mathbb T}^{L}}
   -\epsilon_L\langle O\rangle_{\mathrm{cyl},L}\right|\\
 &\le\frac{\|\widehat{\mathbb T}^{m}-P\|_F\,\|\mathcal E_O\|_F}{\tr\widehat{\mathbb T}^{L}}
   +\epsilon_L\|O\|.
 \end{aligned}
\]
Here $\|\cdot\|_F$ is the Frobenius norm defined above, whereas $\|O\|$ is the largest absolute eigenvalue of $O$. We now bound the two norms on the right. First, Hermiticity, $\widehat{\mathbb T}P=P$, and $2m=L$ give
\[
 \|\widehat{\mathbb T}^{m}-P\|_F^2
 =\tr\widehat{\mathbb T}^{L}-1
 =\frac{\epsilon_L}{1-\epsilon_L}.
\]
To bound $\|\mathcal E_O\|_F$, we again choose eigenstates of $O$, with eigenvalues $o_s$. The matrices $K_s$ now contain amplitudes for these $m$ columns, so
\[
 \mathcal E_O=\lambda_0^{-m}\sum_s o_s K_s\otimes\overline{K_s},
 \qquad
 \mathcal E_I=\lambda_0^{-m}\sum_s K_s\otimes\overline{K_s}
 =\widehat{\mathbb T}^{m}.
\]
Expanding $\|\mathcal E_O\|_F^2=\tr(\mathcal E_O^\dagger\mathcal E_O)$ and using $\tr(A\otimes B)=\tr A\,\tr B$, we obtain
\[
 \begin{aligned}
 \|\mathcal E_O\|_F^2
 &=\lambda_0^{-L}\sum_{s,t}o_so_t
   \left|\tr(K_s^\dagger K_t)\right|^2\\
 &\le\|O\|^2\lambda_0^{-L}\sum_{s,t}
   \left|\tr(K_s^\dagger K_t)\right|^2\\
 &=\|O\|^2\|\mathcal E_I\|_F^2
 =\|O\|^2\tr\widehat{\mathbb T}^{L}
 =\frac{\|O\|^2}{1-\epsilon_L}.
 \end{aligned}
\]
The inequality uses $o_so_t\le\|O\|^2$ and the nonnegative squared overlaps. The last line uses Hermiticity and $2m=L$.
Substituting these two bounds into the difference of expectations gives
\begin{equation}
 |\langle O\rangle_{L\times L}-\langle O\rangle_{\mathrm{cyl},L}|
 \le\|O\|\bigl(\sqrt{\epsilon_L}+\epsilon_L\bigr).
 \label{eq:peps-finite-cylinder}
\end{equation}

It remains to show that $\epsilon_L$ is exponentially small in $L$. Choose $j$ with $L_j\le L\le2L_j$. Equation~\eqref{eq:peps-interpolation} bounds the leading weight of $\mathbb T_x(L)^{L_j}$ below by $1-3u_j$. Increasing the even power from $L_j$ to $L$ can only increase this weight. Therefore
\begin{equation}
 \epsilon_L=1-\frac{\lambda_0^L}{\tr\mathbb T_x(L)^L}
 \le3u_j=\frac35\,20^{-2^j}\le\frac35\,20^{-L/(2L_0)}.
 \label{eq:peps-square-tail}
\end{equation}
The last inequality uses $2^j\ge L/(2L_0)$. Thus the square and cylinder expectations approach one another exponentially with $L$.

Applying this to $O_1,O_2,O_1O_2$ bounds the change in their connected correlation by $3\|O_1\|\|O_2\|(\sqrt{3u_j}+3u_j)$. Since $2^j\ge L/(2L_0)$, this proves Eq.~\eqref{eq:peps-square-correlation}.
Here the two disjoint supports must together fit within $L/2$ consecutive columns. Every fixed pair of local observables satisfies this condition for sufficiently large $L$. Exchanging the axes gives the corresponding bound for separation in rows.

\subsection{A unique thermodynamic limit}\label{app:peps-limit}

Finally, we show that local expectation values converge to the same limit regardless of how the two even side lengths grow. For an observable supported in a square of side at most $L_j/2$, the comparison in Eq.~\eqref{eq:peps-finite-cylinder} applies to rectangles as well. If $L_j\le L_y\le2L_j$ and $L_x\ge L_j$, with both lengths even, Eq.~\eqref{eq:peps-interpolation} and increasing the even transfer power again bound the total subleading weight by $3u_j$. Thus the finite expectation differs from that on the infinite cylinder of circumference $L_y$ by at most $\|O\|(\sqrt{3u_j}+3u_j)$.

For even $L_x,L_y\in[L_j,2L_j]$, compare
\[
 L_x\times L_x\ \longrightarrow\ L_x\times L_y\
 \longrightarrow\ L_y\times L_y.
\]
The first pair shares the cylinder obtained by extending the $y$ direction to infinity, with circumference $L_x$. The second shares the cylinder obtained by extending the $x$ direction, with circumference $L_y$. Each finite expectation is close to its corresponding cylinder expectation, so
\begin{equation}
 |\langle O\rangle_{L_x\times L_x}-\langle O\rangle_{L_y\times L_y}|
 \le4\|O\|\bigl(\sqrt{3u_j}+3u_j\bigr).
 \label{eq:peps-square-comparison}
\end{equation}
Taking $L_x=L_j$, $L_y=L_{j+1}$ and summing these bounds proves convergence along the doubled squares, since $u_j=\frac15\,20^{-2^j}$ makes the sum finite. The same comparison shows that every even size between doublings approaches this limit as well.

To include arbitrary growing rectangles, choose $j$ with $L_j\le L_x\le2L_j$ and suppose $L_y\ge L_x$. The comparison $L_x\times L_y\to L_x\times(2L_j)\to(2L_j)\times(2L_j)$ gives the same bound. Exchanging the axes covers $L_x>L_y$. Thus all local expectations, and hence the reduced density matrix on any fixed region, converge to one common limit as both even side lengths tend to infinity. The limiting state satisfies Eq.~\eqref{eq:peps-cylinder-correlation}.

\section{Details of the applications}\label{app:applications}
We give the calculations supporting the applications at the ends of Secs.~\ref{sec:higher-d}--\ref{sec:peps}. We first explain the partition-function bounds common to the four Hamiltonians, then check the transfer conditions and finite inputs for each model. The final subsection treats the Jian--Zaletel state.
\subsection{Bounds used in the finite tests}\label{app:finite-bounds}
\paragraph{Bounds on the two finite partition functions.}
For the $2+1$-dimensional Hamiltonians considered below, we want a lower bound on $Q(L,\beta)=Z(2L,2L,2\beta)/Z(L,L,\beta)^8$ from Eq.~\eqref{eq:2d-cube}. For each model, we specify the states and boundary conditions included in the trace $Z$. A bound $Q\ge97/98$ is sufficient to give all three initial purity bounds.\footnote{\label{sec:numerical-updates}In these examples the initial bound on $Q$ can be below $97/98$. We first apply Eqs.~\eqref{eq:2d-Rx}, \eqref{eq:2d-Ry}, and \eqref{eq:2d-Rt} and the squaring inequality with the computed bounds. After finitely many doublings, all three purities exceed $97/98$, and the theorem applies. We denote this starting scale by $(L_0,\beta_0)$. This refinement uses no additional partition-function calculations.}

To bound $Q$ from below, we need an upper bound on $Z(L,L,\beta)$ and a lower bound on $Z(2L,2L,2\beta)$. For the latter, it suffices to find an energy $E_{\rm tr}\ge E_0(2L,2L)$, where $E_0$ is the lowest energy included in this trace. We obtain such an upper bound from a normalized trial state, checking any sector restriction in the corresponding application. Then
\[
 Z(2L,2L,2\beta)\ge e^{-2\beta E_0(2L,2L)}\ge e^{-2\beta E_{\rm tr}}.
\]
Combining this with an upper bound $Z_U$ on the $L\times L$ partition function gives
\begin{equation}
 Z(L,L,\beta)\le Z_U
 \quad\Longrightarrow\quad
 Q(L,\beta)\ge e^{-D},\qquad
 D:=8\log Z_U+2\beta E_{\rm tr}.
 \label{eq:trial}
\end{equation}
Thus we can check the finite test using just a partition-function bound and a trial energy. In the examples below these calculations use $L=4$: we bound $Z(4,4,\beta)$ and choose a trial state on $8\times8$. An upper bound on $Z(4,4,\beta)$ must account for every state included in its trace.

\paragraph{Bounds for a parameter regime using only end points}\label{sec:coupling-intervals}
It would be convenient if we could certify every coupling between two values $s_1$ and $s_2$ by checking only these endpoints. Suppose $H(s)=H_0+sV$.   We keep the system sizes and inverse temperature $\beta$ fixed, and choose one normalized trial state for the whole interval $[s_1,s_2]$. In particular, both endpoint trial energies must be evaluated in this same
state. Then $\log Z$ is convex in $s$, and the trial energy $E_{\rm tr}(s)$ is linear in $s$. Therefore, their combination $8\log
Z+2\beta E_{\rm tr}$ is also convex in $s$. Consequently,
\begin{equation}
8\log Z(L,L,\beta;s)+2\beta E_{\rm tr}(s)
\le \max\{D(s_1),D(s_2)\},
\qquad s_1\le s\le s_2,
\label{eq:coupling-interval}
\end{equation}
where $D(s_1),D(s_2)$ are the computed upper bounds from Eq.~\eqref{eq:trial} at the two endpoints. If both bounds are below the threshold required by the finite test, the
entire interval is certified.  In the applications below, we subdivide the coupling range into subintervals and make these fixed choices separately for each subinterval.

\subsection{Finite partition-function bounds for the Heisenberg bilayer}\label{app:bilayer}
For the Hamiltonian in Sec.~\ref{sec:bilayer}, we use the ordinary partition function $Z_{\rm bi}=\tr e^{-\beta H_{\rm bi}}$ with periodic boundaries. We verify its transfer conditions and the finite bounds below.

\paragraph{Verifying the two spatial transfer representations.}\label{sec:spin-transfer}
We first check that this partition function has Hermitian spatial transfer operators in both directions, as required in Sec.~\ref{sec:higher-d}. Time reversal squares to $+1$ on a rung, so we can choose a rung basis in which it acts by complex conjugation. In that basis the rung Hamiltonian is real and symmetric, and every $iS_a^\alpha$ is real. A bond within a layer therefore has the form
\[
 J S_{i,a}^\alpha S_{j,a}^\alpha=-A_{i,a}^\alpha A_{j,a}^\alpha,
 \qquad A_a^\alpha=i\sqrt J\,S_a^\alpha,
\]
with the same real matrix at both ends.

To see why this gives a Hermitian transfer, we group rungs into columns and expand the bonds between columns in imaginary time. A column contributes a real trace of its internal evolution and the matrices inserted by its left and right neighbors. Exchanging the left and right lists leaves this trace unchanged, because the two ends of every bond carry the same matrix. Thus the transfer kernel is real and symmetric. The identical construction with rows gives the second spatial representation. This argument also applies below whenever the cell Hamiltonian is real symmetric and bonds have the form $-\sum_\alpha A_{\alpha,i}A_{\alpha,j}$ with real $A_\alpha$.

To bound the ratio in Eq.~\eqref{eq:2d-cube}, we need an upper bound on the $4\times4$ partition function and a lower bound on the $8\times8$ partition function at twice the inverse temperature. On both sizes, it is convenient to measure energies relative to the singlet product. For $N$ rungs, we write
\[
 \widehat H_{\rm bi}=H_{\rm bi}+\tfrac34J_\perp N I
                   =J_\perp\mathcal N+V,
\]
where $\mathcal N$ counts triplet rungs and $V$ is the interaction within the layers. We denote the shifted partition function by $\widehat Z_{\rm bi}$. This constant per rung cancels from the ratio, so
\[
Q(4,\beta)=\frac{\widehat Z_{\rm bi}(8,8,2\beta)}{\widehat Z_{\rm bi}(4,4,\beta)^8}.
\]
All energies below refer to $\widehat H_{\rm bi}$.

\paragraph{An upper bound on $\widehat Z_{\rm bi}(4,4,\beta)$.}
Since $E\ge\mu$ implies $e^{-\beta E}\le e^{-\beta\mu}$, we need lower bounds on the energies and a count of the levels they cover. The $4\times4$ system has $N=16$ rungs, with $m_k=\binom Nk3^k$ basis states containing $k$ triplets. Although $V$ mixes different triplet numbers, these dimensions will still let us count the exact energy levels.

We first bound energies on the subspace with at least $k$ triplets, for $1\le k\le N$. In the singlet--triplet basis, $V$ preserves triplet number or changes it by two. Its number-preserving part is bounded below by $-2J\ell$ on the $\ell$-triplet subspace: each bond contributes at least $-J(\mathcal N_i+\mathcal N_j)/2$, and each rung has four neighbors. For the part creating two triplets, a bond creates three possible pairs with amplitude $J/2$.

Let $e_\ell$ be the largest number of bonds with both ends among any $\ell$ rungs of the $4\times4$ torus; we find these numbers by enumerating its subsets. From an $\ell$-triplet configuration, pair creation can use at most $e_{N-\ell}$ singlet--singlet bonds, with three possible pairs per bond. In a resulting configuration, at most $e_{\ell+2}$ bonds can have supplied the pair. Thus the creation matrix has absolute column and row sums at most $3Je_{N-\ell}/2$ and $Je_{\ell+2}/2$, respectively. Its norm is bounded by the square root of their product.

We collect these bounds in a matrix indexed by triplet number $\ell=1,\ldots,N$:
\begin{equation}
 C_{\ell\ell}=(J_\perp-2J)\ell,\qquad
 C_{\ell,\ell+2}=C_{\ell+2,\ell}
 =-\frac{\sqrt3J}{2}\sqrt{e_{N-\ell}e_{\ell+2}},
 \label{eq:bilayer-comparison}
\end{equation}
with all other off-diagonal entries zero. Let $\mu_k$ be a lower bound on the smallest eigenvalue of its submatrix with indices $k,\ldots,N$. For any normalized state $\psi=\sum_{\ell=k}^N\psi_\ell$, where $\psi_\ell$ has exactly $\ell$ triplets, the preceding estimates give
\begin{equation}
 \langle\psi,\widehat H_{\rm bi}\psi\rangle
 \ge\sum_{\ell,m=k}^N C_{\ell m}\|\psi_\ell\|\,\|\psi_m\|
 \ge\mu_k\sum_{\ell=k}^N\|\psi_\ell\|^2=\mu_k.
 \label{eq:bilayer-energy-comparison}
\end{equation}
Thus matrices with at most 16 rows give energy bounds on subspaces of the full $4^{16}$-dimensional Hilbert space.

We next count how many exact energy levels obey each bound. The states with fewer than $k$ triplets span $\sum_{j<k}m_j$ dimensions. By the variational principle, at most that many levels can lie below $\mu_k$: any larger space of low-energy states would contain a state orthogonal to all those configurations, contradicting Eq.~\eqref{eq:bilayer-energy-comparison}. We can therefore bound the next $m_k$ Gibbs weights by $e^{-\beta\mu_k}$. Taking $k=1,\ldots,N$ accounts for every level except the ground state.

The remaining contribution is $e^{-\beta E_0}$. To bound it from above, we need a lower bound on $E_0$, for which we use a trial state on this same $4\times4$ system. Starting from the singlet product, we apply $V$, retain the components at each triplet number, and repeat up to four applications. This gives 16 linearly independent vectors. Their overlap and Hamiltonian entries are sums of products of single-rung matrix elements. Minimizing in this span gives a normalized state with energy $E_R$ and variance $\sigma^2=\langle(\widehat H_{\rm bi}-E_R)^2\rangle$. The counting argument above gives $E_1\ge\mu_1$. If $E_R<\mu_1$, Temple's inequality~\cite{temple1928} then yields
\begin{equation}
 E_-:=E_R-\frac{\sigma^2}{\mu_1-E_R}\le E_0,\qquad
 \widehat Z_{\rm bi}(4,4,\beta)
 \le e^{-\beta E_-}+\sum_{k=1}^{16}\binom{16}{k}3^k e^{-\beta\mu_k}.
 \label{eq:temple}
\end{equation}
The first inequality follows by taking the expectation of $(\widehat H_{\rm bi}-E_0)(\widehat H_{\rm bi}-\mu_1)$, which is nonnegative on every eigenstate. The second inequality now bounds the full partition function, including all $4^{16}$ states.

\paragraph{A lower bound on $\widehat Z_{\rm bi}(8,8,2\beta)$.}
We now need only an upper bound on the $8\times8$ ground-state energy. We use the same trial-state construction on this size, again obtaining a 16-dimensional span. If its optimized energy is $E_{\rm tr}$, the variational principle gives
\[
\widehat Z_{\rm bi}(8,8,2\beta)
\ge e^{-2\beta E_0(8,8)}\ge e^{-2\beta E_{\rm tr}}.
\]

The upper bound in Eq.~\eqref{eq:temple} and this lower bound supply the two inputs to Eq.~\eqref{eq:trial}. At $\beta J_\perp=67/4$, they yield a certificate throughout $0\le J/J_\perp\le1/10$, using Eq.~\eqref{eq:coupling-interval} to cover the interval. This gives a unique ground state and a size-independent gap for $J_\perp/J\ge10$ on sufficiently large even periodic squares.

\subsection{Gap and exponentially small splitting for the \texorpdfstring{$2+1$}{2+1}-dimensional Ising model}\label{app:tfim}
We verify the conditions and finite inputs for Sec.~\ref{sec:tfim}, then derive the gap and an upper bound on the splitting exponentially small in the system volume $L^2$.

\paragraph{Sector partition functions.}
To apply the spatial-transfer argument within a symmetry sector, we sum over boundary twists as in Sec.~\ref{sec:ising-argument}. The symmetry is $U=\prod_iX_i$. There are two twists, $t_x,t_y=\pm1$, and we define
\begin{equation}
 \mathcal Z_k(L_x,L_y,\beta)
 =\sum_{t_x,t_y=\pm1}\tr\!\left[\frac{I+kU}{2}
 e^{-\beta H^{(t_x,t_y)}}\right],\qquad k=\pm1.
 \label{eq:tfim-traces}
\end{equation}
Recall that a negative twist reverses the sign of the Ising bonds crossing the corresponding boundary seam. The original Hamiltonian has $t_x=t_y=+1$.

Following Sec.~\ref{sec:ising-argument}, we first show that the combined $k=+1$ spectrum has a gap. For this, in Eq.~\eqref{eq:2d-cube} we use $\mathcal Z_{+1}$ in place of $Z$. To proceed, we need Hermitian spatial transfer operators for this same $\mathcal Z_{+1}$ in both directions. For the $x$ direction, we regard a column as one cell. Its Hamiltonian is real and commutes with the product of $X$ operators in that column. Bonds between columns have the form $-JZ_y\otimes Z_y$, and conjugating $Z_y$ by the column symmetry changes its sign. The construction in Appendix~\ref{sec:ising-checks} therefore applies. Summing over $t_x$, then taking direct sums of the operators for $t_y=\pm1$, gives
$\mathcal Z_k(L_x,L_y,\beta)=\tr A_0^{L_x}+k\tr A_1^{L_x}.$
Here $A_0,A_1$ depend on $L_y,\beta$ but not $L_x$. Repeating the construction with rows gives the corresponding representation in $y$, with the same sector label $k$. Also, the untwisted bonds have nonpositive off-diagonal entries in the product $X$ basis. The energy-ordering argument of Appendix~\ref{app:ising-bounds} therefore puts each sector's lowest energy among the four twists in the original, untwisted Hamiltonian.

\paragraph{An upper bound on $\mathcal Z_{+1}(4,4,\beta)$.}
We bound the ratio in Eq.~\eqref{eq:2d-cube} using $4\times4$ and $8\times8$ systems. On $4\times4$, we diagonalize the Hamiltonian in sector $k=+1$ for each of the four twists. Summing all these Gibbs weights gives $\mathcal Z_{+1}(4,4,\beta)$, the denominator input to the finite test.

\paragraph{A lower bound on $\mathcal Z_{+1}(8,8,2\beta)$.}
We use a trial state for the untwisted $8\times8$ Hamiltonian,
\[
 \psi(\sigma)\propto\exp\!\left[\alpha\sum_i\sigma_i
 +\kappa\sum_{\langle ij\rangle}\sigma_i\sigma_j\right],
 \qquad Z_i|\sigma\rangle=\sigma_i|\sigma\rangle,\quad \sigma_i=\pm1.
\]
The two parameters allow a preference for one magnetization and correlations between neighboring spins. Its norm and energy can be summed row by row without truncation. For $h>0$, the untwisted ground state is unique with positive amplitudes in this basis, so the spin-flip symmetry leaves it invariant: $U=+1$. At $h=0$ there is also a ground state with $U=+1$. Thus its ground energy is the lowest energy $E_{0,+1}$ entering $\mathcal Z_{+1}$, and the trial energy gives
\[
\mathcal Z_{+1}(8,8,2\beta)
\ge e^{-2\beta E_{0,+1}(8,8)}\ge e^{-2\beta E_{\rm tr}}.
\]
These two finite bounds enter Eq.~\eqref{eq:trial}. Using Eq.~\eqref{eq:coupling-interval}, they initialize the argument throughout $0\le h/J\le1.8$. We now derive the gap and splitting that follow from them.

\paragraph{A gap above the two low states.}
We define $S_x,S_y,T_{+1}$ from $\mathcal Z_{+1}$ by Eq.~\eqref{eq:2d-purities}. Applying the doubling bounds to the finite input gives a scale $(L_0,\beta_0)$ at which
\[
 S_x(L_0,L_0,\beta_0),\quad S_y(2L_0,L_0,\beta_0),\quad
 T_{+1}(2L_0,2L_0,\beta_0)\ge1-10^{-3}.
\]
Here $L_0,\beta_0$ label the scale reached by the inequalities; the finite calculations still use only $4\times4$ and $8\times8$.
For deviations below $10^{-3}$, the coefficient in Eq.~\eqref{eq:2d-x-update} is less than $25$. The three purities at $L_j=2^j L_0$, $\beta_j=2^j\beta_0$ therefore deviate from one by at most
$u_j=\frac1{25}\,40^{-2^j}.$
For even rectangles $L_j\le L_x,L_y\le2L_j$, the length comparisons of Sec.~\ref{sec:higher-d} give $T_{+1}\ge1-7u_j$. On the same rectangle, $S_x$ is at least $1-3u_j$ at $\beta_j$ and $1-7u_j$ at $2\beta_j$. Equation~\eqref{eq:ising-sector-comparison} therefore gives $\mathcal Z_{-1}/\mathcal Z_{+1}\ge1-6u_j$ and $1-14u_j$ at the respective temperatures. Substituting these ratios into $T_k=\mathcal Z_k(2\beta)/\mathcal Z_k(\beta)^2$, we obtain
\[
 T_{-1}(\beta_j)
 =T_{+1}(\beta_j)
 \frac{\mathcal Z_{-1}(2\beta_j)}{\mathcal Z_{+1}(2\beta_j)}
 \left[\frac{\mathcal Z_{+1}(\beta_j)}{\mathcal Z_{-1}(\beta_j)}\right]^2
 \ge(1-7u_j)(1-14u_j)\ge1-21u_j.
\]
All factors here refer to the same rectangle, and the squared bracket is at least one.

To extract the sector gaps, and later control the excited contribution at lower temperatures, let $E_{a,k}$ denote the ordered energies in sector $k$, including all four twists. Write
\begin{equation}
 \mathcal Z_k=e^{-\beta E_{0,k}}(1+r_k),\qquad
 r_k(\beta)=\sum_{a\ge1}e^{-\beta(E_{a,k}-E_{0,k})}.
 \label{eq:tfim-excited-contribution}
\end{equation}
The ground-state Gibbs weight is $1/(1+r_k)$ and is at least $T_k$. Thus, in both sectors,
\begin{equation}
 r_k(\beta_j)\le\frac{21u_j}{1-21u_j}<40^{-2^j}
 =e^{-\gamma\beta_j},\qquad \gamma=\frac{\log40}{\beta_0}.
 \label{eq:tfim-excited-bound}
\end{equation}
Every term in this positive sum is smaller than the bound on the sum, so $E_{1,k}-E_{0,k}>\gamma$. The comparison of ground-state contributions in Eq.~\eqref{eq:twist-band} also gives
\[
 0\le W_{L_x,L_y}:=E_{0,-1}-E_{0,+1}
 \le\frac{-\log(1-6u_j)-\log(1-21u_j)}{\beta_j}
 <\frac{28u_j}{\beta_j}.
\]
Both lowest states belong to the untwisted Hamiltonian. Their separation is smaller than either sector gap, so they are its two lowest states. Writing $\Gamma_L=E_2(L)-E_1(L)$ for the gap above this pair on a square gives
\begin{equation}
 \Gamma_L>\frac{\log40-0.028}{\beta_0}>0
 \qquad(L\ge L_0,\ L\text{ even}).
 \label{eq:tfim-sector-gap}
\end{equation}

\paragraph{A splitting exponentially small in the system volume.}
We now use the gap within each sector to show that $W_L\equiv W_{L,L}$ is bounded by $Ce^{-cL^2}$, with $C,c>0$ independent of $L$. We first prove this for the squares of side $L_j=2^j L_0$, then include the lengths between them. In each round we double both spatial lengths at fixed inverse temperature, then multiply the inverse temperature by four for the next round.

We first show that doubling each spatial direction squares the relative difference between the sector partition functions. At fixed temperature, we write this difference as
\[
 \delta(L_x,L_y,\beta):=1-\frac{\mathcal Z_{-1}}{\mathcal Z_{+1}}
 =\frac{2\tr A_1^{L_x}}{\mathcal Z_{+1}}.
\]
At fixed $L_y,\beta$, the eigenvalues of $A_1^{L_x}$ are nonnegative because $L_x$ is even. Hence $\tr A_1^{2L_x}\le(\tr A_1^{L_x})^2$. Using $S_x(L_x,L_y,\beta)=\mathcal Z_{+1}(2L_x,L_y,\beta)/\mathcal Z_{+1}(L_x,L_y,\beta)^2$ in the denominator gives
\begin{equation}
 \delta(2L_x,L_y,\beta)
 \le\frac{2(\tr A_1^{L_x})^2}
 {S_x(L_x,L_y,\beta)\mathcal Z_{+1}(L_x,L_y,\beta)^2}
 =\frac{\delta(L_x,L_y,\beta)^2}{2S_x(L_x,L_y,\beta)}.
 \label{eq:tfim-delta-square}
\end{equation}
The calculation in $y$ bounds the same $\delta$, since it compares the same sectors $k=\pm1$. If $S_x(L_j,L_j,\beta)$ and $S_y(2L_j,L_j,\beta)$ both exceed $1/2$, the denominators $2S_x,2S_y$ exceed one. Doubling first $x$ and then $y$ therefore gives
\[
 \delta(2L_j,2L_j,\beta)\le\delta(L_j,L_j,\beta)^4.
\]
Repeated fourth powers would give a suppression exponential in $4^j=L_j^2/L_0^2$, which is the $L_j^2$ dependence we want. We next need to translate this improvement into a bound on the energy splitting.

To connect $\delta$ to $W_{L_x,L_y}$, we substitute the ground-state and excited-state contributions from Eq.~\eqref{eq:tfim-excited-contribution} into the ratio of partition functions:
\begin{equation}
 1-\delta(L_x,L_y,\beta)
 =e^{-\beta W_{L_x,L_y}}\frac{1+r_{-1}(\beta)}{1+r_{+1}(\beta)}.
 \label{eq:tfim-ratio-splitting}
\end{equation}
All quantities here refer to the same rectangle. If the excited-state contributions were zero, we would have $\beta W_{L_x,L_y}=-\log(1-\delta)$, which is close to $\delta$ when $\delta$ is small. More precisely, whenever $r_{\pm1}\le\varepsilon$, the exact relation gives
\begin{equation}
 \begin{split}
 \delta&\le\beta W_{L_x,L_y}+\varepsilon,\\
 \beta W_{L_x,L_y}
 &\le-\log(1-\delta)+\log(1+\varepsilon)
 \le\frac{\delta}{1-\delta}+\varepsilon.
 \end{split}
 \label{eq:tfim-splitting-comparison}
\end{equation}
Thus, to obtain a splitting exponentially small in $L^2$, we also need the excited-state contributions to be exponentially small in $L^2$.

We accomplish this by lowering the temperature, using the gap already proved within each sector. Since every excitation energy is greater than $\gamma=(\log40)/\beta_0$, the definition of $r_k$ in Eq.~\eqref{eq:tfim-excited-contribution} and its bound in Eq.~\eqref{eq:tfim-excited-bound} give
\[
 r_k(\beta)\le e^{-\gamma(\beta-\beta_j)}r_k(\beta_j)
 \le e^{-\gamma\beta},\qquad \beta\ge\beta_j.
\]
We therefore choose the inverse temperature proportional to $L_j^2$:
\[
 \widehat\beta_j=2\cdot4^j\beta_0
 =\frac{2\beta_0}{L_0^2}L_j^2.
\]
Then $r_k(\widehat\beta_j)\le\varepsilon_j:=1600^{-4^j}$, as required. This bound applies on both the $L_j\times L_j$ and $2L_j\times2L_j$ squares at this same temperature, since $\widehat\beta_j\ge\beta_{j+1}$.

For the two spatial squarings to remain valid, we must check that $S_x,S_y>1/2$ at this lower temperature. Since $\widehat\beta_j/\beta_j=2^{j+1}$, the partition-function bounds give
\[
 \begin{split}
 S_x(L_j,L_j,\widehat\beta_j)
 &\ge\bigl[S_x(L_j,L_j,\beta_j)
 T_{+1}(2L_j,L_j,\beta_j)\bigr]^{2^{j+1}}\\
 &\ge(1-u_j)^{2^{j+1}}(1-7u_j)^{2^{j+1}}
 \ge1-16\cdot2^ju_j\ge0.984.
 \end{split}
\]
For the first inequality, we bound the numerator of $S_x$ by its ground-state contribution, whose Gibbs weight at $\beta_j$ is at least $T_{+1}$. For the denominator, we use $\mathcal Z_{+1}(\widehat\beta_j)\le\mathcal Z_{+1}(\beta_j)^{2^{j+1}}$. The same argument bounds $S_y(2L_j,L_j,\widehat\beta_j)$. In particular, both purities exceed $0.9$, so the fourth-power bound applies at the chosen temperature.

We can now combine these bounds and repeat the procedure. Write $v_j=\widehat\beta_jW_{L_j}$. Equation~\eqref{eq:tfim-splitting-comparison} gives $\delta(L_j,L_j,\widehat\beta_j)\le v_j+\varepsilon_j$, so the two spatial doublings give
\[
 \delta(2L_j,2L_j,\widehat\beta_j)\le(v_j+\varepsilon_j)^4.
\]
We apply Eq.~\eqref{eq:tfim-splitting-comparison} again to bound $\widehat\beta_jW_{2L_j}$. For the next round, the inverse temperature is $\widehat\beta_{j+1}=4\widehat\beta_j$, so we multiply this bound by four:
\begin{equation}
 v_{j+1}\le4\left[\frac{(v_j+\varepsilon_j)^4}
 {1-(v_j+\varepsilon_j)^4}+\varepsilon_j\right]
 \le8(v_j+\varepsilon_j)^4+4\varepsilon_j.
 \label{eq:tfim-volume-recursion}
\end{equation}
The last inequality holds for $v_j\le1/4$. Initially $W_{L_0}<28u_0/\beta_0$ gives $v_0<0.056<1/8$. Induction now yields
\[
 v_j\le\tfrac14\,2^{-4^j}.
\]
Indeed, $\varepsilon_j\le0.01\,2^{-4^{j+1}}$, so substituting this and the induction hypothesis into Eq.~\eqref{eq:tfim-volume-recursion} gives
$v_{j+1}\le[8(0.25+0.01)^4+0.04]2^{-4^{j+1}}<\tfrac14\,2^{-4^{j+1}}$.
Since $W_{L_j}=v_j/\widehat\beta_j$, $\widehat\beta_j\ge2\beta_0$, and $4^j=L_j^2/L_0^2$, we have proved
\begin{equation}
 W_{L_j}\le\frac{1}{8\beta_0}
 \exp\!\left[-\frac{\log2}{L_0^2}L_j^2\right],
 \qquad L_j=2^j L_0.
 \label{eq:tfim-volume-doubled}
\end{equation}
Thus we have bounded the splitting by a quantity exponentially small in the system volume, with explicit constants determined by the starting scale $(L_0,\beta_0)$.

We finally extend this result to every even $L\ge L_0$. Fix $L_j\le L\le2L_j$ and keep $\beta=\widehat\beta_j$. Comparing eigenvalue powers, as in Sec.~\ref{sec:higher-d}, gives
\[
 \begin{split}
 \delta(x,L_j)&\le
 2\left[\frac{\delta(L_j,L_j)}{2(0.9)}\right]^{x/L_j}
 \le\frac{\delta(L_j,L_j)}{0.9},\\
 S_y(x,L_j)&\ge
 [S_x(L_j,L_j)^2S_y(2L_j,L_j)]^{x/(2L_j)}
 \ge0.9^3,\qquad L_j\le x\le2L_j.
 \end{split}
\]
In the first line, we bound the numerator using the sum of the $A_1^{L_j}$ eigenvalues. For the denominator, one eigenvalue contributes at least $0.9$ of $\mathcal Z_{+1}(L_j,L_j,\widehat\beta_j)$, because the largest spatial weight is at least $S_x$. Repeating the bound in $y$, using $S_y(x,L_j)\ge0.9^3$, gives $\delta(L,L)\le0.9^{-4}\delta(L_j,L_j)<2\delta(L_j,L_j)$. The preceding bounds relating $\delta$ to $\beta W$ then give
$\widehat\beta_jW_L\le4v_j+5\varepsilon_j<2\cdot2^{-4^j}$. Since $L\le2L_j$, we obtain
\begin{equation}
 W_L\le\frac1{\beta_0}
 \exp\!\left[-\frac{\log2}{4L_0^2}L^2\right],
 \qquad L\ge L_0,\ L\text{ even}.
 \label{eq:tfim-volume-splitting}
\end{equation}
Thus the finite bounds give both the positive gap in Eq.~\eqref{eq:tfim-sector-gap} and the exponential upper bound on the splitting in Eq.~\eqref{eq:tfim-volume-splitting}. The sector gap allows us to suppress excited states by lowering the temperature; the two spatial squarings supply the stronger bound on the splitting.

\subsection{Gap and splitting for the toric code}\label{app:toric}
We prove the gap and splitting bounds stated in Sec.~\ref{sec:toric}. We map the four loop sectors with $A_v=+1$ at every vertex to an Ising model with four boundary conditions and compare their spectra to obtain four low states. Finally, we show that states with $A_v=-1$ at some vertices lie above these four low states, separated from them by a nonzero gap.

\paragraph{Mapping to the Ising model.}
Every star operator $A_v$ is conserved. We first consider states with all $A_v=+1$, which contain the four ground states at $h=0$. Products of $X_e$ on edges crossed by dual-lattice loops winding around the torus in either direction are also conserved. Their eigenvalues $\eta_x,\eta_y=\pm1$ divide this space into four sectors, labeled by $\eta=(\eta_x,\eta_y)$.

We now express the Hamiltonian in each of these sectors as an Ising Hamiltonian, so that we can check the spatial-transfer conditions. We introduce one spin $\tau_p$ per plaquette and use the operator correspondence
$B_p=\tau_p^x,\qquad X_e=\eta_{pp'}\tau_p^z\tau_{p'}^z,$
where $p,p'$ meet at edge $e$. The bond signs $\eta_{pp'}$ have product $+1$ around each elementary dual plaquette and products $\eta_x,\eta_y$ around the two noncontractible cycles. We can put the negative signs on boundary seams. Thus the four physical sectors correspond to four boundary conditions of the same Ising Hamiltonian:
\begin{equation}
 \widetilde H^{(\eta)}
 =-J_m\sum_p\tau_p^x-h\sum_{\langle pp'\rangle}\eta_{pp'}\tau_p^z\tau_{p'}^z.
 \label{eq:ising}
\end{equation}
The identity $\prod_pB_p=I$ requires $U_\tau:=\prod_p\tau_p^x=+1$. Thus each physical sector has the spectrum of only this symmetry sector of the corresponding Ising Hamiltonian:
\begin{equation}
 \left.(H_{\rm TC}+J_eNI)\right|_{A_v=+1,\,\eta}
 \simeq\left.\widetilde H^{(\eta)}\right|_{U_\tau=+1},
 \qquad N=L_xL_y.
 \label{eq:toric-sectors}
\end{equation}
The shift $J_eN$ removes the constant energy of the stars with $A_v=+1$. We can check the equivalence by matching the plaquette eigenvalues to $\tau^x$ eigenvalues: $X_e$ flips the adjacent pair, and both spaces have dimension $2^{N-1}$~\cite{toricbackground}.

We will bound the full Ising spectra first, then restrict to the physical states. Accordingly, our partition functions include both values of $U_\tau$:
\[
 \mathcal Z_\eta(L_x,L_y,\beta)=\tr e^{-\beta\widetilde H^{(\eta)}},
 \qquad \mathcal Z_0:=\mathcal Z_{(+1,+1)}.
\]
A gap in a full Ising spectrum gives a gap in its physical subspace if the ground state has $U_\tau=+1$. This holds for every $\eta$: in the $\tau^z$ basis, the off-diagonal entries $-J_m$ connect all configurations by single-spin flips, so the ground state is unique with positive amplitudes. The symmetry $U_\tau$ permutes the configurations, and uniqueness and positivity therefore give $U_\tau=+1$ for this state. Every other physical state is also an Ising eigenstate and obeys its excitation-gap bound.

\paragraph{The finite test for the untwisted Ising model.}
We first check that Sec.~\ref{sec:higher-d} applies to $\mathcal Z_0$. The on-site matrix $-J_m\tau^x$ is real symmetric, and each bond is $-(\sqrt h\,\tau^z)\otimes(\sqrt h\,\tau^z)$. The construction in Appendix~\ref{sec:spin-transfer} therefore gives Hermitian spatial transfer operators in both directions. We write $S_x,S_y,T_0$ for the three purities formed from $\mathcal Z_0$ by Eq.~\eqref{eq:2d-purities}.

For the finite test in Eq.~\eqref{eq:2d-cube}, we use
\[
 Q(4,\beta)=\frac{\mathcal Z_0(8,8,2\beta)}{\mathcal Z_0(4,4,\beta)^8}.
\]
The $4\times4$ Ising Hilbert space has $2^{16}$ states. We compute its complete spectrum and sum all Gibbs weights to bound the denominator.

To bound the numerator from below, we use an $8\times8$ trial state with positive amplitudes
\[
 \psi(s)\propto\exp\!\left[\sum_{a=1}^6\alpha_aF_a(s)\right],
 \qquad \tau_p^z|s\rangle=s_p|s\rangle,\quad s_p=\pm1.
\]
The first five $F_a$ sum pair products with displacements $(1,0)$, $(0,1)$, $(1,\pm1)$, $(2,0)$, and $(2,\pm1)$, respectively; the sixth sums four-spin products around elementary plaquettes. These terms allow nearby spins to correlate. Taking the second coordinate as the row index, all terms involve a single row or two adjacent rows, so the norm and energy can be summed exactly one row at a time. The resulting trial energy gives $\mathcal Z_0(8,8,2\beta)\ge e^{-2\beta E_{\rm tr}}$, as in Eq.~\eqref{eq:trial}.

At $\beta J_m=11/2$, these finite bounds, together with Eq.~\eqref{eq:coupling-interval}, initialize the purity bounds throughout $0\le h/J_m\le1/7$. We now show how these bounds yield four low states and a gap for the toric code when $J_e=J_m$ (the argument also applies to any fixed $J_e>0$, with the gap bound depending on $J_e$).

\paragraph{Comparing the four partition functions.}
We next want to show that every $\mathcal Z_\eta/\mathcal Z_0$ is close to one when $S_x,S_y$ are close to one. We first compare periodic boundaries with an antiperiodic boundary in $x$, keeping $y$ periodic. We use the bond expansion of Appendix~\ref{sec:ising-checks}, with each column treated as one site. There we averaged the boundary conditions, which removed odd histories. Here we keep the two partition functions separate, so we retain both even and odd histories.

The internal Hamiltonian of column $x$ is
\[
b_x=-J_m\sum_y\tau_{x,y}^x-h\sum_y\tau_{x,y}^z\tau_{x,y+1}^z,
\qquad \tau_{x,L_y+1}^z=\tau_{x,1}^z.
\]
We expand the horizontal interactions $-h\tau_{x,y}^z\tau_{x+1,y}^z$, keeping the evolution under $b_x$ exact. Each insertion places one $\tau^z$ operator in each adjoining column. The list of insertion times and row labels $y$ between two neighboring columns is their bond history. Using the ordinary column trace ($a=0$ in Eq.~\eqref{eq:ising-direct-kernel}), now allowing both even and odd histories, gives a Hermitian transfer operator $X_x$ with $\mathcal Z_0=\tr X_x^{L_x}$.

For any column $x$, conjugation by $\prod_y\tau_{x,y}^x$ preserves $b_x$ and changes the sign of each inserted $\tau_{x,y}^z$. Conjugation leaves a trace unchanged, so the column trace must vanish if the total number of insertions from its left and right is odd. The transfer operator $X_x$ therefore connects only histories of the same parity.

Changing to antiperiodic boundaries reverses every horizontal interaction across the seam between columns $L_x$ and $1$. A history with $N_{\rm ins}$ insertions across this seam acquires the sign $(-1)^{N_{\rm ins}}$. We define $R_x$ to act as $+1$ on even histories and as $-1$ on odd histories. The two boundary conditions consequently give
\[
 \mathcal Z_0=\tr X_x^{L_x},\qquad
 \mathcal Z_{(-1,+1)}=\tr(X_x^{L_x}R_x),
 \qquad [X_x,R_x]=0,\qquad R_x^2=I.
\]
Using $S_x=\tr X_x^{2L_x}/(\tr X_x^{L_x})^2$, the same trace inequality used in Eq.~\eqref{eq:ising-sector-comparison} gives
\[
 0\le1-\frac{\mathcal Z_{(-1,+1)}}{\mathcal Z_0}\le2(1-S_x).
\]
The same calculation in $y$ gives
$0\le1-\mathcal Z_{(+1,-1)}/\mathcal Z_0\le2(1-S_y)$.

To compare the boundary condition with both twists, we expand $e^{-\beta\widetilde H^{(+1,+1)}}$ in its Ising bond interactions, keeping the $-J_m\sum_p\tau_p^x$ evolution exact. In the $\tau^x$ basis, every term contributing to $\mathcal Z_0$ has nonnegative weight. A twist in either direction multiplies each term by a sign $s_x$ or $s_y$, while both twists multiply it by $s_xs_y$. Averaging
$1-s_xs_y\le(1-s_x)+(1-s_y)$ with these weights gives
\[
 1-\frac{\mathcal Z_{(-1,-1)}}{\mathcal Z_0}
 \le\left(1-\frac{\mathcal Z_{(-1,+1)}}{\mathcal Z_0}\right)
   +\left(1-\frac{\mathcal Z_{(+1,-1)}}{\mathcal Z_0}\right)
 \le2(1-S_x)+2(1-S_y).
\]
Thus all four partition functions approach one another as $S_x,S_y$ approach one.

\paragraph{Bounds on the Ising energies.}
We now turn the partition-function comparisons into energy bounds. Applying the doubling inequalities to these finite bounds gives a scale $(L_0,\beta_0)$ at which
\[
 S_x(L_0,L_0,\beta_0),\quad S_y(2L_0,L_0,\beta_0),\quad
 T_0(2L_0,2L_0,\beta_0)\ge1-u,\qquad u\le10^{-3}.
\]
Here $L_0,\beta_0$ denote the scale reached by the inequalities; no additional finite spectra are needed. On any even square $L_0\le L\le2L_0$, the length comparisons of Sec.~\ref{sec:higher-d} give $T_0\ge1-7u$, and give $S_x,S_y\ge1-3u$ at $\beta_0$ and $S_x,S_y\ge1-7u$ at $2\beta_0$.\footnote{For example,
$S_x(L_0,L,\beta_0)\ge[\mathcal Z_0(2L_0,2L_0,\beta_0)/\mathcal Z_0(L_0,L_0,\beta_0)^4]^{L/(2L_0)}\ge(1-u)^3$.
Increasing its first argument from $L_0$ to $L$ increases the purity. At $2\beta_0$, the ratio in brackets is at least $Q(L_0,\beta_0)\ge(1-u)^7$. Interchanging the two lengths gives the corresponding bounds for $S_y$.}
The preceding comparisons therefore imply, on this same square,
\[
 \frac{\mathcal Z_\eta(\beta_0)}{\mathcal Z_0(\beta_0)}\ge1-12u,
 \qquad
 \frac{\mathcal Z_\eta(2\beta_0)}{\mathcal Z_0(2\beta_0)}\ge1-28u.
\]
Both ratios are at most one. Substituting into the Gibbs purity gives
\[
 T_\eta(\beta_0)
 =T_0(\beta_0)\frac{\mathcal Z_\eta(2\beta_0)}{\mathcal Z_0(2\beta_0)}
 \left[\frac{\mathcal Z_0(\beta_0)}{\mathcal Z_\eta(\beta_0)}\right]^2
 \ge(1-7u)(1-28u)\ge1-35u.
\]
Let $E_{0,\eta},E_{1,\eta}$ be the two lowest Ising energies for boundary condition $\eta$. The Gibbs-weight argument in Sec.~\ref{sec:twists} gives the first bound below. Comparing the ground-state contributions to $\mathcal Z_\eta$ and $\mathcal Z_0$ gives the second:
\begin{equation}
 \begin{split}
 E_{1,\eta}-E_{0,\eta}
 &\ge\frac1{\beta_0}\log\frac{1-35u}{35u}=:\gamma,\\
 0\le E_{0,\eta}-E_{0,0}
 &\le\frac{-\log(1-12u)-\log(1-35u)}{\beta_0}=:B.
 \end{split}
 \label{eq:toric-errors}
\end{equation}
For the second line, we used
$e^{-\beta_0(E_{0,\eta}-E_{0,0})}\ge T_\eta\mathcal Z_\eta/\mathcal Z_0$.
The nonnegative energy difference follows from $\mathcal Z_\eta\le\mathcal Z_0$ at every temperature.

\paragraph{The four lowest toric-code states.}
We now return to the physical spectrum of the perturbed toric code. Each Ising ground state has $U_\tau=+1$, as checked above, so Eq.~\eqref{eq:toric-sectors} supplies one physical state in each of the four loop sectors. Their energies differ by at most $B$, and every other state with all $A_v=+1$ lies at least $\gamma$ above its sector's ground state.

It remains to show that states with $A_v=-1$ at some vertices lie above these four states. The identity $\prod_vA_v=I$ requires at least two such vertices, increasing the star energy by at least $4J_e$. In the Ising representation, these star eigenvalues become products of bond signs around the corresponding dual plaquettes. Removing all bond signs makes every weight in the bond expansion nonnegative and gives the untwisted partition function $\mathcal Z_0$. The partition function with any sign pattern is therefore at most $\mathcal Z_0$, and its lowest Ising energy is no lower than $E_{0,0}$. Thus the Ising part cannot offset the star-energy cost: all these physical states lie at least $4J_e$ above the lowest toric-code energy.

The four sector ground energies consequently form a band of width at most $B$. Once $B<\min\{\gamma,4J_e\}$, they are the four lowest states, and
\begin{equation}
 E_3-E_0\le B,\qquad
 E_4-E_3\ge\min\{\gamma,4J_e\}-B.
 \label{eq:toric-band}
\end{equation}

\paragraph{Dependence on system size.}
To see the dependence on size, recall that successive doublings give
$L_j=2^j L_0,\qquad \beta_j=2^j\beta_0,\qquad u_j\le\frac1{25}\,40^{-2^j}.$
We apply Eq.~\eqref{eq:toric-errors} with $\beta_0,u$ replaced by $\beta_j,u_j$, obtaining
$B\le49u_j/\beta_j$ and $\gamma>(\log20)/\beta_0$.
For the first bound, $-\log(1-x)\le x/(1-x)$ and $u_j\le10^{-3}$ give $\beta_j B\le(12/0.988+35/0.965)u_j<49u_j$. The second follows from $35u_j/(1-35u_j)<2\cdot40^{-2^j}\le20^{-2^j}$ and $\beta_j=2^j\beta_0$.
Thus the bound $B$ on the splitting decreases exponentially with $L_j$, while the gap bound stays positive. Substituting into Eq.~\eqref{eq:toric-band} on each interval $L_j\le L\le2L_j$ proves
\[
 E_3(L)-E_0(L)\le Ce^{-cL},\qquad
 E_4(L)-E_3(L)\ge\Delta>0
\]
for all sufficiently large even $L$, with $C,c,\Delta$ independent of $L$.

\subsection{Transfer checks and finite bounds for the SMG model}\label{app:smg}
We verify the transfer conditions and finite test for the Hamiltonian in Sec.~\ref{sec:smg}, first with antiperiodic boundaries and then with periodic boundaries. We write $Z_{\rm A}$ for the trace over all occupation states at zero chemical potential, with both spatial boundaries antiperiodic. Earlier SMG lattice realizations include related honeycomb models~\cite{slagle2015,he2016} and staggered-fermion models on a cubic spacetime lattice~\cite{ayyarchandrasekharan2015,ayyarchandrasekharan2016}.

\paragraph{Spatial transfer operators.}
For the Hamiltonian in Sec.~\ref{sec:smg}, we need Hermitian transfer operators for $Z_{\rm A}$ in both directions. Recall the sufficient condition in Eq.~\eqref{eq:fermion-signed-local}: after grouping a column or row into one cell, an antiunitary must preserve its internal Hamiltonian and multiply its Majoranas by the signs of their couplings to the next cell.

For the $x$ direction, the column Hamiltonian $b$ contains the vertical hopping and the on-site interactions. Writing the four complex modes at each site as eight Majoranas $\gamma_{x,y}^r$, with $r=1,\ldots,8$, the hopping between columns is
\[
 v_{x,x+1}=-\frac{it}{2}\sum_{y=1}^{L_y}\sum_{r=1}^8
 (-1)^y\gamma_{x,y}^r\gamma_{x+1,y}^r.
\]
Thus the signs in Eq.~\eqref{eq:fermion-signed-local} are $\eta_{y,r}=(-1)^y$. We choose the antiunitary particle--hole transformation
\[
 \mathcal C_x c_{y,a,\sigma}\mathcal C_x^{-1}=(-1)^y c_{y,a,\sigma}^\dagger,
 \qquad
 \mathcal C_x\gamma_y^r\mathcal C_x^{-1}=(-1)^y\gamma_y^r.
\]
It reverses both layer spins, leaving their dot product unchanged. In every vertical hopping term, complex conjugation reverses $i$, while the alternating signs at its endpoints supply a second minus sign. For even $L_y$, this also preserves the closing bond. Thus $\mathcal C_xb\mathcal C_x^{-1}=b$, as required.

For the $y$ direction, the gauge change $c_{x,y}\mapsto(-1)^{xy}c_{x,y}$ makes horizontal bond signs $+1$ and vertical signs $(-1)^x$. The same check then applies to rows. At even lengths this gauge change preserves the spatial boundary conditions. We have therefore established
\[
 Z_{\rm A}(L_x,L_y,\beta)
 =\tr X_x(L_y,\beta)^{L_x}
 =\tr X_y(L_x,\beta)^{L_y},
\]
with both transfers Hermitian, so Sec.~\ref{sec:higher-d} applies. We will also need these checks when changing the boundaries. They remain valid with either transverse boundary condition, since reversing a closing bond's sign leaves its invariance under the antiunitary unchanged.

For the finite test in Eq.~\eqref{eq:2d-cube}, we use $4\times4$ and $8\times8$ systems. As in Appendix~\ref{app:bilayer}, we shift the energy to put the singlet product at zero: on $N$ sites, $\widehat H_{\rm A}=H_{\rm A}+3JNI/4$. Writing $\widehat Z_{\rm A}$ for its partition function, the shifts cancel to give
\[
Q(4,\beta)=\frac{\widehat Z_{\rm A}(8,8,2\beta)}{\widehat Z_{\rm A}(4,4,\beta)^8}.
\]
All energies in the following finite bounds refer to $\widehat H_{\rm A}$.

\paragraph{An upper bound on $\widehat Z_{\rm A}(4,4,\beta)$.}
We need lower bounds on the energies and a count of the levels they cover. The shifted single-site spectrum is
\begin{equation}
 \begin{array}{c|ccc}
 \text{energy above the singlet}&0&3J/4&J\\
 \text{multiplicity}&1&12&3.
 \end{array}
 \label{eq:local}
\end{equation}
We call a site excited when its state lies in the 15-dimensional space orthogonal to the singlet. There are $\binom{16}{k}15^k$ configurations with $k$ excited sites on $4\times4$, and their on-site energies are at least $3Jk/4$.

To include hopping, we group states by the number of excited sites, just as we grouped the bilayer states by triplet number. Each hopping bond changes this number by at most two. Its matrix acts on $16^2=256$ two-site states. We bound the absolute row and column sums of this local matrix and count the available bonds in each group on $4\times4$. The square root of the product of the resulting row and column bounds controls the norm of the hopping between groups. Fermionic ordering signs and flux signs do not affect these absolute sums.

Putting the on-site energies and these hopping bounds in a matrix indexed by excitation number gives the same inequality as Eq.~\eqref{eq:bilayer-energy-comparison}. Let $\mu_k$ bound its smallest eigenvalue when restricted to excitation numbers $k,\ldots,16$. Then every state with at least $k$ excited sites has energy at least $\mu_k$. Since the configurations with fewer excited sites span $\sum_{j<k}\binom{16}{j}15^j$ dimensions, at most this many exact levels can lie below $\mu_k$. Thus we can assign this bound to the next $\binom{16}{k}15^k$ Gibbs weights.

We still need a lower bound on the ground-state energy. On $4\times4$, we start from the singlet product, apply the hopping up to twice, and retain the components at each unperturbed on-site energy. This gives seven linearly independent vectors. Their overlap and Hamiltonian matrices can be evaluated exactly. Minimizing in this span gives a trial energy $E_R$ and variance $\sigma^2$. When $E_R<\mu_1\le E_1$, Temple's inequality gives $E_-:=E_R-\sigma^2/(\mu_1-E_R)\le E_0$, and hence
\[
 \widehat Z_{\rm A}(4,4,\beta)
 \le e^{-\beta E_-}
    +\sum_{k=1}^{16}\binom{16}{k}15^k e^{-\beta\mu_k}.
\]
The first term bounds the ground-state weight; the sum bounds every other weight. Together they account for all $16^{16}=2^{64}$ states.

\paragraph{A lower bound on $\widehat Z_{\rm A}(8,8,2\beta)$.}
On $8\times8$, the same trial-state construction again gives seven vectors. Minimizing the energy in their span gives $E_{\rm tr}\ge E_0(8,8)$, so
\[
\widehat Z_{\rm A}(8,8,2\beta)
\ge e^{-2\beta E_0(8,8)}\ge e^{-2\beta E_{\rm tr}}.
\]

At $\beta J=24$, these bounds yield a certificate throughout $0\le t/J\le1/35$, using the endpoint comparison in Eq.~\eqref{eq:coupling-interval}. Section~\ref{sec:higher-d} therefore gives a unique ground state and a size-independent gap on sufficiently large even squares with antiperiodic boundaries.

\paragraph{Periodic spatial boundaries.}
We want to obtain the same conclusion with periodic boundaries. Write $Z_{\rm AA}=Z_{\rm A}$, $Z_{\rm AP}$, and $Z_{\rm PP}$, with the first label specifying the $x$ boundary and the second specifying $y$. Our transfer checks allow us to use Eq.~\eqref{eq:fermion-periodic-ratio} for each boundary change, giving
$0<Z_{\rm PP}\le Z_{\rm AP}\le Z_{\rm AA}$
on any fixed even rectangle. To see what remains to prove, we write the periodic Gibbs purity directly in terms of the antiperiodic one:
\begin{equation}
 \begin{split}
 T_{\rm PP}(L_x,L_y,\beta)
 &=T_{\rm AA}(L_x,L_y,\beta)
 \frac{Z_{\rm PP}(L_x,L_y,2\beta)}{Z_{\rm AA}(L_x,L_y,2\beta)}
 \left[\frac{Z_{\rm AA}(L_x,L_y,\beta)}
 {Z_{\rm PP}(L_x,L_y,\beta)}\right]^2\\
 &\ge T_{\rm AA}(L_x,L_y,\beta)
 \frac{Z_{\rm PP}(L_x,L_y,2\beta)}{Z_{\rm AA}(L_x,L_y,2\beta)}.
 \end{split}
 \label{eq:smg-periodic-identity}
\end{equation}
We already know that $T_{\rm AA}$ approaches one. It therefore suffices to show that $Z_{\rm PP}/Z_{\rm AA}$ also approaches one, at twice the inverse temperature.

We first compare the $x$ boundaries while keeping $y$ periodic. Equation~\eqref{eq:fermion-periodic-ratio} gives
\[
 \frac{Z_{\rm PP}(L_x,L_y,\beta)}{Z_{\rm AP}(L_x,L_y,\beta)}
 \ge\frac{2Z_{\rm AP}(2L_x,L_y,\beta)}
 {Z_{\rm AP}(L_x,L_y,\beta)^2}-1.
\]
Multiplying by $Z_{\rm AP}/Z_{\rm AA}$ and using $0<Z_{\rm AP}\le Z_{\rm AA}$ gives the first line below. In the last line, we change the $y$ boundary on the rectangle with doubled $x$ length:
\begin{equation}
 \begin{split}
 \frac{Z_{\rm PP}(L_x,L_y,\beta)}{Z_{\rm AA}(L_x,L_y,\beta)}
 &\ge\frac{2Z_{\rm AP}(2L_x,L_y,\beta)}
 {Z_{\rm AA}(L_x,L_y,\beta)^2}-1\\
 &=2S_x(L_x,L_y,\beta)
 \frac{Z_{\rm AP}(2L_x,L_y,\beta)}{Z_{\rm AA}(2L_x,L_y,\beta)}-1\\
 &\ge2S_x(L_x,L_y,\beta)\,[2S_y(2L_x,L_y,\beta)-1]-1.
 \end{split}
 \label{eq:smg-periodic-ratio}
\end{equation}
Here both spatial purities are formed from $Z_{\rm AA}$; in particular,
$S_x(L_x,L_y,\beta)=Z_{\rm AA}(2L_x,L_y,\beta)/Z_{\rm AA}(L_x,L_y,\beta)^2$.
If the two purities on the last line are at least $1-\varepsilon$, this bounds the periodic-to-antiperiodic ratio below by $1-6\varepsilon$.

It remains to supply the two spatial purities at inverse temperature $2\beta$, as needed in Eq.~\eqref{eq:smg-periodic-identity}. Let $(L_0,\beta_0)$ be a scale reached by the successive doublings applied to $Z_{\rm AA}$, with
\[
S_x(L_0,L_0,\beta_0),\quad S_y(2L_0,L_0,\beta_0),\quad T_{\rm AA}(2L_0,2L_0,\beta_0)
\ge1-u,\qquad 0<u<1/98.
\]
Choose any even $L_0\le L\le2L_0$. A spatial purity increases with its own transfer length, so we first shorten that length to $L_0$. Comparing the remaining lengths as in Sec.~\ref{sec:higher-d} then gives
\[
 \begin{split}
 S_x(L,L,2\beta_0)
 &\ge S_x(L_0,L,2\beta_0)
 \ge\left[\frac{Z_{\rm AA}(2L_0,2L_0,2\beta_0)}
 {Z_{\rm AA}(L_0,L_0,\beta_0)^8}\right]^{L/(2L_0)}
 \ge(1-u)^7,\\
 S_y(2L,L,2\beta_0)
 &\ge S_y(2L,L_0,2\beta_0)
 \ge\left[\frac{Z_{\rm AA}(4L_0,2L_0,2\beta_0)}
 {Z_{\rm AA}(2L_0,L_0,\beta_0)^8}\right]^{L/(2L_0)}
 \ge(1-u)^7.
 \end{split}
\]
The ratios factor into purities as in Eqs.~\eqref{eq:2d-Rx} and \eqref{eq:2d-Ry}; the second uses the bound on $S_x(2L_0,2L_0,2\beta_0)$, whose deviation is at most $49u^2\le u$. The power $L/(2L_0)\le1$ can only increase these ratios.

Thus both required spatial purities are at least $1-7u$. Equation~\eqref{eq:smg-periodic-ratio} at $2\beta_0$ gives $Z_{\rm PP}/Z_{\rm AA}\ge1-42u$, while Eq.~\eqref{eq:2d-interpolation} gives $T_{\rm AA}(L,L,\beta_0)\ge1-7u$. Substitution into Eq.~\eqref{eq:smg-periodic-identity} yields
\begin{equation}
 T_{\rm PP}(L,L,\beta_0)\ge(1-7u)(1-42u)\ge1-49u.
 \label{eq:smg-periodic-purity}
\end{equation}
Under further doublings, $L_j=2^jL_0$, $u_j\le(49u)^{2^j}/49$, and $\beta_j=2^j\beta_0$. For $L_j\le L\le2L_j$, the Gibbs-weight argument therefore gives
\[
 \Delta_{{\rm P},L}\ge
 \frac1{\beta_j}\log\frac{1-49u_j}{49u_j}
 \ge\frac1{\beta_0}\log\frac{1-49u}{49u}>0.
\]
The second inequality uses $1-q^m\ge(1-q)^m$ for $0\le q\le1$, with $q=49u$ and $m=2^j$. Every even $L\ge L_0$ lies between two successive doubled lengths, so this gap bound holds on all such periodic squares.

The same finite input therefore certifies a periodic gap throughout $0\le t/J\le1/35$. For either boundary condition, uniqueness makes the ground state a spin singlet. Particle--hole symmetry and conservation of particle number in each layer fix the average density to one fermion per layer per site. This completes the conclusion stated in Sec.~\ref{sec:smg}.

\subsection{Finite calculations for the Jian--Zaletel state}\label{app:jz}
We derive the local tensor, verify its Hermitian norm transfers, and obtain the finite norm bounds that imply the correlation result in Sec.~\ref{sec:spin1-peps}.

\subsubsection{The state and its local tensor}\label{app:jz-construction}
We derive the tensor used in Sec.~\ref{sec:spin1-peps} from the state defined by Jian and Zaletel~\cite[Eq.~(7) and Figs.~5--6]{jianzaletel2016}, keeping track of the spin basis and the singlets on the bonds.

\paragraph*{The state.}
Their Fig.~6 gives a simple way to understand the construction. Put an independent spin-one AKLT chain on every row of the square lattice, and another on every column. Each site now contains two spin-one degrees of freedom, one from each chain. Two spin ones can have total spin zero, one, or two. At every site we project onto the total-spin-one subspace and identify its three states with a single physical spin one. Applying this projection at every site gives the Jian--Zaletel state.

\paragraph*{The four virtual spins at a site.}
We now write this state in terms of local tensors. Recall that an AKLT chain can be constructed by assigning two virtual spin halves to every site, forming singlets between virtual spins at neighboring sites, and projecting the two spins at each site onto their triplet. Thus the horizontal AKLT chain supplies the left and right virtual spins, while the vertical chain supplies the down and up virtual spins. These are the four indices $l,r,d,u$ in our PEPS notation.

We label the virtual spin states by $1=\downarrow$ and $2=\uparrow$. The coefficients $T^h_{\alpha\beta}$ express a spin-one triplet with magnetic quantum number $h$ in the product basis $|\alpha,\beta\rangle$ of two virtual spins. Their nonzero entries are
$T^{-1}_{11}=T^{+1}_{22}=1,\qquad T^0_{12}=T^0_{21}=\frac1{\sqrt2}.$
For example, the $h=0$ triplet is $(|1,2\rangle+|2,1\rangle)/\sqrt2$. For horizontal and vertical triplets with magnetic quantum numbers $h$ and $v$, the coefficients at a site are $T^h_{rl}$ and $T^v_{ud}$, respectively.

\paragraph*{The projection to one physical spin.}
We next specify which three states of the two triplets we retain. Write $|h,v\rangle$ for their product basis, with the horizontal spin first and the vertical spin second. We identify the physical spin-one basis with the following total-spin-one states:
\begin{equation}
\begin{aligned}
|+1\rangle_{\rm phys}&\longleftrightarrow
\frac{|+1,0\rangle-|0,+1\rangle}{\sqrt2},\\
|0\rangle_{\rm phys}&\longleftrightarrow
\frac{|+1,-1\rangle-|-1,+1\rangle}{\sqrt2},\\
|-1\rangle_{\rm phys}&\longleftrightarrow
\frac{|0,-1\rangle-|-1,0\rangle}{\sqrt2}.
\end{aligned}
\label{eq:jz-spin-projection}
\end{equation}
The coefficient of $|h,v\rangle$ on the right is $\widetilde T^s_{hv}$, where $s=-1,0,+1$ labels the physical state on the left. All unlisted coefficients vanish. Combining this projection with the two triplet tensors gives
\begin{equation}
D^s_{urdl}=\sum_{h,v=-1}^{1}\widetilde T^s_{hv}T^h_{rl}T^v_{ud}.
\label{eq:jz-original-tensor}
\end{equation}
This is their Eq.~(7): their compass indices north, east, south, west have become $u,r,d,l$, and their physical index $m$ has become $s$. The tensors $T$ and $\widetilde T$ are the Clebsch--Gordan coefficients appearing there.

\paragraph*{Including the bond singlets in the site tensor.}
The tensor $D$ describes the projection at a site. To specify the wave function, we must also include the virtual singlets joining neighboring sites. Orienting each bond to the right or upward, their coefficients are
\[
S=\frac1{\sqrt2}\begin{pmatrix}0&1\\-1&0\end{pmatrix},
\qquad \sum_{\alpha,\beta=1}^{2}S_{\alpha\beta}|\alpha,\beta\rangle
=\frac{|1,2\rangle-|2,1\rangle}{\sqrt2}.
\]
Jian and Zaletel keep $D$ at each site and $S$ on each bond, as in their Fig.~5. In our wave-function formula in Sec.~\ref{sec:peps}, neighboring tensors share a single summed index. To obtain that form, we include the right and upper singlet matrices in the tensor at each site:
\begin{equation}
\begin{aligned}
A^s_{lrdu}
&=\sum_{r',u'=1}^{2}D^s_{u'r'dl}S_{r'r}S_{u'u}\\
&=\sum_{h,v=-1}^{1}\widetilde T^s_{hv}
\left(\sum_{r'=1}^{2}T^h_{r'l}S_{r'r}\right)
\left(\sum_{u'=1}^{2}T^v_{u'd}S_{u'u}\right).
\end{aligned}
\label{eq:jz-bond-absorption}
\end{equation}
For a horizontal bond, $r'$ labels the right virtual spin at this site, and $r$ labels the left virtual spin at its neighbor. Summing the shared index therefore inserts exactly the singlet between them. The same holds for the upper bond, so every bond singlet is included once.

Substitution gives the local tensor $A^s_{lrdu}$, with $s=-1,0,+1$ the physical $S^z$ eigenvalue. Displayed as matrices with rows $(l,r)$ and columns $(d,u)$, both ordered as $11,12,21,22$, its entries are
\begin{equation}
\begin{gathered}
A^{-1}=\frac14
\begin{pmatrix}
0&-1&0&0\\
1&0&0&-1\\
0&0&0&0\\
0&1&0&0
\end{pmatrix},\qquad
A^0=\frac14
\begin{pmatrix}
0&0&0&0\\
0&0&\sqrt2&0\\
0&-\sqrt2&0&0\\
0&0&0&0
\end{pmatrix},\qquad
A^{+1}=\frac14
\begin{pmatrix}
0&0&-1&0\\
0&0&0&0\\
1&0&0&-1\\
0&0&1&0
\end{pmatrix}.
\end{gathered}
\label{eq:jz-local-tensor}
\end{equation}
For example, only $r'=2,u'=1$ contribute to $A^{-1}_{1112}$. Equation~\eqref{eq:jz-original-tensor} gives $D^{-1}_{1211}=\widetilde T^{-1}_{0,-1}T^0_{21}T^{-1}_{11}=1/2$, and hence
\[
A^{-1}_{1112}=D^{-1}_{1211}S_{21}S_{12}
=\frac12\left(-\frac1{\sqrt2}\right)\frac1{\sqrt2}=-\frac14.
\]
This is the first-row, second-column entry of $A^{-1}$, since the row labels $(l,r)$ and column labels $(d,u)$ are both ordered as $11,12,21,22$.

We have used normalized singlets, while their Eq.~(3) omits the factor $1/\sqrt2$. There are $2L_xL_y$ bonds on a periodic rectangle, so this choice multiplies the state by $2^{-L_xL_y}$. It cancels from normalized expectation values and $Q_{\rm PEPS}$.

As an independent check, these tensors reproduce the correlation lengths and entanglement entropies in Fig.~7 of Ref.~\cite{jianzaletel2016}
for all five cylinder circumferences shown, within the precision of the plot.
We can immediately check that the state is nonzero on even rectangles. For the physical configuration with $s_i=0$ at every site, only two alternating assignments of the virtual indices contribute. The displayed entries give
\[
\langle0,\ldots,0|\Psi_{L_x,L_y}\rangle
=2(2\sqrt2)^{-L_xL_y}>0,
\qquad L_x,L_y\ \text{even}.
\]
Thus $\mathcal N(L_x,L_y)=\langle\Psi_{L_x,L_y}|\Psi_{L_x,L_y}\rangle$ is nonzero on every rectangle needed in the argument.

\subsubsection{Hermiticity of the norm transfers}\label{app:jz-transfers}
We form the column matrices $K_{\boldsymbol s}$ and the transfer $\mathbb T_x(L_y)=\sum_{\boldsymbol s}K_{\boldsymbol s}\otimes\overline{K_{\boldsymbol s}}$, as in Eq.~\eqref{eq:peps-transfer}. These matrices are real because the tensor is real. We will show that
\begin{equation}
K_{\boldsymbol s}^{\mathsf T}
=(-1)^{\sum_y s_y}K_{-\boldsymbol s},
\qquad -\boldsymbol s=(-s_1,\ldots,-s_{L_y}).
\label{eq:jz-column-transpose}
\end{equation}
Here $\boldsymbol s=(s_1,\ldots,s_{L_y})$ is the physical configuration of a column. For fixed $s,l,r$, regard $A^s_{lrdu}$ as a two-by-two matrix in its remaining indices:
\[
(A^s_{lr})_{du}=A^s_{lrdu},\qquad
J=\begin{pmatrix}0&1\\-1&0\end{pmatrix}.
\]
Direct substitution of the entries in Eq.~\eqref{eq:jz-local-tensor} gives
\[
A^s_{rl}=(-1)^{1-s}J A^{-s}_{lr}J,\qquad J^2=-I.
\]
The column matrix $K_{\boldsymbol s}$ is obtained by summing the vertical indices around the periodic column, so its matrix elements are
\[
(K_{\boldsymbol s})_{\boldsymbol l,\boldsymbol r}
=\tr\bigl(A^{s_1}_{l_1r_1}\cdots A^{s_{L_y}}_{l_{L_y}r_{L_y}}\bigr).
\]
Transposing $K_{\boldsymbol s}$ exchanges every $l_y,r_y$. Applying the local identity at each site therefore gives
\[
\begin{aligned}
(K_{\boldsymbol s}^{\mathsf T})_{\boldsymbol l,\boldsymbol r}
&=(-1)^{L_y-\sum_y s_y}
\tr\bigl(JA^{-s_1}_{l_1r_1}J\cdots JA^{-s_{L_y}}_{l_{L_y}r_{L_y}}J\bigr)\\
&=(-1)^{\sum_y s_y}
\tr\bigl(A^{-s_1}_{l_1r_1}\cdots A^{-s_{L_y}}_{l_{L_y}r_{L_y}}\bigr)
=(-1)^{\sum_y s_y}(K_{-\boldsymbol s})_{\boldsymbol l,\boldsymbol r}.
\end{aligned}
\]
Each pair $J^2=-I$, including the closing pair inside the trace, supplies one minus sign. This proves Eq.~\eqref{eq:jz-column-transpose}. The sign occurs twice in the norm transfer and cancels, so
\begin{equation}
\mathbb T_x(L_y)^{\mathsf T}
=\sum_{\boldsymbol s}K_{\boldsymbol s}^{\mathsf T}\otimes K_{\boldsymbol s}^{\mathsf T}
=\sum_{\boldsymbol s}K_{-\boldsymbol s}\otimes K_{-\boldsymbol s}
=\mathbb T_x(L_y).
\label{eq:jz-paper-hermitian}
\end{equation}
The last equality relabels the sum over all physical configurations. Thus $\mathbb T_x$ is real symmetric and hence Hermitian.

Since $\widetilde T^s_{hv}=-\widetilde T^s_{vh}$, exchanging the horizontal and vertical factors in Eq.~\eqref{eq:jz-bond-absorption} gives $A^s_{lrdu}=-A^s_{dulr}$. This overall minus sign cancels between the bra and ket in the norm. The row-transfer check is consequently identical.

\subsubsection{Bounds on the two finite norms}\label{app:jz-norms}
 For the finite norm calculation, we use squares of side 12 and 24. To bound $Q_{\rm PEPS}(12)=\mathcal N(24,24)/\mathcal N(12,12)^4$ from below, we need a lower bound on $\mathcal N(24,24)$ and an upper bound on $\mathcal N(12,12)$. We obtain both without directly summing all tensor indices on either square.

For the lower bound, we use the usual variational principle, applied to the largest eigenvalue of $\mathbb T_x(24)$. The expectation in any trial vector is a weighted average of the eigenvalues and therefore cannot exceed the largest one. Thus, if a trial vector $|v\rangle$ gives a positive expectation,
\[
 r:=\frac{\langle v|\mathbb T_x(24)|v\rangle}{\langle v|v\rangle}>0,
 \qquad
 \mathcal N(24,24)=\tr\mathbb T_x(24)^{24}\ge r^{24}.
\]
The last inequality holds because the trace includes the twenty-fourth power of every eigenvalue, all nonnegative.

We choose $|v\rangle=(a+b\mathbb T_x(24)+c\mathbb T_x(24)^2)|I\rangle$, adjusting the real coefficients to increase $r$. Here $|I\rangle=\sum_{\boldsymbol l}|\boldsymbol l\rangle\otimes|\boldsymbol l\rangle$ pairs identical bra and ket virtual configurations on the column boundary. Expanding the numerator and denominator of $r$ requires only $\langle I|\mathbb T_x(24)^p|I\rangle$ for $0\le p\le5$. Each expression involves a strip at most five columns wide. Summing its tensor indices along $y$ therefore uses matrices of dimension at most $4^5=1024$.

We now need an upper bound on $\mathcal N(12,12)$. Evaluating $\tr\mathbb T_x(12)^{12}$ directly is too costly: the matrix has dimension $4^{12}=16{,}777{,}216$. Instead, we compute the norm of an $8\times12$ rectangle and use it to bound the square's norm. Write $\lambda_{x,a}(L_y)$ and $\lambda_{y,a}(L_x)$ for the eigenvalues of the column and row transfers, respectively, counting multiplicities. Then
\[
 \begin{aligned}
 \mathcal N(12,12)&=\sum_a|\lambda_{x,a}(12)|^{12},\\
 \mathcal N(8,12)&=\sum_a|\lambda_{x,a}(12)|^8
 =\sum_a|\lambda_{y,a}(8)|^{12}
 =\tr\mathbb T_y(8)^{12}.
 \end{aligned}
\]
The last expression is computable: $\mathbb T_y(8)$ has dimension $4^8=65{,}536$, and symmetry reduces it to blocks of dimension at most 12870. We multiply these matrices to obtain their twelfth powers and sum their diagonal entries, with bounds on rounding errors. Thus we know $\mathcal N(8,12)$ within a narrow numerical interval, even though we do not know the individual $\lambda_{x,a}(12)$. Since the two norms involve different powers of these same eigenvalues,
\[
 \mathcal N(12,12)=\sum_a\bigl(|\lambda_{x,a}(12)|^8\bigr)^{3/2}
 \le\left(\sum_a|\lambda_{x,a}(12)|^8\right)^{3/2}
 =\mathcal N(8,12)^{3/2}.
\]
The inequality uses nonnegativity of the eighth powers. With our computed numbers, this gives only $Q_{\rm PEPS}(12)>0.6738$, which is insufficient for the subsequent bootstrap estimates.

To improve the upper bound, we need to exclude the possibility that the entire eighth-power sum comes from one eigenvalue. We again use trial vectors, now in separate symmetry sectors of $\mathbb T_x(12)$. For a normalized trial vector $|v_i\rangle$ in a chosen sector, calculate
\[
 b_i:=\bigl|\langle v_i|\mathbb T_x(12)|v_i\rangle\bigr|
 \le\max_{a\ \text{in sector }i}|\lambda_{x,a}(12)|.
\]
Here $i$ labels our chosen sectors. The expectation is a weighted average of the eigenvalues in that sector, so at least one of them has magnitude at least $b_i$. Separate sectors give separate eigenstates. We use spin-one trial vectors at twelve momenta. The transfer commutes with spin rotations, so each spin-one eigenvalue occurs in all three spin projections. This gives 36 bounds $b_i$, counting multiplicities. These expectations are evaluated on narrow strips, as for $r$, and the numerical bounds are rounded down. A spin-zero trial vector gives a larger lower bound on one additional eigenvalue.

We therefore know that 36 separate terms in the eighth-power sum contribute at least $b_i^8$ each. At fixed total $\mathcal N(8,12)$, the twelfth-power sum is largest when as much as possible is concentrated in one term: transferring an amount from a smaller entry to a larger entry increases the sum of their $3/2$ powers. Subject to our lower bounds, the maximum is obtained by leaving the 36 contributions at $b_i^8$ and assigning the remainder to the additional eigenvalue, whose lower bound is largest. Hence
\begin{equation}
 \mathcal N(12,12)
 \le\left(\mathcal N(8,12)-\sum_{i=1}^{36}b_i^8\right)^{3/2}
       +\sum_{i=1}^{36}b_i^{12}.
 \label{eq:jz-paper-norm-upper}
\end{equation}
All other eigenvalues are accounted for through the full sum $\mathcal N(8,12)$.

For the following numbers we use twice the tensor in Eq.~\eqref{eq:jz-local-tensor}; this rescaling cancels from $Q_{\rm PEPS}$ and does not change the normalized state. Enclosing matrix-multiplication rounding errors gives
\[
 \begin{gathered}
 2.8950742411\times10^{-11}<\mathcal N(8,12)<2.8950744545\times10^{-11},\\
 r>0.0022817655775704,\qquad
 \mathcal N(12,12)<1.4341079450\times10^{-16}.
 \end{gathered}
\]
Together these bounds give
\begin{equation}
Q_{\rm PEPS}(12)
=\frac{\mathcal N(24,24)}{\mathcal N(12,12)^4}>0.93794.
\label{eq:jz-paper-test}
\end{equation}

\subsubsection{Obtaining the starting purity bounds}\label{app:jz-purities}
We now use Eq.~\eqref{eq:jz-paper-test} to establish the two starting purity bounds of Sec.~\ref{sec:peps} at $L_0=24$. Write $q=0.93794$. The exact factorizations
\[
 Q_{\rm PEPS}(12)
 =S_y(12,12)^2S_x(12,24)
 =S_x(12,12)^2S_y(24,12)>q
\]
give $S_x(12,24)>q$ and $S_y(24,12)>q$, since every purity is at most one. Doubling the first length in $S_x(12,24)$ squares the normalized transfer weights. Equation~\eqref{eq:square}, with $f(u)=u^2/[2(1-u)^2]$ increasing for $0\le u<1$, therefore gives
\[
 S_x(24,24)>1-f(1-q)=:s=0.997811011742\ldots.
\]
We next want to bound $S_y(48,24)$. The definitions give
\[
 S_y(48,12)
 =\frac{S_y(24,12)^2S_x(24,24)}{S_x(24,12)^2}
 >q^2s=0.8778057218,
\]
where we used that the denominator is at most one. Doubling the second length now squares the corresponding weights, so Eq.~\eqref{eq:square} gives
\[
 S_y(48,24)>1-f(1-q^2s)=0.990311089810\ldots.
\]
Both starting purities exceed $0.99$, so the bootstrap applies with $L_0=24$. Interchanging the axes gives the same bounds in the other direction. No additional norm calculation is needed.

\subsubsection{Decay of correlations and absence of local order}
Having checked the assumptions, we apply Eq.~\eqref{eq:peps-square-correlation} with $L_0=24$:
\begin{equation}
\begin{split}
&|\langle O_1O_2\rangle_{L\times L}
 -\langle O_1\rangle_{L\times L}\langle O_2\rangle_{L\times L}|\\
&\qquad\le\|O_1\|\,\|O_2\|
\left(20^{-r/24}+5\,20^{-L/96}\right),
\qquad L\ge24\ \text{even}.
\end{split}
\label{eq:jz-paper-correlation}
\end{equation}
Here $O_1,O_2$ are bounded local Hermitian observables, separated by $r$ columns between their supports, whose supports together fit in half the torus. For fixed local observables, taking $L\to\infty$ leaves the bound $\|O_1\|\|O_2\|20^{-r/24}$. The convergence argument in Sec.~\ref{sec:peps} gives a unique local limit through even periodic rectangles. It preserves spin and lattice symmetries, which together with this bound excludes N\'eel, quadrupolar, and dimer long-range order.

As an aside, Ref.~\cite{midha2026locality} gives sufficient conditions for exponential decay of connected correlations in PEPS using belief propagation and cluster expansions. However, the sufficient conditions discussed in Ref.~\cite{midha2026locality} are not satisfied by the original Jian--Zaletel tensor. At present, the purity bootstrap method discussed here is the only way we know to rigorously establish exponential decay of connected correlations for arbitrarily large system sizes.

\subsubsection{A parent Hamiltonian with a unique ground state}\label{app:jz-parent}
To obtain a local parent Hamiltonian with this PEPS as its unique ground state, we verify injectivity~\cite{perezgarcia2008peps}. This means that the map $\Gamma_R$ from virtual boundary states to physical states in a region $R$ has linearly independent columns. For a $4\times3$ rectangle, it suffices to check total $S^z=0$. The kernel is SU(2) invariant, and every representation made from its 14 boundary spin halves has an $S^z=0$ member. An invertible physical rescaling makes this block an integer matrix $M$ of size $73789\times3432$. Exact arithmetic gives $\det(M^{\mathsf T}M)\equiv26043\pmod{65521}\ne0$, proving injectivity.

Tile a torus of side $L\in12\mathbb N$ by $4\times3$ rectangles and sum projectors onto the complements of the PEPS images on neighboring pairs of tiles. By Theorem~3 of Ref.~\cite{perezgarcia2008peps}, injectivity makes the PEPS the unique zero-energy state  (however, injectivity alone does not
guarantee a nonzero gap in the thermodynamic limit). Adding all translations and square-lattice rotations preserves uniqueness, since every added projector annihilates the PEPS, and gives a local Hamiltonian with spin and lattice symmetries. We emphasize that injectivity is not used in our correlation-decay proof above. We do not know if this parent Hamiltonian has a nonzero gap in the thermodynamic limit.

\bibliography{references}

\begin{thebibliography}{48}%
\makeatletter
\providecommand \@ifxundefined [1]{%
 \@ifx{#1\undefined}
}%
\providecommand \@ifnum [1]{%
 \ifnum #1\expandafter \@firstoftwo
 \else \expandafter \@secondoftwo
 \fi
}%
\providecommand \@ifx [1]{%
 \ifx #1\expandafter \@firstoftwo
 \else \expandafter \@secondoftwo
 \fi
}%
\providecommand \natexlab [1]{#1}%
\providecommand \enquote  [1]{``#1''}%
\providecommand \bibnamefont  [1]{#1}%
\providecommand \bibfnamefont [1]{#1}%
\providecommand \citenamefont [1]{#1}%
\providecommand \href@noop [0]{\@secondoftwo}%
\providecommand \href [0]{\begingroup \@sanitize@url \@href}%
\providecommand \@href[1]{\@@startlink{#1}\@@href}%
\providecommand \@@href[1]{\endgroup#1\@@endlink}%
\providecommand \@sanitize@url [0]{\catcode `\\12\catcode `\$12\catcode
  `\&12\catcode `\#12\catcode `\^12\catcode `\_12\catcode `\%12\relax}%
\providecommand \@@startlink[1]{}%
\providecommand \@@endlink[0]{}%
\providecommand \url  [0]{\begingroup\@sanitize@url \@url }%
\providecommand \@url [1]{\endgroup\@href {#1}{\urlprefix }}%
\providecommand \urlprefix  [0]{URL }%
\providecommand \Eprint [0]{\href }%
\providecommand \doibase [0]{https://doi.org/}%
\providecommand \selectlanguage [0]{\@gobble}%
\providecommand \bibinfo  [0]{\@secondoftwo}%
\providecommand \bibfield  [0]{\@secondoftwo}%
\providecommand \translation [1]{[#1]}%
\providecommand \BibitemOpen [0]{}%
\providecommand \bibitemStop [0]{}%
\providecommand \bibitemNoStop [0]{.\EOS\space}%
\providecommand \EOS [0]{\spacefactor3000\relax}%
\providecommand \BibitemShut  [1]{\csname bibitem#1\endcsname}%
\let\auto@bib@innerbib\@empty
%</preamble>
\bibitem [{\citenamefont {{OpenAI}}(2026)}]{haldane}%
  \BibitemOpen
  \bibfield  {author} {\bibinfo {author} {\bibnamefont {{OpenAI}}},\ }\href
  {https://github.com/openai/math/blob/fd4aeeb2ee4fc729c18d98444fed42fd0529eeeb/preprints/The-periodic-spin-one-Haldane-gap-September-24-2026/paper.pdf}
  {\bibinfo {title} {{The periodic spin-one Haldane gap}}} (\bibinfo {year}
  {2026}),\ \bibinfo {note} {preprint dated September 24}\BibitemShut {NoStop}%
\bibitem [{\citenamefont {Kirkwood}\ and\ \citenamefont
  {Thomas}(1983)}]{kirkwoodthomas1983}%
  \BibitemOpen
  \bibfield  {author} {\bibinfo {author} {\bibfnamefont {J.~R.}\ \bibnamefont
  {Kirkwood}}\ and\ \bibinfo {author} {\bibfnamefont {L.~E.}\ \bibnamefont
  {Thomas}},\ }\bibfield  {title} {\bibinfo {title} {{Expansions and phase
  transitions for the ground state of quantum Ising lattice systems}},\ }\href
  {https://doi.org/10.1007/BF01211959} {\bibfield  {journal} {\bibinfo
  {journal} {Commun. Math. Phys.}\ }\textbf {\bibinfo {volume} {88}},\ \bibinfo
  {pages} {569} (\bibinfo {year} {1983})}\BibitemShut {NoStop}%
\bibitem [{\citenamefont {Yarotsky}(2006)}]{yarotsky2006}%
  \BibitemOpen
  \bibfield  {author} {\bibinfo {author} {\bibfnamefont {D.~A.}\ \bibnamefont
  {Yarotsky}},\ }\bibfield  {title} {\bibinfo {title} {{Ground states in
  relatively bounded quantum perturbations of classical lattice systems}},\
  }\href@noop {} {\bibfield  {journal} {\bibinfo  {journal} {Commun. Math.
  Phys.}\ }\textbf {\bibinfo {volume} {261}},\ \bibinfo {pages} {799} (\bibinfo
  {year} {2006})},\ \Eprint {https://arxiv.org/abs/math-ph/0412040}
  {arXiv:math-ph/0412040} \BibitemShut {NoStop}%
\bibitem [{\citenamefont {Kennedy}\ and\ \citenamefont
  {Tasaki}(1992)}]{kennedytasaki1992}%
  \BibitemOpen
  \bibfield  {author} {\bibinfo {author} {\bibfnamefont {T.}~\bibnamefont
  {Kennedy}}\ and\ \bibinfo {author} {\bibfnamefont {H.}~\bibnamefont
  {Tasaki}},\ }\bibfield  {title} {\bibinfo {title} {{Hidden symmetry breaking
  and the Haldane phase in $S=1$ quantum spin chains}},\ }\href
  {https://doi.org/10.1007/BF02097239} {\bibfield  {journal} {\bibinfo
  {journal} {Commun. Math. Phys.}\ }\textbf {\bibinfo {volume} {147}},\
  \bibinfo {pages} {431} (\bibinfo {year} {1992})}\BibitemShut {NoStop}%
\bibitem [{\citenamefont {Lapa}\ and\ \citenamefont
  {Levin}(2023)}]{lapalevin2023}%
  \BibitemOpen
  \bibfield  {author} {\bibinfo {author} {\bibfnamefont {M.~F.}\ \bibnamefont
  {Lapa}}\ and\ \bibinfo {author} {\bibfnamefont {M.}~\bibnamefont {Levin}},\
  }\bibfield  {title} {\bibinfo {title} {{Stability of ground state degeneracy
  to long-range interactions}},\ }\href@noop {} {\bibfield  {journal} {\bibinfo
   {journal} {J. Stat. Mech.: Theory Exp.}\ }\textbf {\bibinfo {volume}
  {2023}}\bibfield  {number} {\bibinfo  {number} { (01)},\ \bibinfo {pages}
  {013102}},\ }\Eprint {https://arxiv.org/abs/2107.11396} {arXiv:2107.11396}
  \BibitemShut {NoStop}%
\bibitem [{\citenamefont {Aizenman}\ and\ \citenamefont
  {Nachtergaele}(1994)}]{aizenmannachtergaele1994}%
  \BibitemOpen
  \bibfield  {author} {\bibinfo {author} {\bibfnamefont {M.}~\bibnamefont
  {Aizenman}}\ and\ \bibinfo {author} {\bibfnamefont {B.}~\bibnamefont
  {Nachtergaele}},\ }\bibfield  {title} {\bibinfo {title} {{Geometric aspects
  of quantum spin states}},\ }\href@noop {} {\bibfield  {journal} {\bibinfo
  {journal} {Commun. Math. Phys.}\ }\textbf {\bibinfo {volume} {164}},\
  \bibinfo {pages} {17} (\bibinfo {year} {1994})},\ \Eprint
  {https://arxiv.org/abs/cond-mat/9310009} {arXiv:cond-mat/9310009}
  \BibitemShut {NoStop}%
\bibitem [{\citenamefont {Knabe}(1988)}]{knabe1988}%
  \BibitemOpen
  \bibfield  {author} {\bibinfo {author} {\bibfnamefont {S.}~\bibnamefont
  {Knabe}},\ }\bibfield  {title} {\bibinfo {title} {{Energy gaps and elementary
  excitations for certain VBS-quantum antiferromagnets}},\ }\href
  {https://doi.org/10.1007/BF01019721} {\bibfield  {journal} {\bibinfo
  {journal} {J. Stat. Phys.}\ }\textbf {\bibinfo {volume} {52}},\ \bibinfo
  {pages} {627} (\bibinfo {year} {1988})}\BibitemShut {NoStop}%
\bibitem [{\citenamefont {Nachtergaele}(1996)}]{nachtergaele1996}%
  \BibitemOpen
  \bibfield  {author} {\bibinfo {author} {\bibfnamefont {B.}~\bibnamefont
  {Nachtergaele}},\ }\bibfield  {title} {\bibinfo {title} {{The spectral gap
  for some spin chains with discrete symmetry breaking}},\ }\href@noop {}
  {\bibfield  {journal} {\bibinfo  {journal} {Commun. Math. Phys.}\ }\textbf
  {\bibinfo {volume} {175}},\ \bibinfo {pages} {565} (\bibinfo {year}
  {1996})},\ \Eprint {https://arxiv.org/abs/cond-mat/9410110}
  {arXiv:cond-mat/9410110} \BibitemShut {NoStop}%
\bibitem [{\citenamefont {Nachtergaele}\ \emph {et~al.}(2018)\citenamefont
  {Nachtergaele}, \citenamefont {Sims},\ and\ \citenamefont
  {Young}}]{nachtergaelesimsyoung2018}%
  \BibitemOpen
  \bibfield  {author} {\bibinfo {author} {\bibfnamefont {B.}~\bibnamefont
  {Nachtergaele}}, \bibinfo {author} {\bibfnamefont {R.}~\bibnamefont {Sims}},\
  and\ \bibinfo {author} {\bibfnamefont {A.}~\bibnamefont {Young}},\ }\bibfield
   {title} {\bibinfo {title} {{Lieb--Robinson bounds, the spectral flow, and
  stability of the spectral gap for lattice fermion systems}},\ }in\ \href@noop
  {} {\emph {\bibinfo {booktitle} {Mathematical Problems in Quantum
  Physics}}},\ \bibinfo {series} {Contemp. Math.}, Vol.\ \bibinfo {volume}
  {717}\ (\bibinfo  {publisher} {American Mathematical Society},\ \bibinfo
  {year} {2018})\ pp.\ \bibinfo {pages} {93--115},\ \Eprint
  {https://arxiv.org/abs/1705.08553} {arXiv:1705.08553} \BibitemShut {NoStop}%
\bibitem [{\citenamefont {Wang}\ and\ \citenamefont {You}(2022)}]{wangyou2022}%
  \BibitemOpen
  \bibfield  {author} {\bibinfo {author} {\bibfnamefont {J.}~\bibnamefont
  {Wang}}\ and\ \bibinfo {author} {\bibfnamefont {Y.-Z.}\ \bibnamefont {You}},\
  }\bibfield  {title} {\bibinfo {title} {{Symmetric mass generation}},\ }\href
  {https://doi.org/10.3390/sym14071475} {\bibfield  {journal} {\bibinfo
  {journal} {Symmetry}\ }\textbf {\bibinfo {volume} {14}},\ \bibinfo {pages}
  {1475} (\bibinfo {year} {2022})}\BibitemShut {NoStop}%
\bibitem [{\citenamefont {Jian}\ and\ \citenamefont
  {Zaletel}(2016)}]{jianzaletel2016}%
  \BibitemOpen
  \bibfield  {author} {\bibinfo {author} {\bibfnamefont {C.-M.}\ \bibnamefont
  {Jian}}\ and\ \bibinfo {author} {\bibfnamefont {M.}~\bibnamefont {Zaletel}},\
  }\bibfield  {title} {\bibinfo {title} {{Existence of featureless paramagnets
  on the square and the honeycomb lattices in $2+1$ dimensions}},\ }\href
  {https://doi.org/10.1103/PhysRevB.93.035114} {\bibfield  {journal} {\bibinfo
  {journal} {Phys. Rev. B}\ }\textbf {\bibinfo {volume} {93}},\ \bibinfo
  {pages} {035114} (\bibinfo {year} {2016})}\BibitemShut {NoStop}%
\bibitem [{\citenamefont {Tasaki}(2026{\natexlab{a}})}]{tasaki2026notes}%
  \BibitemOpen
  \bibfield  {author} {\bibinfo {author} {\bibfnamefont {H.}~\bibnamefont
  {Tasaki}},\ }\href
  {https://haltasaki.github.io/writings/AI/OpenAI_Haldane.html} {\bibinfo
  {title} {{OpenAI's proof of the Haldane gap in the $S=1$ antiferromagnetic
  Heisenberg chain}}} (\bibinfo {year} {2026}{\natexlab{a}}),\ \bibinfo {note}
  {expository notes, October}\BibitemShut {NoStop}%
\bibitem [{\citenamefont {Tasaki}(2026{\natexlab{b}})}]{tasaki2026uniqueness}%
  \BibitemOpen
  \bibfield  {author} {\bibinfo {author} {\bibfnamefont {H.}~\bibnamefont
  {Tasaki}},\ }\href@noop {} {\bibinfo {title} {{On the uniqueness of
  translation-invariant ground states of the $S=1$ Heisenberg and $J$--$D$
  antiferromagnetic chains}}} (\bibinfo {year} {2026}{\natexlab{b}}),\ \Eprint
  {https://arxiv.org/abs/2610.34330v2} {arXiv:2610.34330v2} \BibitemShut
  {NoStop}%
\bibitem [{\citenamefont {Feynman}(1948)}]{feynman1948}%
  \BibitemOpen
  \bibfield  {author} {\bibinfo {author} {\bibfnamefont {R.~P.}\ \bibnamefont
  {Feynman}},\ }\bibfield  {title} {\bibinfo {title} {{Space-Time Approach to
  Non-Relativistic Quantum Mechanics}},\ }\href
  {https://doi.org/10.1103/RevModPhys.20.367} {\bibfield  {journal} {\bibinfo
  {journal} {Rev. Mod. Phys.}\ }\textbf {\bibinfo {volume} {20}},\ \bibinfo
  {pages} {367} (\bibinfo {year} {1948})}\BibitemShut {NoStop}%
\bibitem [{\citenamefont {Kac}(1949)}]{kac1949}%
  \BibitemOpen
  \bibfield  {author} {\bibinfo {author} {\bibfnamefont {M.}~\bibnamefont
  {Kac}},\ }\bibfield  {title} {\bibinfo {title} {{On distributions of certain
  Wiener functionals}},\ }\href
  {https://doi.org/10.1090/S0002-9947-1949-0027960-X} {\bibfield  {journal}
  {\bibinfo  {journal} {Trans. Amer. Math. Soc.}\ }\textbf {\bibinfo {volume}
  {65}},\ \bibinfo {pages} {1} (\bibinfo {year} {1949})}\BibitemShut {NoStop}%
\bibitem [{\citenamefont {Ginibre}(1968)}]{ginibre1968}%
  \BibitemOpen
  \bibfield  {author} {\bibinfo {author} {\bibfnamefont {J.}~\bibnamefont
  {Ginibre}},\ }\bibfield  {title} {\bibinfo {title} {{Reduced density matrices
  of the anisotropic Heisenberg model}},\ }\href
  {https://doi.org/10.1007/BF01654238} {\bibfield  {journal} {\bibinfo
  {journal} {Commun. Math. Phys.}\ }\textbf {\bibinfo {volume} {10}},\ \bibinfo
  {pages} {140} (\bibinfo {year} {1968})}\BibitemShut {NoStop}%
\bibitem [{\citenamefont {Suzuki}(1985)}]{suzuki1985}%
  \BibitemOpen
  \bibfield  {author} {\bibinfo {author} {\bibfnamefont {M.}~\bibnamefont
  {Suzuki}},\ }\bibfield  {title} {\bibinfo {title} {{Transfer-matrix method
  and Monte Carlo simulation in quantum spin systems}},\ }\href
  {https://doi.org/10.1103/PhysRevB.31.2957} {\bibfield  {journal} {\bibinfo
  {journal} {Phys. Rev. B}\ }\textbf {\bibinfo {volume} {31}},\ \bibinfo
  {pages} {2957} (\bibinfo {year} {1985})}\BibitemShut {NoStop}%
\bibitem [{\citenamefont {Betsuyaku}(1985)}]{betsuyaku1985}%
  \BibitemOpen
  \bibfield  {author} {\bibinfo {author} {\bibfnamefont {H.}~\bibnamefont
  {Betsuyaku}},\ }\bibfield  {title} {\bibinfo {title} {{Study of
  One-Dimensional Quantum Spin Systems by the Transfer-Matrix Method}},\ }\href
  {https://doi.org/10.1143/PTP.73.319} {\bibfield  {journal} {\bibinfo
  {journal} {Prog. Theor. Phys.}\ }\textbf {\bibinfo {volume} {73}},\ \bibinfo
  {pages} {319} (\bibinfo {year} {1985})}\BibitemShut {NoStop}%
\bibitem [{\citenamefont {Bertini}\ \emph {et~al.}(2018)\citenamefont
  {Bertini}, \citenamefont {Kos},\ and\ \citenamefont
  {Prosen}}]{bertinikosprosen2018}%
  \BibitemOpen
  \bibfield  {author} {\bibinfo {author} {\bibfnamefont {B.}~\bibnamefont
  {Bertini}}, \bibinfo {author} {\bibfnamefont {P.}~\bibnamefont {Kos}},\ and\
  \bibinfo {author} {\bibfnamefont {T.}~\bibnamefont {Prosen}},\ }\bibfield
  {title} {\bibinfo {title} {{Exact Spectral Form Factor in a Minimal Model of
  Many-Body Quantum Chaos}},\ }\href
  {https://doi.org/10.1103/PhysRevLett.121.264101} {\bibfield  {journal}
  {\bibinfo  {journal} {Phys. Rev. Lett.}\ }\textbf {\bibinfo {volume} {121}},\
  \bibinfo {pages} {264101} (\bibinfo {year} {2018})}\BibitemShut {NoStop}%
\bibitem [{\citenamefont {Lu}\ and\ \citenamefont
  {Grover}(2021)}]{lugrover2021}%
  \BibitemOpen
  \bibfield  {author} {\bibinfo {author} {\bibfnamefont {T.-C.}\ \bibnamefont
  {Lu}}\ and\ \bibinfo {author} {\bibfnamefont {T.}~\bibnamefont {Grover}},\
  }\bibfield  {title} {\bibinfo {title} {{Spacetime Duality between
  Localization Transitions and Measurement-Induced Transitions}},\ }\href
  {https://doi.org/10.1103/PRXQuantum.2.040319} {\bibfield  {journal} {\bibinfo
   {journal} {PRX Quantum}\ }\textbf {\bibinfo {volume} {2}},\ \bibinfo {pages}
  {040319} (\bibinfo {year} {2021})}\BibitemShut {NoStop}%
\bibitem [{\citenamefont {Ippoliti}\ \emph {et~al.}(2022)\citenamefont
  {Ippoliti}, \citenamefont {Rakovszky},\ and\ \citenamefont
  {Khemani}}]{ippolitirakovszkykhemani2022}%
  \BibitemOpen
  \bibfield  {author} {\bibinfo {author} {\bibfnamefont {M.}~\bibnamefont
  {Ippoliti}}, \bibinfo {author} {\bibfnamefont {T.}~\bibnamefont
  {Rakovszky}},\ and\ \bibinfo {author} {\bibfnamefont {V.}~\bibnamefont
  {Khemani}},\ }\bibfield  {title} {\bibinfo {title} {{Fractal, Logarithmic,
  and Volume-Law Entangled Nonthermal Steady States via Spacetime Duality}},\
  }\href {https://doi.org/10.1103/PhysRevX.12.011045} {\bibfield  {journal}
  {\bibinfo  {journal} {Phys. Rev. X}\ }\textbf {\bibinfo {volume} {12}},\
  \bibinfo {pages} {011045} (\bibinfo {year} {2022})}\BibitemShut {NoStop}%
\bibitem [{\citenamefont {Sandvik}\ and\ \citenamefont
  {Scalapino}(1994)}]{sandvikscalapino1994}%
  \BibitemOpen
  \bibfield  {author} {\bibinfo {author} {\bibfnamefont {A.~W.}\ \bibnamefont
  {Sandvik}}\ and\ \bibinfo {author} {\bibfnamefont {D.~J.}\ \bibnamefont
  {Scalapino}},\ }\bibfield  {title} {\bibinfo {title} {{Order-disorder
  transition in a two-layer quantum antiferromagnet}},\ }\href
  {https://doi.org/10.1103/PhysRevLett.72.2777} {\bibfield  {journal} {\bibinfo
   {journal} {Phys. Rev. Lett.}\ }\textbf {\bibinfo {volume} {72}},\ \bibinfo
  {pages} {2777} (\bibinfo {year} {1994})}\BibitemShut {NoStop}%
\bibitem [{\citenamefont {Cardy}(1986)}]{cardy1986boundary}%
  \BibitemOpen
  \bibfield  {author} {\bibinfo {author} {\bibfnamefont {J.~L.}\ \bibnamefont
  {Cardy}},\ }\bibfield  {title} {\bibinfo {title} {{Effect of boundary
  conditions on the operator content of two-dimensional conformally invariant
  theories}},\ }\href {https://doi.org/10.1016/0550-3213(86)90596-1} {\bibfield
   {journal} {\bibinfo  {journal} {Nucl. Phys. B}\ }\textbf {\bibinfo {volume}
  {275}},\ \bibinfo {pages} {200} (\bibinfo {year} {1986})}\BibitemShut
  {NoStop}%
\bibitem [{\citenamefont {Zuber}(1986)}]{zuber1986discrete}%
  \BibitemOpen
  \bibfield  {author} {\bibinfo {author} {\bibfnamefont {J.-B.}\ \bibnamefont
  {Zuber}},\ }\bibfield  {title} {\bibinfo {title} {{Discrete symmetries of
  conformal theories}},\ }\href {https://doi.org/10.1016/0370-2693(86)90936-6}
  {\bibfield  {journal} {\bibinfo  {journal} {Phys. Lett. B}\ }\textbf
  {\bibinfo {volume} {176}},\ \bibinfo {pages} {127} (\bibinfo {year}
  {1986})}\BibitemShut {NoStop}%
\bibitem [{\citenamefont {Shao}(2023)}]{shao2023tasi}%
  \BibitemOpen
  \bibfield  {author} {\bibinfo {author} {\bibfnamefont {S.-H.}\ \bibnamefont
  {Shao}},\ }\href@noop {} {\bibinfo {title} {{What's Done Cannot Be Undone:
  TASI Lectures on Non-Invertible Symmetries}}} (\bibinfo {year} {2023}),\
  \Eprint {https://arxiv.org/abs/2308.00747} {arXiv:2308.00747} \BibitemShut
  {NoStop}%
\bibitem [{\citenamefont {Koma}\ and\ \citenamefont
  {Tasaki}(1994)}]{komatasaki1994}%
  \BibitemOpen
  \bibfield  {author} {\bibinfo {author} {\bibfnamefont {T.}~\bibnamefont
  {Koma}}\ and\ \bibinfo {author} {\bibfnamefont {H.}~\bibnamefont {Tasaki}},\
  }\bibfield  {title} {\bibinfo {title} {{Symmetry breaking and finite-size
  effects in quantum many-body systems}},\ }\href@noop {} {\bibfield  {journal}
  {\bibinfo  {journal} {J. Stat. Phys.}\ }\textbf {\bibinfo {volume} {76}},\
  \bibinfo {pages} {745} (\bibinfo {year} {1994})},\ \Eprint
  {https://arxiv.org/abs/cond-mat/9708132} {arXiv:cond-mat/9708132}
  \BibitemShut {NoStop}%
\bibitem [{\citenamefont {Kitaev}(2003)}]{kitaev2003}%
  \BibitemOpen
  \bibfield  {author} {\bibinfo {author} {\bibfnamefont {A.~{\relax Yu}.}\
  \bibnamefont {Kitaev}},\ }\bibfield  {title} {\bibinfo {title}
  {{Fault-tolerant quantum computation by anyons}},\ }\href
  {https://doi.org/10.1016/S0003-4916(02)00018-0} {\bibfield  {journal}
  {\bibinfo  {journal} {Ann. Phys.}\ }\textbf {\bibinfo {volume} {303}},\
  \bibinfo {pages} {2} (\bibinfo {year} {2003})}\BibitemShut {NoStop}%
\bibitem [{\citenamefont {Hou}\ and\ \citenamefont {You}(2023)}]{houyou2023}%
  \BibitemOpen
  \bibfield  {author} {\bibinfo {author} {\bibfnamefont {W.}~\bibnamefont
  {Hou}}\ and\ \bibinfo {author} {\bibfnamefont {Y.-Z.}\ \bibnamefont {You}},\
  }\bibfield  {title} {\bibinfo {title} {{Variational Monte Carlo study of
  symmetric mass generation in a bilayer honeycomb lattice model}},\ }\href
  {https://doi.org/10.1103/PhysRevB.108.125130} {\bibfield  {journal} {\bibinfo
   {journal} {Phys. Rev. B}\ }\textbf {\bibinfo {volume} {108}},\ \bibinfo
  {pages} {125130} (\bibinfo {year} {2023})}\BibitemShut {NoStop}%
\bibitem [{\citenamefont {He}\ \emph {et~al.}(2026)\citenamefont {He},
  \citenamefont {You},\ and\ \citenamefont {Xu}}]{heyouxu2026}%
  \BibitemOpen
  \bibfield  {author} {\bibinfo {author} {\bibfnamefont {C.-H.}\ \bibnamefont
  {He}}, \bibinfo {author} {\bibfnamefont {Y.-Z.}\ \bibnamefont {You}},\ and\
  \bibinfo {author} {\bibfnamefont {X.~Y.}\ \bibnamefont {Xu}},\ }\href@noop {}
  {\bibinfo {title} {{Symmetric mass generation in a bilayer honeycomb lattice
  with $\mathrm{SU}(2)\times\mathrm{SU}(2)\times\mathrm{SU}(2)/\mathbb Z_2$
  symmetry}}} (\bibinfo {year} {2026}),\ \Eprint
  {https://arxiv.org/abs/2603.18278} {arXiv:2603.18278} \BibitemShut {NoStop}%
\bibitem [{\citenamefont {Verstraete}\ \emph {et~al.}(2006)\citenamefont
  {Verstraete}, \citenamefont {Wolf}, \citenamefont {P{\'e}rez-Garc{\'i}a},\
  and\ \citenamefont {Cirac}}]{verstraete2006criticality}%
  \BibitemOpen
  \bibfield  {author} {\bibinfo {author} {\bibfnamefont {F.}~\bibnamefont
  {Verstraete}}, \bibinfo {author} {\bibfnamefont {M.~M.}\ \bibnamefont
  {Wolf}}, \bibinfo {author} {\bibfnamefont {D.}~\bibnamefont
  {P{\'e}rez-Garc{\'i}a}},\ and\ \bibinfo {author} {\bibfnamefont {J.~I.}\
  \bibnamefont {Cirac}},\ }\bibfield  {title} {\bibinfo {title} {{Criticality,
  the area law, and the computational power of projected entangled pair
  states}},\ }\href {https://doi.org/10.1103/PhysRevLett.96.220601} {\bibfield
  {journal} {\bibinfo  {journal} {Phys. Rev. Lett.}\ }\textbf {\bibinfo
  {volume} {96}},\ \bibinfo {pages} {220601} (\bibinfo {year}
  {2006})}\BibitemShut {NoStop}%
\bibitem [{\citenamefont {Lieb}\ \emph {et~al.}(1961)\citenamefont {Lieb},
  \citenamefont {Schultz},\ and\ \citenamefont
  {Mattis}}]{liebschultzmattis1961}%
  \BibitemOpen
  \bibfield  {author} {\bibinfo {author} {\bibfnamefont {E.}~\bibnamefont
  {Lieb}}, \bibinfo {author} {\bibfnamefont {T.}~\bibnamefont {Schultz}},\ and\
  \bibinfo {author} {\bibfnamefont {D.}~\bibnamefont {Mattis}},\ }\bibfield
  {title} {\bibinfo {title} {{Two soluble models of an antiferromagnetic
  chain}},\ }\href {https://doi.org/10.1016/0003-4916(61)90115-4} {\bibfield
  {journal} {\bibinfo  {journal} {Ann. Phys.}\ }\textbf {\bibinfo {volume}
  {16}},\ \bibinfo {pages} {407} (\bibinfo {year} {1961})}\BibitemShut
  {NoStop}%
\bibitem [{\citenamefont {Oshikawa}(2000)}]{oshikawa2000}%
  \BibitemOpen
  \bibfield  {author} {\bibinfo {author} {\bibfnamefont {M.}~\bibnamefont
  {Oshikawa}},\ }\bibfield  {title} {\bibinfo {title} {{Commensurability,
  excitation gap, and topology in quantum many-particle systems on a periodic
  lattice}},\ }\href {https://doi.org/10.1103/PhysRevLett.84.1535} {\bibfield
  {journal} {\bibinfo  {journal} {Phys. Rev. Lett.}\ }\textbf {\bibinfo
  {volume} {84}},\ \bibinfo {pages} {1535} (\bibinfo {year}
  {2000})}\BibitemShut {NoStop}%
\bibitem [{\citenamefont {Hastings}(2004)}]{hastings2004}%
  \BibitemOpen
  \bibfield  {author} {\bibinfo {author} {\bibfnamefont {M.~B.}\ \bibnamefont
  {Hastings}},\ }\bibfield  {title} {\bibinfo {title} {{Lieb--Schultz--Mattis
  in higher dimensions}},\ }\href {https://doi.org/10.1103/PhysRevB.69.104431}
  {\bibfield  {journal} {\bibinfo  {journal} {Phys. Rev. B}\ }\textbf {\bibinfo
  {volume} {69}},\ \bibinfo {pages} {104431} (\bibinfo {year}
  {2004})}\BibitemShut {NoStop}%
\bibitem [{\citenamefont {Kimchi}\ \emph {et~al.}(2013)\citenamefont {Kimchi},
  \citenamefont {Parameswaran}, \citenamefont {Turner}, \citenamefont {Wang},\
  and\ \citenamefont {Vishwanath}}]{kimchi2013}%
  \BibitemOpen
  \bibfield  {author} {\bibinfo {author} {\bibfnamefont {I.}~\bibnamefont
  {Kimchi}}, \bibinfo {author} {\bibfnamefont {S.~A.}\ \bibnamefont
  {Parameswaran}}, \bibinfo {author} {\bibfnamefont {A.~M.}\ \bibnamefont
  {Turner}}, \bibinfo {author} {\bibfnamefont {F.}~\bibnamefont {Wang}},\ and\
  \bibinfo {author} {\bibfnamefont {A.}~\bibnamefont {Vishwanath}},\ }\bibfield
   {title} {\bibinfo {title} {{Featureless and nonfractionalized Mott
  insulators on the honeycomb lattice at $1/2$ site filling}},\ }\href
  {https://doi.org/10.1073/pnas.1307245110} {\bibfield  {journal} {\bibinfo
  {journal} {Proc. Natl. Acad. Sci. U.S.A.}\ }\textbf {\bibinfo {volume}
  {110}},\ \bibinfo {pages} {16378} (\bibinfo {year} {2013})}\BibitemShut
  {NoStop}%
\bibitem [{\citenamefont {Parameswaran}\ \emph {et~al.}(2013)\citenamefont
  {Parameswaran}, \citenamefont {Kimchi}, \citenamefont {Turner}, \citenamefont
  {Stamper-Kurn},\ and\ \citenamefont {Vishwanath}}]{parameswaran2013}%
  \BibitemOpen
  \bibfield  {author} {\bibinfo {author} {\bibfnamefont {S.~A.}\ \bibnamefont
  {Parameswaran}}, \bibinfo {author} {\bibfnamefont {I.}~\bibnamefont
  {Kimchi}}, \bibinfo {author} {\bibfnamefont {A.~M.}\ \bibnamefont {Turner}},
  \bibinfo {author} {\bibfnamefont {D.~M.}\ \bibnamefont {Stamper-Kurn}},\ and\
  \bibinfo {author} {\bibfnamefont {A.}~\bibnamefont {Vishwanath}},\ }\bibfield
   {title} {\bibinfo {title} {{Wannier permanent wave functions for featureless
  bosonic Mott insulators on the $1/3$-filled kagome lattice}},\ }\href
  {https://doi.org/10.1103/PhysRevLett.110.125301} {\bibfield  {journal}
  {\bibinfo  {journal} {Phys. Rev. Lett.}\ }\textbf {\bibinfo {volume} {110}},\
  \bibinfo {pages} {125301} (\bibinfo {year} {2013})}\BibitemShut {NoStop}%
\bibitem [{\citenamefont {Kim}\ \emph {et~al.}(2016)\citenamefont {Kim},
  \citenamefont {Lee}, \citenamefont {Jiang}, \citenamefont {Ware},
  \citenamefont {Jian}, \citenamefont {Zaletel}, \citenamefont {Han},\ and\
  \citenamefont {Ran}}]{kim2016}%
  \BibitemOpen
  \bibfield  {author} {\bibinfo {author} {\bibfnamefont {P.}~\bibnamefont
  {Kim}}, \bibinfo {author} {\bibfnamefont {H.}~\bibnamefont {Lee}}, \bibinfo
  {author} {\bibfnamefont {S.}~\bibnamefont {Jiang}}, \bibinfo {author}
  {\bibfnamefont {B.}~\bibnamefont {Ware}}, \bibinfo {author} {\bibfnamefont
  {C.-M.}\ \bibnamefont {Jian}}, \bibinfo {author} {\bibfnamefont
  {M.}~\bibnamefont {Zaletel}}, \bibinfo {author} {\bibfnamefont {J.~H.}\
  \bibnamefont {Han}},\ and\ \bibinfo {author} {\bibfnamefont {Y.}~\bibnamefont
  {Ran}},\ }\bibfield  {title} {\bibinfo {title} {{Featureless quantum
  insulator on the honeycomb lattice}},\ }\href
  {https://doi.org/10.1103/PhysRevB.94.064432} {\bibfield  {journal} {\bibinfo
  {journal} {Phys. Rev. B}\ }\textbf {\bibinfo {volume} {94}},\ \bibinfo
  {pages} {064432} (\bibinfo {year} {2016})}\BibitemShut {NoStop}%
\bibitem [{\citenamefont {Affleck}\ \emph {et~al.}(1987)\citenamefont
  {Affleck}, \citenamefont {Kennedy}, \citenamefont {Lieb},\ and\ \citenamefont
  {Tasaki}}]{aklt1987}%
  \BibitemOpen
  \bibfield  {author} {\bibinfo {author} {\bibfnamefont {I.}~\bibnamefont
  {Affleck}}, \bibinfo {author} {\bibfnamefont {T.}~\bibnamefont {Kennedy}},
  \bibinfo {author} {\bibfnamefont {E.~H.}\ \bibnamefont {Lieb}},\ and\
  \bibinfo {author} {\bibfnamefont {H.}~\bibnamefont {Tasaki}},\ }\bibfield
  {title} {\bibinfo {title} {{Rigorous results on valence-bond ground states in
  antiferromagnets}},\ }\href {https://doi.org/10.1103/PhysRevLett.59.799}
  {\bibfield  {journal} {\bibinfo  {journal} {Phys. Rev. Lett.}\ }\textbf
  {\bibinfo {volume} {59}},\ \bibinfo {pages} {799} (\bibinfo {year}
  {1987})}\BibitemShut {NoStop}%
\bibitem [{\citenamefont {Shankar}(2017)}]{shankar2017}%
  \BibitemOpen
  \bibfield  {author} {\bibinfo {author} {\bibfnamefont {R.}~\bibnamefont
  {Shankar}},\ }\href {https://doi.org/10.1017/9781139044349} {\emph {\bibinfo
  {title} {{Quantum Field Theory and Condensed Matter: An Introduction}}}}\
  (\bibinfo  {publisher} {Cambridge University Press},\ \bibinfo {year}
  {2017})\ \bibinfo {note} {sec.~6.10}\BibitemShut {NoStop}%
\bibitem [{\citenamefont {Akiyama}\ and\ \citenamefont
  {Kadoh}(2021)}]{akiyamakadoh2021}%
  \BibitemOpen
  \bibfield  {author} {\bibinfo {author} {\bibfnamefont {S.}~\bibnamefont
  {Akiyama}}\ and\ \bibinfo {author} {\bibfnamefont {D.}~\bibnamefont
  {Kadoh}},\ }\bibfield  {title} {\bibinfo {title} {{More about the Grassmann
  tensor renormalization group}},\ }\href@noop {} {\bibfield  {journal}
  {\bibinfo  {journal} {JHEP}\ }\textbf {\bibinfo {volume} {2021}}\bibfield
  {number} {\bibinfo  {number} { (10)},\ \bibinfo {pages} {188}},\ }\Eprint
  {https://arxiv.org/abs/2005.07570} {arXiv:2005.07570} \BibitemShut {NoStop}%
\bibitem [{\citenamefont {Jaffe}\ and\ \citenamefont
  {Janssens}(2016)}]{jaffejanssens2016}%
  \BibitemOpen
  \bibfield  {author} {\bibinfo {author} {\bibfnamefont {A.}~\bibnamefont
  {Jaffe}}\ and\ \bibinfo {author} {\bibfnamefont {B.}~\bibnamefont
  {Janssens}},\ }\bibfield  {title} {\bibinfo {title} {{Characterization of
  Reflection Positivity: Majoranas and Spins}},\ }\href@noop {} {\bibfield
  {journal} {\bibinfo  {journal} {Commun. Math. Phys.}\ }\textbf {\bibinfo
  {volume} {346}},\ \bibinfo {pages} {1021} (\bibinfo {year} {2016})},\ \Eprint
  {https://arxiv.org/abs/1506.04197} {arXiv:1506.04197} \BibitemShut {NoStop}%
\bibitem [{\citenamefont {Temple}(1928)}]{temple1928}%
  \BibitemOpen
  \bibfield  {author} {\bibinfo {author} {\bibfnamefont {G.}~\bibnamefont
  {Temple}},\ }\bibfield  {title} {\bibinfo {title} {{The theory of Rayleigh's
  principle as applied to continuous systems}},\ }\href
  {https://doi.org/10.1098/rspa.1928.0098} {\bibfield  {journal} {\bibinfo
  {journal} {Proc. R. Soc. Lond. A}\ }\textbf {\bibinfo {volume} {119}},\
  \bibinfo {pages} {276} (\bibinfo {year} {1928})}\BibitemShut {NoStop}%
\bibitem [{\citenamefont {Vidal}\ \emph {et~al.}(2009)\citenamefont {Vidal},
  \citenamefont {Dusuel},\ and\ \citenamefont {Schmidt}}]{toricbackground}%
  \BibitemOpen
  \bibfield  {author} {\bibinfo {author} {\bibfnamefont {J.}~\bibnamefont
  {Vidal}}, \bibinfo {author} {\bibfnamefont {S.}~\bibnamefont {Dusuel}},\ and\
  \bibinfo {author} {\bibfnamefont {K.~P.}\ \bibnamefont {Schmidt}},\
  }\bibfield  {title} {\bibinfo {title} {{Low-energy effective theory of the
  toric code model in a parallel magnetic field}},\ }\href@noop {} {\bibfield
  {journal} {\bibinfo  {journal} {Phys. Rev. B}\ }\textbf {\bibinfo {volume}
  {79}},\ \bibinfo {pages} {033109} (\bibinfo {year} {2009})},\ \Eprint
  {https://arxiv.org/abs/0807.0487} {arXiv:0807.0487} \BibitemShut {NoStop}%
\bibitem [{\citenamefont {Slagle}\ \emph {et~al.}(2015)\citenamefont {Slagle},
  \citenamefont {You},\ and\ \citenamefont {Xu}}]{slagle2015}%
  \BibitemOpen
  \bibfield  {author} {\bibinfo {author} {\bibfnamefont {K.}~\bibnamefont
  {Slagle}}, \bibinfo {author} {\bibfnamefont {Y.-Z.}\ \bibnamefont {You}},\
  and\ \bibinfo {author} {\bibfnamefont {C.}~\bibnamefont {Xu}},\ }\bibfield
  {title} {\bibinfo {title} {{Exotic quantum phase transitions of strongly
  interacting topological insulators}},\ }\href
  {https://doi.org/10.1103/PhysRevB.91.115121} {\bibfield  {journal} {\bibinfo
  {journal} {Phys. Rev. B}\ }\textbf {\bibinfo {volume} {91}},\ \bibinfo
  {pages} {115121} (\bibinfo {year} {2015})}\BibitemShut {NoStop}%
\bibitem [{\citenamefont {He}\ \emph {et~al.}(2016)\citenamefont {He},
  \citenamefont {Wu}, \citenamefont {You}, \citenamefont {Xu}, \citenamefont
  {Meng},\ and\ \citenamefont {Lu}}]{he2016}%
  \BibitemOpen
  \bibfield  {author} {\bibinfo {author} {\bibfnamefont {Y.-Y.}\ \bibnamefont
  {He}}, \bibinfo {author} {\bibfnamefont {H.-Q.}\ \bibnamefont {Wu}}, \bibinfo
  {author} {\bibfnamefont {Y.-Z.}\ \bibnamefont {You}}, \bibinfo {author}
  {\bibfnamefont {C.}~\bibnamefont {Xu}}, \bibinfo {author} {\bibfnamefont
  {Z.~Y.}\ \bibnamefont {Meng}},\ and\ \bibinfo {author} {\bibfnamefont
  {Z.-Y.}\ \bibnamefont {Lu}},\ }\bibfield  {title} {\bibinfo {title} {{Quantum
  critical point of Dirac fermion mass generation without spontaneous symmetry
  breaking}},\ }\href {https://doi.org/10.1103/PhysRevB.94.241111} {\bibfield
  {journal} {\bibinfo  {journal} {Phys. Rev. B}\ }\textbf {\bibinfo {volume}
  {94}},\ \bibinfo {pages} {241111} (\bibinfo {year} {2016})}\BibitemShut
  {NoStop}%
\bibitem [{\citenamefont {Ayyar}\ and\ \citenamefont
  {Chandrasekharan}(2015)}]{ayyarchandrasekharan2015}%
  \BibitemOpen
  \bibfield  {author} {\bibinfo {author} {\bibfnamefont {V.}~\bibnamefont
  {Ayyar}}\ and\ \bibinfo {author} {\bibfnamefont {S.}~\bibnamefont
  {Chandrasekharan}},\ }\bibfield  {title} {\bibinfo {title} {{Massive fermions
  without fermion bilinear condensates}},\ }\href
  {https://doi.org/10.1103/PhysRevD.91.065035} {\bibfield  {journal} {\bibinfo
  {journal} {Phys. Rev. D}\ }\textbf {\bibinfo {volume} {91}},\ \bibinfo
  {pages} {065035} (\bibinfo {year} {2015})}\BibitemShut {NoStop}%
\bibitem [{\citenamefont {Ayyar}\ and\ \citenamefont
  {Chandrasekharan}(2016)}]{ayyarchandrasekharan2016}%
  \BibitemOpen
  \bibfield  {author} {\bibinfo {author} {\bibfnamefont {V.}~\bibnamefont
  {Ayyar}}\ and\ \bibinfo {author} {\bibfnamefont {S.}~\bibnamefont
  {Chandrasekharan}},\ }\bibfield  {title} {\bibinfo {title} {{Origin of
  fermion masses without spontaneous symmetry breaking}},\ }\href
  {https://doi.org/10.1103/PhysRevD.93.081701} {\bibfield  {journal} {\bibinfo
  {journal} {Phys. Rev. D}\ }\textbf {\bibinfo {volume} {93}},\ \bibinfo
  {pages} {081701} (\bibinfo {year} {2016})}\BibitemShut {NoStop}%
\bibitem [{\citenamefont {Midha}\ \emph {et~al.}(2026)\citenamefont {Midha},
  \citenamefont {Zhang}, \citenamefont {Malz}, \citenamefont {Abanin},\ and\
  \citenamefont {Gopalakrishnan}}]{midha2026locality}%
  \BibitemOpen
  \bibfield  {author} {\bibinfo {author} {\bibfnamefont {S.}~\bibnamefont
  {Midha}}, \bibinfo {author} {\bibfnamefont {Y.~F.}\ \bibnamefont {Zhang}},
  \bibinfo {author} {\bibfnamefont {D.}~\bibnamefont {Malz}}, \bibinfo {author}
  {\bibfnamefont {D.~A.}\ \bibnamefont {Abanin}},\ and\ \bibinfo {author}
  {\bibfnamefont {S.}~\bibnamefont {Gopalakrishnan}},\ }\href@noop {} {\bibinfo
  {title} {{Algorithmic Locality of Quantum Tensor Network Belief
  Propagation}}} (\bibinfo {year} {2026}),\ \Eprint
  {https://arxiv.org/abs/2604.21919} {arXiv:2604.21919} \BibitemShut {NoStop}%
\bibitem [{\citenamefont {P{\'e}rez-Garc{\'i}a}\ \emph
  {et~al.}(2008)\citenamefont {P{\'e}rez-Garc{\'i}a}, \citenamefont
  {Verstraete}, \citenamefont {Cirac},\ and\ \citenamefont
  {Wolf}}]{perezgarcia2008peps}%
  \BibitemOpen
  \bibfield  {author} {\bibinfo {author} {\bibfnamefont {D.}~\bibnamefont
  {P{\'e}rez-Garc{\'i}a}}, \bibinfo {author} {\bibfnamefont {F.}~\bibnamefont
  {Verstraete}}, \bibinfo {author} {\bibfnamefont {J.~I.}\ \bibnamefont
  {Cirac}},\ and\ \bibinfo {author} {\bibfnamefont {M.~M.}\ \bibnamefont
  {Wolf}},\ }\bibfield  {title} {\bibinfo {title} {{PEPS as unique ground
  states of local Hamiltonians}},\ }\href@noop {} {\bibfield  {journal}
  {\bibinfo  {journal} {Quantum Inf. Comput.}\ }\textbf {\bibinfo {volume}
  {8}},\ \bibinfo {pages} {650} (\bibinfo {year} {2008})},\ \Eprint
  {https://arxiv.org/abs/0707.2260} {arXiv:0707.2260} \BibitemShut {NoStop}%
\end{thebibliography}%
\end{document}